\documentclass[ amsfonts, amssymb, amsmath, superscriptaddress, nofootinbib,onecolumn]{revtex4-2}
\usepackage{array}[=2016-10-06]
\usepackage[english]{babel}
\usepackage[utf8]{inputenc}
\usepackage[T1]{fontenc}
\usepackage[pdftex, pdftitle={Article},colorlinks=true, allcolors=blue ]{hyperref} % For hyperlinks in the PDF
\usepackage{pgfplots}
\pgfplotsset{compat=1.18}
\usepackage{graphicx}% Include figure files
\AtBeginDocument{\pagenumbering{arabic}}
\usepackage{quantikz}
\usepackage{bm}% bold math
\usepackage{booktabs}
\usepackage{multirow}
\usepackage{siunitx}

\usepackage{amsmath,amssymb}
\usepackage{braket}
\usepackage[italicdiff]{physics}
\usepackage[toc,page,title]{appendix}
\usepackage{subcaption}
\usepackage{mwe}
\usepackage{tikz}
\usepackage{adjustbox}
\usetikzlibrary{arrows, hobby}
\usepackage{wrapfig}
\usepackage{ mathrsfs }
\usepackage{indentfirst}
\usepackage{color}
\usepackage{empheq}
\usetikzlibrary{shapes}
\usetikzlibrary{plotmarks}
\usepackage{mathptmx}
\usetikzlibrary{arrows, hobby}
\usepackage{amsthm}
\newtheorem{lemma}{Lemma}
\newtheorem{definition}{Definition}
\newtheorem{theorem}{Theorem}

\usetikzlibrary{shapes}
\usetikzlibrary{plotmarks}

    \theoremstyle{definition}
    \newtheorem{assumption}{Assumption}

    \theoremstyle{plain}

    \newtheorem{corollary}{Corollary}
    
    \theoremstyle{remark}
    \newtheorem{remark}{Remark}
\usetikzlibrary{backgrounds,fit,decorations.pathreplacing}  % TikZ libraries

\tikzset{pics/.cd,
collector/.style={code={
\draw[fill=gray!20] (0,0.5) arc(90:-90:0.75cm and 0.5cm) -- cycle;}},
splitter/.style={code={\draw[ultra thick] (#1:{sqrt(1/2)}) --
(#1+180:{sqrt(1/2)});}},splitter/.default=135}

\tikzstyle{beamsplitter}=[fill=blue, fill opacity=0.2]

\usetikzlibrary{positioning,fit,arrows.meta}

\usetikzlibrary{arrows.meta, positioning, calc, fit}
\tikzset{
  box/.style = {draw, rounded corners, inner sep=6pt, outer sep=0pt, line width=0.6pt, fill=white},
  ghost/.style = {inner sep=0pt, outer sep=0pt},
  arrow/.style = {-{Latex[length=2.2mm,width=1.6mm]}, line width=0.6pt},
  thinarrow/.style = {-{Latex[length=2mm,width=1.2mm]}, line width=0.5pt},
  note/.style = {font=\footnotesize, inner sep=1pt, outer sep=0pt},
  eqnote/.style = {font=\footnotesize\itshape, inner sep=1pt, outer sep=0pt},
  every picture/.style = {line cap=round, line join=round}
}
\newcommand{\HKDF}{\mathrm{HKDF}}

\makeatletter
\newcommand{\romannum}[1]{\romannumeral #1}
\DeclareRobustCommand{\Romannum}[1]{\MakeUppercase{\romannum{#1}}}
\makeatother
\begin{document}

\author{S. P. Kish}\email{sebastian.kish@uqconnect.edu.au}
\affiliation{
Centre for Quantum Software and Information, University of Technology Sydney, Sydney, New South Wales 2007,
Australia}
 \affiliation{Data61, CSIRO, Marsfield, NSW, Australia.}
\author{H. J. Vallury}
 \affiliation{Data61, CSIRO, Marsfield, NSW, Australia.}
\author{J. Pieprzyk}
\affiliation{Institute of Computer Science, Polish Academy of Sciences, Warsaw, Poland.}
 \affiliation{Data61, CSIRO, Marsfield, NSW, Australia.}
 
\author{C. Thapa}
 \affiliation{Data61, CSIRO, Marsfield, NSW, Australia.}
\author{S. Camtepe}
 \affiliation{Data61, CSIRO, Marsfield, NSW, Australia.}
%\altaffiliation[Also at ]{School of Physical and Mathematical Sciences, Nanyang Technological University, Singapore 639673, Republic of Singapore.}

%\affil[2]{Affiliation, department, city, postcode, country}

%\affil[+]{these authors contributed equally to this work}

%\keywords{Keyword1, Keyword2, Keyword3}

\title{Quantum Spectral Authentication: Entity Authentication and Key Derivation from a Hidden Eigenstate of a Public Unitary Challenge}
% QSA abstract --- 199 words, within the Scientific Reports 200-word cap.
% RevTeX form (matches current manuscript). For sn-article.cls, replace the
% environment with \abstract{ ... }.

% QSA abstract --- 199 words, within the Scientific Reports 200-word cap.
% RevTeX form (matches current manuscript). For sn-article.cls, replace the
% environment with \abstract{ ... }.

\begin{abstract}
We introduce Quantum Spectral Authentication (QSA), a symmetric-key
entity-authentication and key-derivation protocol in which a remote endpoint
proves it still holds a hidden planted state, an eigenstate of the challenge
it can prepare, without revealing it. Each round issues a
fresh public unitary challenge with its own planted state, and the endpoint
returns an eigenphase feature from which both parties derive transcript-bound
session material. QSA runs in two provisioning
modes with materially different security: a \emph{seed mode}, storing a
classical seed that regenerates the planted state, and a \emph{state mode},
holding only the physical register carrying it, non-copyable and consumed on
use. We give a formal security model in which QSA is an identification
scheme, prove mutual authentication under a single label-hiding assumption,
and prove unconditionally that the published spectrum is near-uniform at the
readout resolution, leaving at least $m-1$ bits of min-entropy per instance;
state mode satisfies the assumption by construction. We develop a symmetric
verifier-driven compiler compatible with low-depth quantum phase estimation,
which also admits a fast path that recovers the label by an
inverse-compiler measurement. Simulations show it is more
noise-tolerant than an asymmetric alternative, and experiments on IBM
\texttt{ibm\_fez} provide a hardware sanity check.
\end{abstract}

\maketitle

\section{Introduction}
Deployments of networked quantum modules, including cloud-accessed quantum processing units (QPUs) and entanglement-enabled links, face a distinct systems problem: provisioning a remote device with an inherently quantum credential and then authenticating that it is actually held. The credential may be a planted state or a compact state-preparation capability, such as a planted-state circuit or seed. In many settings, this quantum credential is delivered by quantum communication, for example by teleporting a state using pre-shared entanglement~\cite{PhysRevLett.70.1895} or by distributing and transporting entanglement resources within quantum-network testbeds. What is missing is a lightweight, application-facing mechanism that converts ``the right quantum credential is present at the endpoint'' into an authentication token without revealing the state description to the network. This question is no longer purely hypothetical. Teleportation and entanglement distribution have already been demonstrated over installed urban fibre links, including dark-fibre and coexistence settings with conventional traffic~\cite{kucera2024teleportation,thomas2024teleportationCoexisting}. In parallel, quantum-network testbeds increasingly target system-integration questions such as control, metadata, orchestration, and interoperability~\cite{monga2023quantnet}. These trends shift the security question from only ``can we distribute quantum states?'' to ``can we authenticate that a particular quantum credential is present and usable at a remote endpoint after provisioning, using a lightweight control-plane token rather than tomography or full computation verification?''

Existing approaches address related but different goals. QKD establishes correlated secret material between endpoints, but it does not answer the post-provisioning question studied here: whether a particular quantum credential, such as a planted state or secret state-preparation capability, is actually present and operational at a designated remote endpoint after commissioning or quantum delivery~\cite{BB84,bozzio2024beyondqkd}. Quantum message-authentication and signature protocols protect communicated data~\cite{barnum2002authqmsg,portmann2017authrecycle,amiri2016qdsreview}, while quantum identity and entity-authentication protocols certify the communicating party or its shared authentication resource~\cite{dutta2021qiareview}. Challenge-response schemes based on physical unclonable functions or quantum-readout optical keys authenticate a hard-to-clone object through its characteristic response~\cite{goorden2014qsa,nikolopoulos2017cvpuk}. At the other end of the spectrum, verifiable delegated quantum computation and proofs of quantumness provide stronger evidence that a prover executed a genuinely quantum process~\cite{mahadev2018,brakerski2018poq}. None of these directly targets the narrower systems task considered here: validating that a previously provisioned hidden quantum credential is still present at the endpoint and can be turned into fresh authentication material without exposing its description.

This paper addresses that missing primitive. We introduce \emph{Quantum Spectral Authentication} (QSA), a spectral challenge-response mechanism that converts possession of a hidden planted quantum credential into fresh session material under public unitary challenges. The verifier supplies, or both parties deterministically reconstruct, fresh public instances of $k$ unitaries. Only a device provisioned with the same hidden resource can efficiently reproduce the corresponding spectral feature response and thereby derive the correct session material, after which explicit confirmation yields an application-level authentication token. The public challenges are chosen so that, across independently seeded instances, their eigenbases are expected to appear generic and decorrelated. Unitary designs, local random circuits, and related pseudorandom unitary constructions provide intuition for how efficiently specified circuits can approximate Haar-like behaviour for broad classes of tests~\cite{dankert2009designs,brandao2016localdesigns,haferkamp2022designs}, thereby limiting adversarial leverage from propagating approximate eigenstate information between distinct challenges. In this sense, QSA is a lightweight control-plane primitive: the unitaries are public, the secret is the state-preparation description, and the output is conventional symmetric key material consumable by higher-level protocols.

Concretely, QSA publishes a small family of $n$-qubit unitaries $U_1,\ldots,U_k$. Honest parties share a planted state-preparation circuit $P$ defining a planted state $\ket{\psi}=P^\dagger\ket{0^n}$, or a schedule of planted states derived from a provisioned seed. They compute a short eigenphase feature vector $\boldsymbol{\Theta}$ by applying quantum phase estimation (QPE) routines to $\{U_i\}$ on $\ket{\psi}$, and then feed $\boldsymbol{\Theta}$ into a standard key derivation function (KDF) to obtain symmetric session material. An adversary sees the full public circuit descriptions of $U_1,\ldots,U_k$ and may run arbitrary classical or quantum algorithms on these unitaries, but does not know the secret state-preparation circuit or seed and does not have copy access to the planted state. Breaking the mechanism therefore amounts to forging the spectral response for fresh public challenges, either by reproducing $\boldsymbol{\Theta}$ with non-negligible probability or by producing an alternative witness that yields the same derived key and passes explicit confirmation. We analyse attack families that try to compute features from $\{U_i\}$ alone, that attempt to propagate eigenstate information across challenge instances, and that exploit leakage across repeated sessions.

QSA does not solve the initial provisioning problem. The planted state, or the seed or circuit that defines it, may be established by secure manufacturing enrolment, out-of-band commissioning, pre-shared commissioning keys, or quantum communication, for example by teleporting a witness instance or distributing entanglement resources. QSA begins after that step and turns the provision into per-session authentication tokens by answering fresh public spectral challenges.

We instantiate the primitive in three regimes to separate conceptual requirements from implementation constraints. QSA-M is matrix based with dense $2^n\times 2^n$ unitaries and serves as a reference model. QSA-C is circuit based with classical evaluation, where each $U_i$ is an expressive near-Haar circuit and features are extracted by deterministic simulation for moderate $n$. QSA-Q is circuit based with quantum-hardware evaluation, where the $U_i$ are engineered circuits executed on a QPU and features are extracted using low-depth phase estimation (LDQPE) at larger $n$. Orthogonally to these evaluation regimes, QSA runs in two provisioning modes that differ in what the endpoint stores and hence in what it can guarantee: a seed mode requiring a variational compilation step, and a state mode that needs none. We return to this distinction in Sec.~\ref{sec:secmodel}, where it determines the security claims.

To make the QSA-Q regime usable on realistic hardware, honest evaluation must remain tractable on noisy devices. The main systems challenge is therefore to generate public circuit families that appear generic from their gate descriptions while still admitting stable feature extraction by a prover that can prepare the planted state. Our main implementation route is a verifier-driven symmetric compiler of the form \(U = VDV^\dagger\), which preserves structured low-depth phase extraction and supports the hardware pathway studied in this paper.

In summary, this paper:
\begin{enumerate}
  \item introduces QSA as a post-provisioning spectral authentication primitive that converts possession of a hidden quantum credential into transcript-bound symmetric session material under fresh public unitary challenges;
  \item develops a formal security model for this interface: a mutual-authentication experiment, a single label-hiding assumption, an unconditional feature-uniformity lemma, and a resulting authentication theorem and separates it from an operational cryptanalysis (online forgery, eigenstate propagation, repeated-session leakage, reference spectrum attacks) that serves as evidence for the assumption rather than proof;
  \item identifies the condition under which the construction is sound that the planted-state ensemble match the compiler ansatz and shows that the state-mode provisioning route satisfies it with no residual assumption, yielding label hiding by construction;
  \item presents and evaluates a practical QSA-Q pathway based on a symmetric compiler \(U = VDV^\dagger\), supported by asymptotic analysis, noisy LDQPE simulations, and small-instance executions on IBM \texttt{ibm\_fez}.
\end{enumerate}

% =====================================================================
%  QSA revision: comparative table for the Introduction (Reviewer 2, pt 2)
%  Also carries the Lou-Wang citation requested by Reviewer 1, pt 4.
%
%  PLACEMENT: end of the Introduction, after the contributions paragraph.
%  Uses revtex4-1 `ruledtabular`; table* spans both columns.
%
%  The Lou-Wang row is now checked against the published paper (TIFS 21,
%  2061--2075, 2026). Their credential is a shared sequence of classical
%  verification codes refreshed each round; the code is encoded per session
%  onto a quantum register via RY(theta), and entanglement plus classical
%  feedforward carries both the identity test and a session key drawn from
%  fresh quantum randomness. They claim PARTIAL forward secrecy (their
%  Sec. V-D) -- see the note in the paragraph below, which is the one place
%  this table works against us.
%
%  Citation keys resolve against sample__5_.bib as it stands, EXCEPT
%  LouWang2026 and Wiesner1983, which are in qsa_bib_additions.bib.
% =====================================================================

\begin{table}[t!]
\caption{QSA against representative authentication primitives. ``Credential''
is what the endpoint must hold between sessions; ``acceptance evidences'' is what
a verifier learns from a successful run beyond the bare fact of authentication.
The classical baseline is included deliberately: in seed mode QSA matches it in
every security column, and the security differentiator appears only in state mode.
FS $=$ forward secrecy.
$^{a}$~Claimed in Ref.~\cite{LouWang2026}, Sec.~V-D: session keys are drawn from
fresh randomness rather than derived from the long-term secret, so exposure of the
verification codes does not reveal past session keys; the authors do not claim full PFS.
$^{b}$~Forward-secure against \emph{prover} compromise, not verifier compromise: the
endpoint holds no $S_0$ and measurement consumes $\ket{\psi_i}$, so states spent in past
sessions cannot be recovered from a seized endpoint. See Sec.~\ref{sec:secmodel}.}
\label{tab:compare}
%\begin{tabular}{\textwidth}{YYYYY}
\begin{tabular}{p{2.9cm}p{2.9cm}p{3.4cm}p{3.2cm}p{1.0cm}}
\toprule
Primitive & Endpoint credential & Security rests on & Acceptance evidences & FS \\
\hline
Classical challenge-response, $K=\mathrm{PRF}(S_0,\textit{sid})$ (HMAC, AES-CMAC)
 & classical seed $S_0$
 & PRF security (standard, well-tested)
 & possession of $S_0$
 & no \\[2pt]
\hline
QKD-based authenticated key establishment~\cite{BB84,Scarani2009,portmann2017authrecycle}
 & classical pre-shared authentication key
 & information-theoretic channel security $+$ classical authentication of the classical channel
 & possession of the pre-shared key; a working QKD link
 & yes \\[2pt]
\hline
Quantum PUF / quantum token~\cite{Wiesner1983,goorden2014qsa,nikolopoulos2017cvpuk}
 & physical device or unclonable state
 & no-cloning $+$ device-specific unpredictability (average-case, device-dependent)
 & physical possession of the device or token
 & no \\[2pt]
\hline
Quantum-assisted mutual AKA, e.g.\ Lou-Wang~\cite{LouWang2026}
 & classical verification codes, refreshed every round
 & secrecy of the shared codes $+$ no-cloning and measurement disturbance of the per-session encoded states
 & possession of the current codes; a few-qubit NISQ device
 & partial$^{a}$ \\[2pt]
\hline
\textbf{QSA, seed mode} (this work)
 & classical seed $S_0$
 & Assumptions~\ref{ass:lh} and~\ref{ass:fu} (average-case, ensemble-specific)
 & possession of $S_0$; an operational QPU for $n$ beyond classical simulation
 & no \\[2pt]
\hline
\textbf{QSA, state mode} (this work)
 & physical planted register $\ket{\psi_i}$, consumed on use; \emph{no classical seed at the endpoint}
 & Assumptions~\ref{ass:lh} and~\ref{ass:fu} $+$ no-cloning $+$ confidentiality of the provisioning layer
 & possession of the planted resource; an operational QPU as above
 & prover$^{b}$ \\
\bottomrule
\end{tabular}
\end{table}

% ---------------------------------------------------------------------
%  Accompanying paragraph. Place immediately before or after the table.
% ---------------------------------------------------------------------

Table~\ref{tab:compare} places QSA against representative alternatives, including a purely classical baseline, since the comparison is only informative with one. Against $K=\mathrm{PRF}(S_0,\mathit{sid})$, \emph{seed-mode QSA offers no security advantage}: both store a classical secret, both fall to endpoint compromise, and the adversary's task is the unpredictability of a keyed public function, for which the PRF is the better-tested assumption. What seed mode adds is \emph{evaluability}: once $n$ exceeds classical simulability the honest response cannot be produced without a working quantum device, so acceptance evidences an operational QPU. This is a completeness property, not a security bound, and we do not use it as one (Sec.~\ref{sec:secmodel}). The genuine differentiator is \emph{state mode}, where the endpoint holds only the physical planted register: no classical credential exists to read out of memory, and by no-cloning an adversary who compromises the endpoint cannot silently duplicate it while leaving the prover functioning. This is why confidentiality of the provisioning layer is load-bearing rather than incidental.

Relative to QKD-based key establishment the two are complementary: QKD authenticates a channel using a pre-shared classical key, whereas QSA authenticates an endpoint's continued possession of a provisioned resource and emits key material for a classical channel. Relative to quantum PUFs and tokens, QSA shares the reliance on unclonability in state mode, but its challenge is a published, auditable circuit rather than a device-specific physical stimulus, so the verifier's trapdoor is a compilation secret rather than a fabrication artefact.

Relative to quantum-assisted mutual AKA schemes such as Lou and Wang~\cite{LouWang2026}, the layering is inverted: there the persistent credential is classical, with verification codes refreshed each round and the quantum resource consumed per session, whereas in QSA the persistent credential is the quantum resource and the derivation is classical. The two therefore place forward secrecy on different axes. Because their session key is drawn from fresh randomness, Lou-Wang claim partial forward secrecy (Ref.~\cite{LouWang2026}, Sec.~V-D); QSA claims none in seed mode, but in state mode is structurally forward-secure against \emph{prover} compromise, since measurement consumes the register and no-cloning forbids a retained copy. Neither offers forward secrecy against \emph{verifier} compromise, which retains $S_0$; since the endpoint is what is exposed to physical capture in our target deployments, this is the asymmetry worth having. Both require quantum hardware at the endpoint, so the remaining distinction is scale: Lou-Wang operate in a classically simulable few-qubit regime, whereas QSA targets $n$ beyond classical simulation, where evaluability becomes a usable security property. This is a difference in operating regime, not a shortcoming of either scheme.

Several existing primitives share individual ingredients with QSA, and it is worth stating precisely which. Closest in mechanism is the observation that quantum phase estimation applied to a random unitary can serve as a quantum physical unclonable function by realising a von~Neumann measurement in the unitary's eigenbasis~\cite{ghosh2024qpuf}; QSA differs in that the secret is a planted index into a \emph{public, compiled} $U=VDV^\dagger$ rather than the unitary itself, that a single designated eigenphase is read out rather than the whole eigenbasis sampled, and that the readout is wrapped in a symmetric identification model with a transcript-bound key-derivation step. The consume-on-use structure of state mode mirrors tokenized message-authentication codes, in which single-use quantum tokens issued from a secret key each authenticate at most once and cannot be duplicated~\cite{bendavid2023qtokens,behera2021tmac}; here the token is a hidden-label eigenstate whose classical response is a spectral feature rather than a signature, and the challenges are public reusable circuits. Using a pseudorandom-looking quantum state as an unclonable credential is the defining feature of pseudorandom-state cryptography~\cite{ji2018prs}, of which our randomly compiled $V\ket{b}$ is a concrete instance; the difference is that QSA extracts a classical spectral feature from the state rather than using the state itself as the tag. Finally, hiding a marked index inside a public family of states is the structure of quantum money from hidden subspaces and its hidden-coset descendants~\cite{aaronson2012hiddensubspaces,coladangelo2021hiddencosets}, which verify by a projective membership test and target public-key bearer instruments, whereas QSA verifies by an eigenphase read-off and is a symmetric endpoint-authentication primitive.

We therefore do not claim a new cryptographic mechanism: reading a spectral or eigenbasis feature off a hidden unclonable state under a public unitary, and compressing it into key material, draws on all of the above. What is new is the specification of QSA at a boundary these works do not target, namely lightweight re-confirmation that a remotely provisioned quantum credential is still held at an endpoint using a public auditable challenge rather than a device stimulus or a bearer-instrument mint; a security model in which the single required assumption (label hiding) is isolated, proved unconditionally at the level of the published spectrum (Lemma~\ref{lem:unif}), and eliminated entirely in state mode; and a NISQ-feasible instantiation with an honest fast path. The contribution is in the scoping and the rigour, not in the quantum machinery of the response.

\section{Results}

\label{sec:results}
\begin{figure}[ht!]
\centering
\resizebox{0.75\linewidth}{!}{%
\begin{tikzpicture}[
    node distance = 18pt and 18pt,
    box/.style = {draw, rounded corners, align=center, inner sep=4pt, text width=4.6cm},
    smallbox/.style = {draw, rounded corners, align=center, inner sep=4pt, text width=4.2cm},
    arrow/.style = {->, >=Latex, thick}
]

%==================== TOP: COMMON SECRET/PUBLIC INPUTS ====================
% \node[box, text width=5.0cm] (secret)
%   {Provisioning secret\\
%    planted-state seed \(S_0\) and/or circuit \(P\)\\
%    (or per-instance derived preparations \(P_i\))};

% \node[box, right=20pt of secret, text width=5.0cm] (public)
%   {Public challenge schedule\\
%    \(S_1,\ldots,S_k\) determining challenge instances};

%==================== QSA-M COLUMN ====================
\node[box, xshift=-4.8cm,text width=4.4cm] (Msetup)
  {\textbf{QSA-M: challenge setup}\\
   Dense-matrix public challenges \(\mathcal{U}_i\in\mathbb{C}^{2^n\times 2^n}\)\\
   sampled from a public seeded procedure};

\node[smallbox, below=14pt of Msetup] (Mver)
  {\textbf{Verifier eval}\\
   Dense eigendecomposition + overlap scan\\
   \(O(2^{3n})\) per challenge};

\node[smallbox, below=14pt of Mver] (Mprov)
  {\textbf{Prover eval}\\
   Dense eigendecomposition + overlap scan\\
   quantize dominant eigenphase to \(m\) bits};

%==================== QSA-C COLUMN ====================
\node[box,text width=4.4cm] (Csetup)
  {\textbf{QSA-C: challenge setup}\\
   Public circuit challenges \(U_i\)\\
   from a seeded expressive random-circuit ensemble};

\node[smallbox, below=14pt of Csetup] (Cver)
  {\textbf{Verifier eval}\\
   Classical autocorrelation / spectroscopy\\
   \(Z_t^{(i)}=\langle\psi_i|U_i^t|\psi_i\rangle\) + FFT};

\node[smallbox, below=14pt of Cver] (Cprov)
  {\textbf{Prover eval}\\
   Same classical spectral evaluation\\
   quantize dominant eigenphase to \(m\) bits};

%==================== QSA-Q COLUMN ====================
\node[box, xshift=4.8cm,text width=4.4cm] (Qsetup)
  {\textbf{QSA-Q: challenge setup}\\
   Public compiled circuit challenges \(U_i\)\\
   symmetric \(U_i=V_iD_iV_i^\dagger\) or asymmetric \(U_i=V_{L,i}V_{R,i}^\dagger\)};

\node[smallbox, below=14pt of Qsetup] (Qver)
  {\textbf{Verifier eval}\\
   Symmetric: closed-form phase read-off\\
   \(\arg\langle b_i|D_i|b_i\rangle\)\\
   Asymmetric: LDQPE evaluation};

\node[smallbox, below=14pt of Qver] (Qprov)
  {\textbf{Prover eval}\\
   Quantum-hardware phase extraction on \(|\psi_i\rangle\)\\
   LDQPE / QPE \; };

%==================== OUTPUT ====================
\node[box, below=38pt of Cprov, text width=7.8cm] (theta)
  {Eigenphase feature vector
   \[
   \boldsymbol{\Theta}=(\theta_1^*,\ldots,\theta_k^*),\qquad
   \theta_i^*\in\left\{0,\frac{2\pi}{2^m},\ldots,\frac{2\pi(2^m-1)}{2^m}\right\}
   \]};

\node[box, below=14pt of theta, text width=6.4cm] (hkdf)
  {Classical KDF (e.g.\ HKDF)\\
   \(\Longrightarrow\) transcript-bound session keys};

%==================== ARROWS ====================
%\draw[arrow] (secret.south) |- (Msetup.north);
%\draw[arrow] (secret.south) |- (Csetup.north);
%\draw[arrow] (secret.south) |- (Qsetup.north);

%\draw[arrow] (public.south) |- (Msetup.north);
%\draw[arrow] (public.south) |- (Csetup.north);
%\draw[arrow] (public.south) |- (Qsetup.north);

\draw[arrow] (Msetup) -- (Mver);
\draw[arrow] (Mver) -- (Mprov);

\draw[arrow] (Csetup) -- (Cver);
\draw[arrow] (Cver) -- (Cprov);

\draw[arrow] (Qsetup) -- (Qver);
\draw[arrow] (Qver) -- (Qprov);

\draw[arrow] (Mprov.south) |- (theta.west);
\draw[arrow] (Cprov.south) -- (theta.north);
\draw[arrow] (Qprov.south) |- (theta.east);

\draw[arrow] (theta) -- (hkdf);

\end{tikzpicture}%
}
\caption{QSA implementation regimes separated by challenge setup, verifier evaluation, and prover evaluation. In all regimes, a provisioning secret such as a planted-state seed \(S_0\) and/or preparation circuit \(P\) defines the planted state resource, while a public challenge schedule determines the per-instance public challenges. In QSA-M, the challenges are dense matrices and both parties evaluate them by eigendecomposition. In QSA-C, the challenges are public circuits evaluated classically through autocorrelation spectroscopy. In QSA-Q, the challenges are compiled public circuits; for the symmetric construction \(U_i=V_iD_iV_i^\dagger\), the verifier can read off the intended phase directly, while the prover performs phase extraction on hardware using LDQPE or QPE, with LDQPE as the main focus in this work. In all cases, the resulting \(m\)-bit phase features are aggregated into \(\boldsymbol{\Theta}\) and compressed by a classical KDF into session material.}
\label{fig:QSA-implementations}
\end{figure}
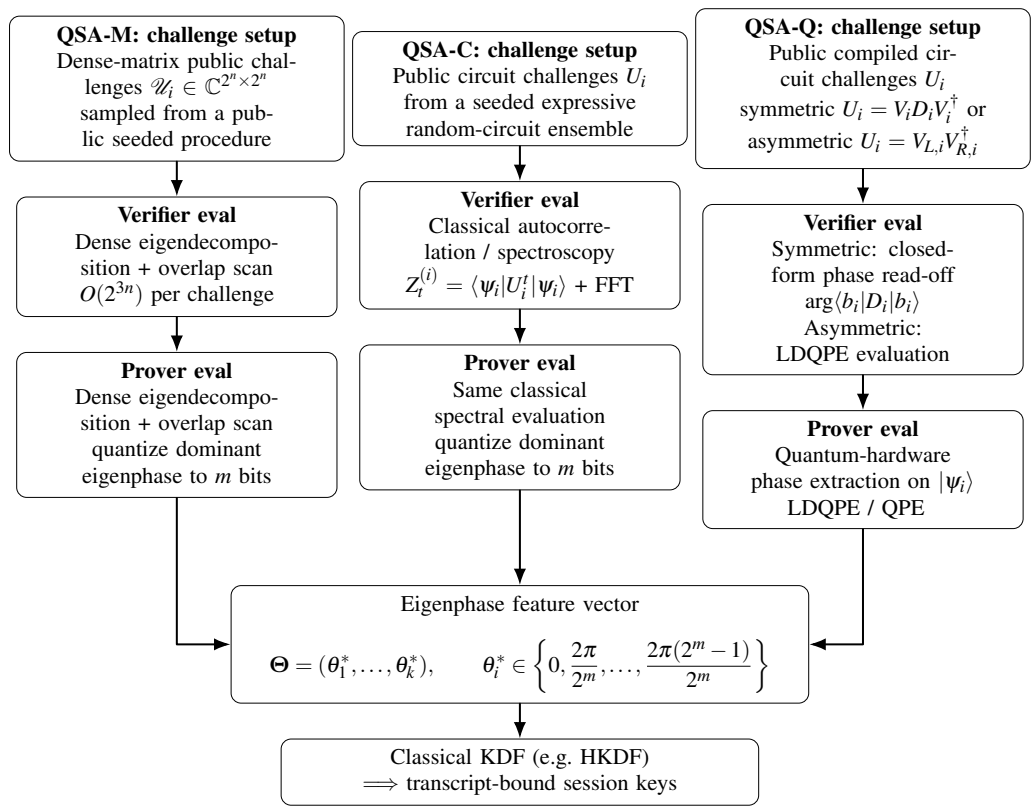
\subsection*{Main implementations}
\label{subsec:main-implementations}

The computational security and practicality of QSA depend crucially on how the public unitary challenges are represented and, most importantly, how an honest prover evaluates the resulting spectral features. Figure~\ref{fig:QSA-implementations} summarizes the common structure: a master planted state seed $S_0$ (OTP-level secret with length comparable to a circuit/state description) induces planted states, a public seed schedule $S_1,\ldots,S_k$ determines the public unitary challenges, and the prover extracts a low-precision eigenphase feature vector that is compressed into a session key by a classical KDF.

\smallskip
\noindent\textbf{Key parameters.}
We write $n$ for the number of qubits, so the Hilbert-space dimension is $d=2^n$.
The protocol publishes $k$ public unitaries $U_1,\ldots,U_k$.
From each unitary, the honest parties extract an eigenphase feature at \emph{$m$-bit precision}, i.e.\
$\theta_i^*\in\{0,\tfrac{2\pi}{2^m},\ldots,\tfrac{2\pi(2^m-1)}{2^m}\}$, forming
$\boldsymbol{\Theta}=(\theta_1^*,\ldots,\theta_k^*)$.
A classical KDF (e.g.\ HKDF \cite{rfc5869}) maps $\boldsymbol{\Theta}$ to a final session key of length $\ell_K$ bits (and optionally additional subkeys), with $\ell_K$ chosen by the application/security level.
Implementation-specific knobs include the public-circuit depth $D(n)$ (QSA-C), the LDQPE precision set $\mathcal{T}$ and compiler tolerances (QSA-Q), and the number of repetitions/shots used in phase extraction.

\smallskip
We consider three evaluation regimes:
\emph{QSA-M} (matrix diagonalization),
\emph{QSA-C} (classical time-signal evaluation and spectrum reconstruction),
and \emph{QSA-Q} (quantum-hardware evaluation via LDQPE/QPE).
Throughout, ``QSA'' refers to the underlying spectral primitive, and QSA-M/C/Q denote particular evaluation regimes. Figure~\ref{fig:QSA-implementations} emphasizes the regime split used throughout this section: the public challenges $\{U_i\}$ and the prover’s \emph{evaluation method} changes (dense diagonalisation in QSA-M, classical circuit evaluation/spectroscopy in QSA-C, or LDQPE on hardware in QSA-Q). The role of the confirmation wrapper is simply to turn agreement on $\boldsymbol{\Theta}$ into a standard short-lived token suitable for control-plane use, without changing the underlying spectral primitive.

\paragraph*{\textbf{QSA-M: matrix-based reference evaluation (dense diagonalization).}}
QSA-M is a purely classical \emph{reference} instantiation in which public unitaries are explicit dense matrices.
For each $i$, the verifier samples (from a public seeded procedure) a dense unitary $\mathcal{U}_i\in\mathbb{C}^{2^n\times 2^n}$ and publishes it, while honest parties derive $\ket{\psi_i}$ from $S_0$ as above.
Given $(\mathcal{U}_i,\ket{\psi_i})$, the prover computes an eigendecomposition $\mathcal{U}_i=V_i\Lambda_iV_i^\dagger$, identifies the eigenvector $v_{i^\star}$ maximizing $|\langle v_{i^\star}|\psi_i\rangle|^2$, and records the corresponding eigenphase $\theta_i$. The phase is quantized to $m$ bits to form $\theta_i^*$, and $\boldsymbol{\Theta}$ is formed by concatenation over $i$.

The cost of QSA-M is dominated by dense linear algebra: eigendecomposition of a $2^n\times 2^n$ matrix scales as $O(2^{3n})$, hence honest evaluation scales as $k\times O(2^{3n})$ (plus $k\times O(2^n)$ to scan overlaps) \cite{golub2013matrix}. Generating dense Haar-random unitaries has the same order of complexity (e.g.\ via QR-based Haar sampling \cite{mezzadri2007generate}), so QSA-M is not intended as a practical deployment regime. It is included as a baseline that cleanly separates the spectral primitive from circuit representations (Table~\ref{tab:QSA-costs-cvp}).

\paragraph*{\textbf{QSA-C: classical evaluation from circuit challenges via autocorrelation spectroscopy.}}
In QSA-C, each public unitary $U_i$ is published as a quantum circuit (gate list) sampled from an expressive random-circuit ensemble on $n$ qubits, using the public seed schedule $\{S_i\}$.
The planted state $\ket{\psi}$ is derived from $S_0$, and the prover extracts a dominant eigenphase feature using a classical ``time-signal $\rightarrow$ spectrum'' routine.

Let $\{\ket{u_j}\}$ be an eigenbasis of $U_i$ with $U_i\ket{u_j}=e^{i\theta_j}\ket{u_j}$ and expand
$\ket{\psi}=\sum_j \alpha_j\ket{u_j}$.
Repeated application of $U_i$ generates the complex autocorrelation sequence
\begin{equation}
  Z_t^{(i)} \;:=\;\langle\psi_i|U_i^{t}|\psi_i\rangle
  \;=\;\sum_j |\alpha_j|^2 e^{i\theta_j t},
  \qquad t=0,1,\ldots,T-1.
  \label{eq:QSA_autocorr}
\end{equation}
Given $\{Z_t^{(i)}\}_{t=0}^{T-1}$, the prover estimates the dominant eigenphase by matched filtering over a grid,
\begin{equation}
  \mathcal{S}_i(\omega)\;=\;\left|\sum_{t=0}^{T-1} Z_t^{(i)}\, e^{-i\omega t}\right|,
  \label{eq:QSA_periodogram}
\end{equation}
implemented efficiently via an FFT (optionally with zero-padding and local refinement), and sets $\widehat{\theta}_i=\arg\max_\omega \mathcal{S}_i(\omega)$.
The resulting $\widehat{\theta}_i$ is quantized to $m$ bits to obtain $\theta_i^*$.
This classical spectral viewpoint is directly analogous to line-spectral estimation~\cite{stoica2005spectral,kay1988modern} and closely related to hybrid eigenvalue routines such as quantum filter diagonalization~\cite{cohn2021qfd}. 

Computing $Z_t^{(i)}$ requires $T-1$ sequential applications of $U_i$ to a single $2^n$-dimensional state vector (or tensor-network representation when applicable) plus $T$ inner products with $\ket{\psi_i}$. For state-vector simulation, the per-unitary cost scales as $
O\!\left(T\cdot 2^n\,\mathrm{poly}(n)\,D(n)\right),
$ with memory $O(2^n)$; the FFT post-processing is $O(T\log T)$ and is negligible for the modest $m$ regimes of interest (typically $T\approx 2^m$).

\paragraph*{\textbf{QSA-Q: QPE evaluation on a QPU with compiled circuit challenges.}}
In QSA-Q, the prover evaluates each public unitary challenge $U_i$ using quantum phase estimation (QPE) on a quantum device \cite{Cleve1998}. The only additional requirement beyond access to the QPU is that the client and server share the planted state resource associated with the $i$th challenge. We keep this state-distribution mechanism generic: in one setting, independent planted states $\{\ket{\psi_i}\}$ are \emph{distributed online} via quantum teleportation using pre-shared entanglement resources~\cite{PhysRevLett.70.1895}; in another, the parties rely on \emph{provisioned seeds} (e.g.\ factory-installed $S_0$) and deterministically derive the corresponding preparations via HKDF, so that $\ket{\psi_i}$ (or its preparation circuit $P_i^\dagger$) can be regenerated locally without any online quantum communication. After establishing access to $\ket{\psi_i}$, the verifier publishes $U_i$ and the prover extracts an $m$-bit eigenphase feature $\theta_i^*$. For the symmetric compiler $U_i=V_iD_iV_i^\dagger$, this extraction admits a fast path that is cheaper than QPE (Methods): since $V_i$ is public and $\ket{\psi_i}$ concentrates on the signal eigenstate $V_i\ket{b}$, the prover applies $V_i^\dagger$ and measures in the computational basis to recover the label $b$ (a single shot in state mode, a few otherwise), then computes $\theta_i^*=\theta(b)$ in $O(n)$ classical time from the public angles, using one ancilla-free circuit of depth $\mathrm{depth}(P_i)+\mathrm{depth}(V_i)$ with no controlled operations. This is the cheapest honest evaluation; the low-depth QPE (LDQPE) primitive is retained as the general route required for QSA-C, QSA-M, and the asymmetric compiler. Cost scalings for all routes are summarised in Table~\ref{tab:QSA-costs-cvp}.

To promote a dominant-eigenphase structure under $\ket{\psi_i}$ (improving LDQPE stability at modest precision), the verifier compiles public challenges $U_i$ using two strategies.
(i) A fast planted-eigenstate construction of the form $U_i = V_i D_i V_i^\dagger$, where $D_i$ is a diagonal $R_z$-layer, and the verifier can compute the intended signal eigenphase in closed form as $\arg\langle b|D_i|b\rangle$ for a hidden computational-basis label $\ket{b}$ (Sec.~\ref{subsec:compilation-fast-eval-QSAq}).
(ii) An asymmetric construction $U_i = V_{L,i} V_{R,i}^\dagger$ in which two independently optimised expressive circuits are learned and composed, removing the diagonal phase layer and the associated ``read-off'' eigenphase shortcut; in this case, the verifier (and prover) obtain $\theta_i^*$ by running LDQPE on $(U_i,\ket{\psi_i})$.

Once the public circuits $\{U_i\}$ are fixed, honest evaluation consists of preparing $\ket{\psi_i}$ and running LDQPE (Algorithm~2 of Ni-Li-Ying~\cite{ni2023lowdepthqpe}) to obtain $\theta_i^*$.
Concretely, Algorithm~2 of~\cite{ni2023lowdepthqpe} targets a \emph{dominant} eigenphase by estimating a small collection of \emph{power moments}
\begin{equation}
Z_t^{(i)} \;:=\; \langle \psi_i| U_i^t | \psi_i \rangle,
\qquad t\in\mathcal{T},
\label{eq:ldqpe-moments}
\end{equation}
via Hadamard-test circuits, followed by classical post-processing to recover an eigenphase estimate which is finally quantized to $m$ bits. When $\ket{\psi_i}$ has most of its spectral weight on a single eigencomponent of $U_i$, the moment sequence $\{Z_t^{(i)}\}$ behaves approximately as a single complex exponential, enabling reliable dominant-eigenphase recovery at modest precision. For this particular algorithm, this overlap with the dominant eigencompoment must satisfy $p_0=|\bra{u_0}\psi\rangle|^2\ge 4-2\sqrt{3}$.

\smallskip

The dominant cost driver depends on the evaluation regime (Table~\ref{tab:QSA-costs-cvp}).
QSA-M is purely a reference baseline, as both challenge generation and honest evaluation require dense $2^n\times 2^n$ eigendecomposition.
QSA-C shifts the burden to classical simulation: the verifier can publish seeded random circuits cheaply, while evaluation requires $O(2^m)$ sequential moment computations on a $2^n$-dimensional state representation, making cost scale as $O(2^m\cdot 2^n)$ up to polynomial and depth factors.

QSA-Q instead targets a QPU-native prover, where evaluation cost is dominated by controlled applications of $U^{2^j}$ for $j=0,\ldots,m-1$.
Here the compiler choice qualitatively changes the $m$-dependence.
For the symmetric construction $U=VDV^\dagger$, we exploit the fast-power identity $U^{2^j}=VD^{2^j}V^\dagger$; because $D$ is diagonal (a tensor product of $R_z$ layers), the controlled-$D^{2^j}$ block has essentially the same gate structure for all $j$ (angles rescale modulo $2\pi$), and the dominant controlled-entangling work from $V$ and $V^\dagger$ is incurred once per moment rather than $2^j$ times.
As a result, the prover’s LDQPE evaluation scales as $O(\mathrm{poly}(n)\cdot m)$ moment evaluations (linear in $m$), rather than $O(\mathrm{poly}(n)\cdot(2^m-1))$.

This symmetry also yields a substantial verifier advantage. For the symmetric compiler, the intended signal eigenphase can be read off directly in $O(n)$ time from $\arg\langle b|D|b\rangle$, so the verifier need not run LDQPE or QPE at all in the usual verification pathway. The prover may instead use either structured LDQPE, with linear-in-$m$ cost, or standard QPE when the planted overlap is too small to satisfy the LDQPE success condition. In the latter case, the coherent circuit cost remains polynomial because $U^{2^j}=VD^{2^j}V^\dagger$, but the practical sampling cost is multiplied by the repetition factor $
N_{\mathrm{rep}}=O\!\left(\frac{\log(1/\varepsilon_{\mathrm{fail}})}{p_0\eta_m}\right),$ where $\varepsilon_{\mathrm{fail}}\in(0,1)$ is the tolerated failure probability, so that $1-\varepsilon_{\mathrm{fail}}$ is the desired confidence level, $p_0$ is the overlap weight of the planted state with the target eigenvector, and $\eta_m$ is the finite-resolution success factor of the $m$-bit phase register, with the standard QPE guarantee $\eta_m\ge 4/\pi^2$ \cite{Cleve1998}. Thus, unlike LDQPE, QPE does not require a large constant overlap threshold, but if $p_0$ is small, then repeated executions are needed before the correct bucket is observed with high confidence. By contrast, the asymmetric construction $U=V_LV_R^\dagger$ has neither closed-form phase read-off nor fast powering in general, so both prover and verifier must rely on LDQPE, with the usual $O(\mathrm{poly}(n)(2^m-1))$ controlled-power dependence.

% This symmetry also yields a substantial verifier advantage: the intended ``signal'' eigenphase can be computed directly in $O(n)$ times via $\arg\langle b|D|b\rangle$ (Eq.~\eqref{eq:phase_closed_form}), so the verifier does not need to run LDQPE at all.
% By contrast, the asymmetric construction $U=V_LV_R^\dagger$ removes the read-off structure and does not admit fast powering in general; consequently, both prover and verifier must run LDQPE with the usual exponential-in-$m$ moment-depth dependence, scaling as $O(\mathrm{poly}(n)\cdot(2^m-1))$ controlled applications.
% Operationally, this makes the symmetric $VDV^\dagger$ family substantially faster than the asymmetric $V_LV_R^\dagger$ family for the same $(n,m,k)$, and it enables pushing to higher $m$ on NISQ devices without incurring the prohibitive $2^m$ controlled-depth blow-up.
%\subsection*{Practical security and attacks}
\begin{figure}[ht!]
  \centering
  \includegraphics[width=0.93\linewidth]{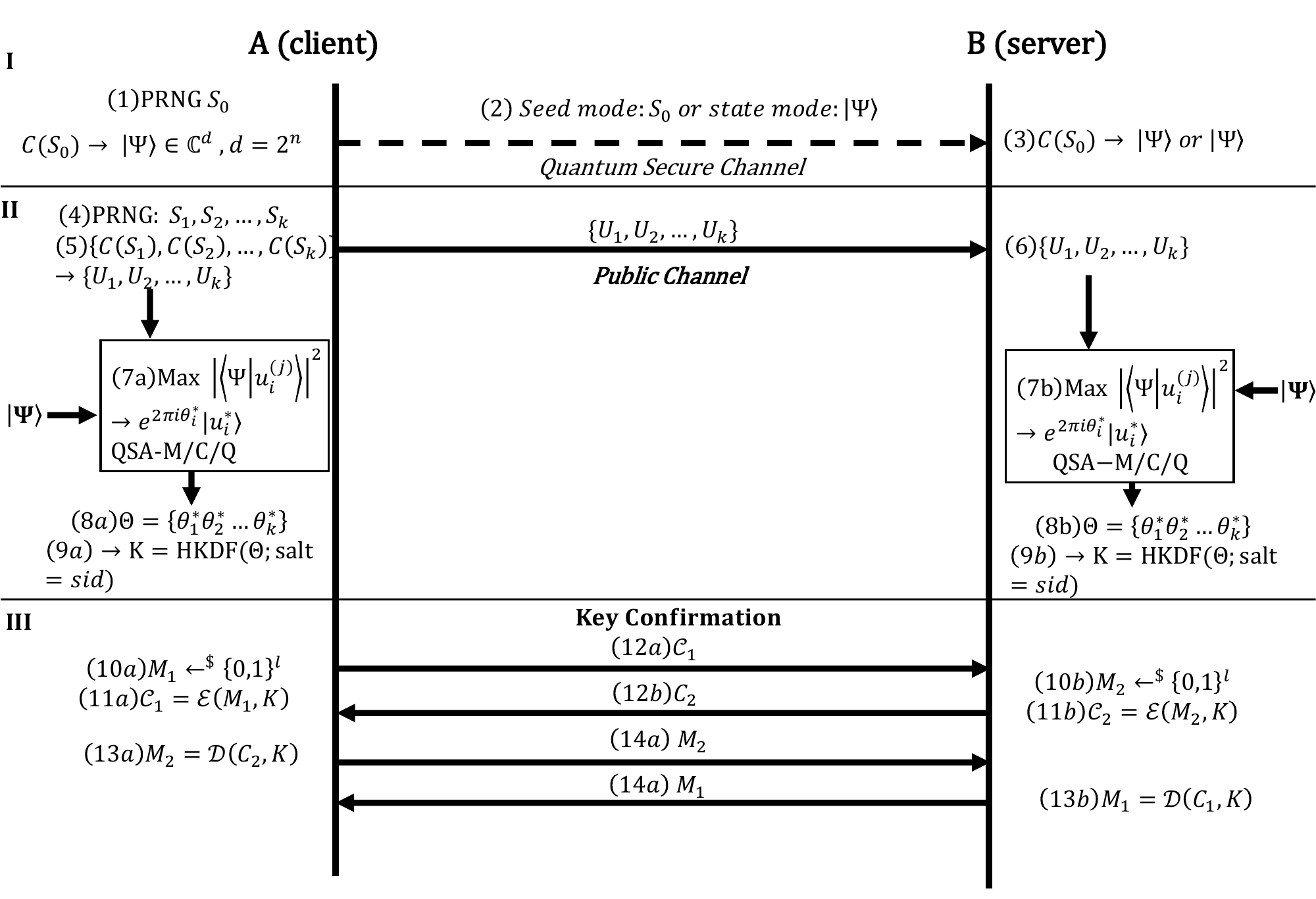}
  \caption{QSA with an explicit key-confirmation wrapper. A client (A) and server (B) share provisioning material (e.g.\ a planted state seed $S_0$ or a securely distributed witness/state-preparation capability) that defines a planted state $\ket{\psi}$. A public seed schedule (or public circuit descriptions) determines the challenge family $\{U_i\}_{i=1}^k$, which is distributed over the public channel. Each side evaluates the same challenges under $\ket{\psi}$ using its chosen evaluation regime (QSA-M/C/Q) to obtain a quantised eigenphase feature vector $\boldsymbol{\Theta}=(\theta_{1^\star},\ldots,\theta_{k^\star})$ and derives session material $K=h(\boldsymbol{\Theta})$. A lightweight mutual challenge-response under a symmetric authenticated primitive (shown abstractly as encryption/decryption $\mathcal{E}/\mathcal{D}$) provides an application-facing confirmation token: if the planted provision is missing, substituted, or inconsistent between endpoints, the parties disagree on $\boldsymbol{\Theta}$ and the confirmation fails except with probability set by the response length.}
  \label{fig:qke-key-confirmation}
\end{figure}

\subsection*{Threat model and adversary objective}
\label{sec:impl-security}

QSA is a symmetric-key identification (entity-authentication) scheme with
session-key derivation, whose long-term secret is a provisioned quantum resource.
In the protocol flow of Fig.~\ref{fig:qke-key-confirmation}, our security analysis is
restricted to the QSA portion itself, namely steps (4) through (13b): fresh public
challenge generation, spectral feature extraction, transcript-bound derivation, and
key confirmation. The analysis does \emph{not} cover how the underlying quantum
provision is initially established or delivered. Teleportation, entanglement
transport, QKD-delivered seeds, manufacturing enrolment, secure commissioning, and
protection of the root provisioning secret \(S_0\) belong to a separate provisioning
layer and are assumed secure before QSA begins. Likewise, any downstream use of the
derived session key, for example under AES-256, lies outside the present threat model.

This section makes two passes over that boundary. The present subsection is
operational: it fixes the parameter regimes and the secret distribution, then works
through the concrete attack families: chained-QPE eigenstate propagation
(Attack~\Romannum{1}), multi-session leakage (Attack~\Romannum{2}), and online forgery
against key confirmation (Attack~\Romannum{3}) that motivate the design choices of
QSA. The following subsection, ``Security model, claims, and what is assumed'', is
formal: it states the authentication experiment and the success event, isolates the
single non-standard assumption (label hiding) under which the advantage is bounded,
establishes an \emph{unconditional} entropy lemma for the symmetric compiler, and
gives the resulting mutual-authentication theorem, with session-key
indistinguishability as a corollary. The two passes are separated deliberately, so
that the boundary between what is proved, what is assumed, and what is cryptanalytic
evidence is visible: the attacks below are evidence for the assumption and calibration
of its slack, not proofs of it, and no bound in the formal subsection depends on the
attack catalogue being exhaustive.

Within this boundary, the verifier exposes public unitary descriptions, or public seeds that define them, together with associated metadata. The prover holds a hidden provisioning secret enabling preparation of a planted state $\ket{\psi_i}=P_i^\dagger\ket{0^n}$.
From each instance, the honest device extracts an \(m\)-bit phase feature using phase-estimation-style routines, producing across \(k\) instances a feature vector \(\boldsymbol{\Theta}\in\{0,1\}^{mk}\). This feature vector is then compressed and transcript-bound through a standard KDF such as HKDF, and the resulting response is validated through standard key confirmation.

The adversary sees the public unitaries and may perform arbitrary classical or quantum computations on them, but does not know the hidden state-preparation secret and does not have copy access to the planted state. The core security goal is therefore false acceptance: cause the verifier to accept without holding the relevant quantum credential. There are two operational ways to approach this goal. One is to exploit quantum or structural information in the public challenges so as to reconstruct the honest spectral response. The other is to attack the final transcript-bound acceptance condition directly. The first route is protocol-specific and determines the main design choices of QSA; the second is the generic baseline once no useful structure remains.

Two parameter regimes matter. In the practically most relevant \emph{token regime}, \(m\lesssim n\), the planted state lives in a large Hilbert space while only modest phase precision is extracted per instance. This is the intended regime for LDQPE-based operation and for QSA-Q in particular. Here the dominant protocol-specific risks are cross-instance eigenstate reuse and repeated-session leakage against reused planted states. By contrast, when \(m\gtrsim n\), one enters a \emph{spectrum-covering regime} in which classical diagonalisation, full-spectrum QPE coupon collecting, or other exhaustive spectral baselines become more meaningful as reference models. We treat those as appendix-level upper bounds rather than as the main operational threats in the intended deployment model.

This leads to three main attack families in the present section. Attack~\Romannum{1} studies chained-QPE or eigenstate propagation across public unitaries and motivates the need for cross-instance decorrelation. Attack~\Romannum{2} studies leakage and multi-session accumulation and motivates renewal policies for any regime in which planted states persist. Attack~\Romannum{3} studies direct online forgery against transcript-bound key confirmation and serves as the generic baseline once the protocol-specific structure has been neutralised. Other spectrum-covering strategies are deferred to Appendix~\ref{app:reference-attacks}, where they serve as calibration tools and conservative upper bounds rather than the dominant attacks inside the QSA boundary.

The key structural point is that the public interface exposes the unitaries, but not the selector that determines which phase bucket is realised. That selector is induced by the hidden planted state. Consequently, public spectral information alone does not reveal the honest response. Any attack stronger than generic one-shot forgery must therefore exploit either cross-instance correlations or repeated-session leakage. This is why the main design levers of QSA are diversification of planted states and challenges, transcript binding, rate-limited confirmation, and explicit renewal where fixed planted states are reused.

\paragraph*{\textbf{Attack \Romannum{1}: Chained-QPE / eigenstate propagation on a QPU.}}
\label{attack:chained-qpe}
In \textbf{Attack \Romannum{1}}, the adversary has quantum computational resources and can run QPE. The attack attempts to avoid the combinatorial blow-up implicit in naive per-instance guessing by \emph{reusing} approximate eigenstates across the public unitaries. As illustrated in Fig.~\ref{fig:chained-qpe-attack}, the adversary runs QPE on \(U_1\) to obtain a measured phase \(\widetilde{\theta}_1\) and a post-measurement eigenstate \(\ket{u_1^\ast}\), then feeds \(\ket{u_1^\ast}\) into QPE for \(U_2\), and continues along \(U_1\to U_2\to\cdots\to U_k\), hoping to accumulate an eigenphase vector that matches the honest \(\boldsymbol{\Theta}\).

To isolate the dependence on cross-instance structure, let \(\ket{u_i^\ast}\) denote the \emph{signal} eigenstate that drives the honest feature extraction for \(U_i\) under the relevant estimator. In QSA-Q, \(\ket{u_i^\ast}\) is the designated high-overlap eigenstate produced by compilation around the planted state for that instance; in QSA-C, \(\ket{u_i^\ast}\) is defined \emph{a posteriori} by the classical feature extractor, namely, the eigenstate or small eigenstate set that dominates the line-spectral estimator or autocorrelation statistic used to define the extracted feature.

Even under the optimistic assumption that the adversary can postselect the signal eigenstate for \(U_1\), the total success probability of chaining across \(k\) unitaries is bounded by
\begin{equation}
\label{eq:ptot-chained-threat}
    p_E^{\mathrm{tot}}
    \;\lesssim\; p_E^{(1)}\prod_{i=1}^{k-1} \bigl|\!\langle u^\ast_{i+1}|u^\ast_i \rangle \!\bigr|^2,
\end{equation}
where \(p_E^{(1)}\) is the probability that the adversary locks onto the signal eigenstate of \(U_1\). Equation~\eqref{eq:ptot-chained-threat} isolates the only possible advantage of chaining: it can beat exponential scaling only when successive signal eigenstates retain non-negligible overlap.

In QSA-C, the public circuits \(U_i\) are independently seeded and sufficiently expressive that signal eigenstates decorrelate across instances. Across independent seeds, typical overlaps follow the Haar benchmark
\begin{equation}
\label{eq:haar-overlap-threat}
    \mathbb{E}\bigl[|\langle u^\ast_{i+1}|u^\ast_i\rangle|^2\bigr] \;\approx\; 2^{-n}.
\end{equation}
Substituting Eq.~\eqref{eq:haar-overlap-threat} into Eq.~\eqref{eq:ptot-chained-threat} gives
\begin{equation}
\label{eq:ptot-2nk-threat}
    \mathbb{E}\bigl[p_E^{\mathrm{tot}}\bigr]
    \;\lesssim\; \mathbb{E}[p_E^{(1)}]\,(2^{-n})^{k-1},
\end{equation}
and identifying the correct signal eigenstate for \(U_1\) costs another factor of order \(2^{-n}\), yielding the pessimistic bound
\begin{equation}
\label{eq:ptot-2nk-final-threat}
    \mathbb{E}\bigl[p_E^{\mathrm{tot}}\bigr] \;\lesssim\; 2^{-nk}.
\end{equation}
Thus chained QPE collapses to the same scaling as independent guessing once the \(U_i\) are sufficiently diversified. Empirical proxies for this decorrelation are shown in Fig.~\ref{fig:topeigenvector:a}.

In QSA-Q, chained QPE is the main reason to avoid reuse of a single planted state across instances. If one reused a common planted state \(\ket{\psi}\) and compiled each \(U_i\) so that it admitted a signal eigenstate $\ket{u_i^\ast}=\sqrt{1-\delta_i}\ket{\psi}+\sqrt{\delta_i}\ket{\psi_{i,\perp}},$
then successive signal eigenstates could remain far more aligned than Haar-random vectors, making Eq.~\eqref{eq:ptot-chained-threat} much larger than \(2^{-nk}\). This motivates per-instance planted-state diversification. A simple construction derives independent planted states from a master provisioning secret via HKDF:
\[
  \sigma_i = \mathrm{HKDF}(S_0,\ \mathrm{info}=\texttt{``QSA-states''}\,\|\,i,\ \ell),
\qquad
  P_i \leftarrow \mathrm{PRG}(\sigma_i),
\qquad
  |\psi_i\rangle = P_i^\dagger|0^n\rangle.
\]
Each \(U_i\) is then compiled around its own planted state so that it admits a signal eigenstate \(\ket{u_i^\ast}\) with
\begin{equation}
     \bigl| \langle \psi_i\,|\,u_i^\ast \rangle \bigr|^2 \ge 1-\delta.
\end{equation}

From the adversary’s perspective, the \(\sigma_i\) behave as pseudorandom and independent across \(i\) without knowledge of \(S_0\), so the planted states \(\{|\psi_i\rangle\}\) and signal eigenstates \(\{\ket{u_i^\ast}\}\) decorrelate back toward the Haar scale, recovering the \(2^{-nk}\) behaviour. Empirical proxies are shown in Fig.~\ref{fig:topeigenvector:b}.

For the symmetric compiled form \(U_i=V_iD_iV_i^\dagger\), the public unitary reveals the eigenbasis \(\{V_i\ket{x}\}_{x\in\{0,1\}^n}\), but not the hidden selector \(b_i\) such that \(\ket{\psi_i}=V_i\ket{b_i}\). Thus Eve may know all \(2^n\) candidate eigenvectors for each instance while still not knowing which one is planted. To reproduce the honest response she must identify the correct \(b_i\), hence the correct eigenphase contribution, for every instance \(i=1,\dots,k\). Without exploitable cross-instance structure, this still leaves a \(2^{nk}\)-scale ambiguity at the level of planted-eigenvector selection. Symmetric compilation therefore does not defeat the chaining bound; it merely makes explicit the candidate eigenbasis among which the hidden planted eigenstate must be found.

\paragraph*{\textbf{Attack \Romannum{2}: Leakage and multi-session accumulation.}}
\label{attack:leakage}
Attack~\Romannum{2} captures settings in which the adversary obtains auxiliary information correlated with the extracted phase features or with the planted state across many runs. This can arise through confirmation side channels, inadvertent leakage of partial correctness of \(\boldsymbol{\Theta}\), or implementation leakage correlated with feature bits. The concern is not that a single run leaks the full response, but that repeated sessions may accumulate information against a fixed planted state.

Consider a pessimistic model in which the adversary learns, for each public unitary instance \(U_i\), the honest phase feature output or some representation strongly correlated with it. Given such leakage, Eve can attempt to prepare an eigenstate consistent with the leaked feature by running QPE on \(U_i\) and post-selecting on the leaked bucket, or by amplitude amplification toward eigenstates whose phases lie in that bin. Even if this succeeds only with small probability, it can yield a sequence of approximate eigenstates \(\{\ket{\tilde{u}_i}\}\) that are partially aligned with the honest hidden structure.

The security problem is accumulation. If the same planted state \(\ket{\psi}=P^\dagger\ket{0^n}\), or equivalently a fixed state-preparation seed, persists across many sessions, then these noisy and biased “views” can gradually become informative about that one hidden state. In the most pessimistic interpretation, sufficiently many such views approach a tomography-style reconstruction problem in dimension \(2^n\), eventually enabling an equivalent responder.

This motivates an explicit \emph{state-renewal policy} whenever planted states persist across many authentications, notably in QSA-M and some QSA-C deployments. The planted-state circuit \(P\), or the seed defining it, should be refreshed after a bounded number of uses and immediately after any suspected compromise or anomalous confirmation behaviour, so that an adversary never accumulates a large corpus of leakage aligned to a single fixed \(\ket{\psi}\).

In QSA-Q, this countermeasure is built in. The intended construction uses per-instance planted states $
\ket{\psi_i}=P_i^\dagger\ket{0^n},$ derived from a master secret via an HKDF schedule or supplied anew by quantum communication, so there is no long-lived fixed planted state against which leakage can accumulate. Even if features leak repeatedly, Eve faces \(k\) essentially independent leakage instances rather than a single reconstruction target. Thus leakage in QSA-Q collapses back to per-session impersonation risk rather than long-horizon state recovery.

\paragraph*{\textbf{Attack \Romannum{3}: Online response forgery under key confirmation.}}
\label{attack:online-forgery}
Attack~\Romannum{3} is the baseline cryptographic attack once the protocol-specific structure has been neutralised. Here the adversary targets the online acceptance condition directly. Let the session transcript include fresh nonces, challenge identifiers, and the public unitary descriptions \(\{U_i\}_{i=1}^k\). Honest parties derive a response key \(K\), or confirmation material, by compressing the phase-feature vector \(\boldsymbol{\Theta}\in\{0,1\}^{mk}\) under a standard KDF with transcript binding, and then run standard key confirmation such as MAC verification or AEAD decryption success. The adversary succeeds if she causes acceptance in a live session without holding the planted provision.

Even granting arbitrary classical or quantum computation on the public unitaries, the spectrum alone does not reveal the selected phase bucket without the hidden planted-state selector. In the intended deployment, fresh challenges, transcript binding, and rate-limited confirmation prevent the adversary from obtaining a high-rate verification oracle. Each session therefore affords essentially one acceptance attempt, or at most a small bounded number before lockout or detection, so the correct abstraction is one-shot forgery probability.

If the effective min-entropy of the response before compression is \(mk\) bits, then any generic online forger has per-attempt success probability at most \(2^{-mk}\), up to negligible bias from extraction and leakage. Equivalently, the online work factor is \(2^{mk}\) in the sense that the adversary must guess the correct response before observing acceptance. If one granted an unrealistically strong oracle permitting \(Q\) adaptive confirmation queries on the same transcript, then classical search would require \(O(2^{mk})\) queries and Grover search would require \(O(2^{mk/2})\). But such oracle access is excluded by the intended protocol boundary: key confirmation is transcript-bound, fresh per session, and operationally rate-limited.

Attack~\Romannum{3} therefore serves primarily as the generic baseline for the final acceptance condition. It motivates choosing \(mk\) as a genuine token-length security parameter and enforcing fresh challenges, transcript binding, bounded retries, and protocol behaviour that does not leak partial correctness of \(\boldsymbol{\Theta}\).

\emph{ Reference spectrum attacks (Appendix).} For completeness and calibration, especially in the \(m\gtrsim n\) spectrum-covering regime, Appendix~\ref{app:reference-attacks} collects reference attacks that assume stronger oracle access than is available inside the QSA boundary. These include Appendix Attack A.1, spectrum-oracle diagonalisation or dense EVD; Appendix Attack A.2, full-spectrum QPE coupon collecting; and Appendix Attack A.3, Hilbert-space or ansatz-manifold state guessing. These are not the dominant operational threats for the intended QSA deployment, but they remain useful as conservative upper bounds and calibration baselines.

% % =====================================================================
%  QSA revision: new MAIN-TEXT security model + replacement Appendix A
%  Replaces old Appendix A (PSP) and Appendix B (key-indistinguishability game).
%
%  PREAMBLE: nothing to add. Your preamble already defines every environment
%  used below -- lemma, definition, theorem, conjecture, assumption,
%  corollary, proposition, remark. Do NOT re-issue those \newtheorem lines;
%  a duplicate gives "Command \lemma already defined".
%
%  Two cosmetic notes on the existing preamble, neither blocking:
%   - lemma/definition/theorem/conjecture are declared before any
%     \theoremstyle, so they inherit "plain" (italic body). If you want
%     Definition 1 and 2 set in upright text like the assumptions, move
%     \newtheorem{definition}{Definition} below your \theoremstyle{definition}.
%   - amsthm is \usepackage'd three times. Harmless, but tidy it.
%
%  PLACEMENT: insert the block below immediately after the existing
%  "Threat model and adversary objective" subsection in Sec. II (Results).
%  Delete old Appendix A and Appendix B; their content is subsumed.
%
%  The Appendix A block at the end assumes the existing \appendix is in
%  force; revtex4-1 then letters it "Appendix A: Proofs" automatically.
%  Verified to compile against revtex4-1 with your appendix package loaded.
% =====================================================================
 
\subsection*{Security model, claims, and assumptions}
\label{sec:secmodel}
 
The preceding subsection described the adversary operationally. This subsection states the
security goal as a formal experiment, isolate the assumptions under which the
advantage is bounded, and give the theorem that this assumption implies. We also
state explicitly which parts of this paper are theorems, which are assumptions,
and which are cryptanalytic evidence or engineering measurements.
 
\paragraph*{The primitive.}
QSA is a two-party \emph{symmetric-key identification (entity-authentication)
scheme with session-key derivation}. Its long-term secret is a provisioned
quantum resource; its per-session response is a publicly parameterised,
quantum-evaluable function of that resource. It is \emph{not} a message
authentication code (it authenticates possession of a resource, not a message),
and it is \emph{not} an authenticated key exchange with forward secrecy (see
the non-claims below). Relative to a purely classical challenge-response
of the form $K=\mathrm{PRF}(S_0,\textit{sid})$, QSA adds two things, and we state
both precisely because only the first is a security property:
(i)~\emph{state mode}, in which the responder holds a physical, non-copyable
planted register and no classical seed is stored at the endpoint, so the credential
cannot be exfiltrated as a bitstring; and
(ii)~\emph{evaluability}: for $n$ beyond classical simulability, an honest response
requires a working quantum device at the endpoint, so acceptance is evidence of an
operational QPU. Property~(ii) is a statement about \emph{correctness cost}, not
about the adversary's advantage, and we do not use it in any security bound.
 
\paragraph*{Provisioning modes and who holds what.}
The security model requires an unambiguous statement of the secret distribution.
The verifier holds the root secret $S_0$ and, per session, the private
compilation trapdoor: the planting circuits $\{P_i\}$, the hidden labels
$\{b_i\}\subset\{0,1\}^n$, and the diagonal angles $\{\beta_{i,q}\}$ (the last of
which are published as part of $U_i$). The prover holds \emph{only} the planted
resource, in one of two modes:
\begin{itemize}
\item \textbf{Seed mode.} The prover holds $S_0$ and derives $P_i$, hence
$\ket{\psi_i}=P_i^\dagger\ket{0^n}$, locally. The labels $b_i$ are sampled from
verifier-private randomness and are \emph{not} derivable from $S_0$. This is
essential: were $b_i$ seed-derived, the prover could evaluate $\theta(b_i)$ in
$O(n)$ classical time by Eq.~\ref{eq:phase_closed_form} and the quantum evaluation would be redundant.
\item \textbf{State mode.} The prover holds only the physical registers of
$\{\ket{\psi_i}\}$, delivered by the upstream provisioning layer (teleportation,
transported memory, or commissioning). No classical description of the credential
is stored at the endpoint.
\end{itemize}
In both modes the prover's only route to $\Theta$ is to evaluate the public
challenge against the planted resource.
 
\paragraph*{Notation.}
Table~\ref{tab:notation-sec} collects the symbols used in this subsection and in
Appendix~\ref{app:proofs}. Three deserve comment in the text. The
\emph{session identifier} $\textit{sid}$ is the public string that
names one protocol run; it is what makes ``the same session'' well defined for two
parties who share no other state, and it is the object that must enter the key
derivation if $K$ is to be bound to this run rather than merely to the challenge
schedule (Remark~\ref{rem:binding}). An \emph{instance} $\pi_P^s$ is the $s$-th
execution of the protocol at party $P$; instances rather than parties are the unit
of the experiment, because the adversary may run many concurrent sessions against
one long-term secret. The \emph{$\mathrm{MA}$-advantage}
$\mathrm{Adv}^{\mathrm{MA}}_{\mathrm{QSA}}(\mathcal{A})$ is simply the probability
that $\mathcal{A}$ wins the mutual-authentication (MA) experiment of
Definition~\ref{def:ma}; there is no baseline to subtract, since the trivial
adversary wins with probability $0$, and ``advantage'' is used here in the standard
sense of a success probability in a security experiment rather than a distinguishing
gap. (The $\mathrm{KI}$-advantage of Corollary~\ref{cor:ki} \emph{is} a
distinguishing gap, and is defined there as
$|\Pr[\mathcal{A}\to1\mid b=1]-\Pr[\mathcal{A}\to1\mid b=0]|$ for the $\mathsf{Test}$
bit $b$.)

\paragraph*{Syntax.}
$\mathrm{QSA}=(\mathsf{Provision},\mathsf{Chal},\mathsf{Eval},\mathsf{VEval},
\mathsf{Derive},\mathsf{Confirm})$, with $\mathsf{Provision}(1^\lambda)\to S_0$;
$\mathsf{Chal}(S_0,r)\to(\textsf{pub},\textsf{priv})$ where
$\textsf{pub}=(\{U_i\}_{i=1}^k,\textit{sid})$ and $\textsf{priv}=\{b_i\}$;
$\mathsf{Eval}(\text{resource},\textsf{pub})\to\Theta\in\{0,1\}^{mk}$ (prover,
quantum); $\mathsf{VEval}(\textsf{priv},\textsf{pub})\to\Theta$ (verifier, $O(n)$
read-off via Eq.~\ref{eq:phase_closed_form} for the symmetric compiler);
$\mathsf{Derive}: K\leftarrow H(\Theta,\textit{sid})$ with $H$ a KDF; and
$\mathsf{Confirm}$ the two-flow mutual challenge-response of Fig.~\ref{fig:qke-key-confirmation}.
 
\begin{remark}[Transcript binding]
\label{rem:binding}
We make the session identifier explicit. Before $\mathsf{Derive}$, the parties
exchange nonces $N_A,N_B\leftarrow\$\,\{0,1\}^{2\lambda}$ and set
$\textit{sid}=(N_A,N_B,\text{digest}(U_1\|\cdots\|U_k))$, with
$K=\mathrm{HKDF}(\Theta;\ \mathrm{salt}=\textit{sid})$. Deriving the key from the
feature vector alone, $K=h(\Theta)$, would make it a deterministic function of
the challenge schedule and admit replay of a schedule whose key has
leaked and leaves the partnering function undefined. Nonce-salted derivation is
required for Theorem~\ref{thm:ma} and costs one round trip.
\end{remark}
 
\paragraph*{The authentication experiment.}
We use the standard Bellare-Rogaway formulation, restricted to the QSA boundary
(steps (4) through (13b) of Fig.~\ref{fig:qke-key-confirmation}). Let $\pi_P^s,\pi_V^t$ denote instances (sessions) of
the prover and verifier. A quantum polynomial-time (QPT) adversary $\mathcal{A}$
controls the network and has access to:
\begin{itemize}
\item $\mathsf{Send}(\pi,\textsf{msg})$: deliver an arbitrary message to an
instance and receive its reply. This subsumes injection, modification, dropping,
reordering, and man-in-the-middle; it is how online impersonation is modelled.
\item $\mathsf{Execute}(P,V)$: obtain the complete transcript of an honest
session. This is how \emph{repeated sessions} and passive transcript accumulation
enter the model.
\item $\mathsf{RevealKey}(\pi)$: obtain $K$ of a completed session. Models
downstream key misuse; forces key independence across sessions.
\item $\mathsf{RevealFeature}(\pi)$: obtain $\Theta$ of a \emph{completed}
session. This is the pessimistic leakage model of Attack~II, and is optional; we
report bounds with and without it.
\item $\mathsf{Corrupt}()$: obtain $S_0$ and \textsf{priv}. Ends freshness for all
instances (see the non-claims below).
\item $\mathsf{Test}(\pi)$: for the key-indistinguishability variant only.
\end{itemize}
$\mathcal{A}$ additionally receives, \emph{without any oracle}, the complete
classical circuit descriptions of every published $U_i$. Because these are classical
descriptions, coherent access is implied and need not be granted separately:
$\mathcal{A}$ may implement controlled-$U_i^{2^j}$, run QPE or LDQPE, diagonalise,
recover the factorisation $U_i=V_iD_iV_i^\dagger$ (which is recoverable in
$O(\mathrm{depth})$ by testing candidate split points), and run any QPT algorithm on
the result. We grant all of this. $\mathcal{A}$ is \emph{not} given $S_0$, the
labels $b_i$, or copies of $\ket{\psi_i}$; the absence of a state-copy oracle is a
load-bearing assumption about the provisioning layer and we flag it as such below.
The KDF $H$ is modelled as a random oracle to which $\mathcal{A}$ has quantum
superposition access (QROM), with $q_H$ queries; $q_s$ bounds the number of
$\mathsf{Send}/\mathsf{Execute}$ sessions.
 
\begin{definition}[Partnering and freshness]
Two instances are \emph{partners} if they hold the same $\textit{sid}$ and opposite
roles. An instance is \emph{fresh} if $\mathsf{Corrupt}$ was never queried and
neither it nor its partner was queried to $\mathsf{RevealKey}$.
\end{definition}
 
\begin{definition}[Success event; MA-advantage]
\label{def:ma}
$\mathcal{A}$ wins $\mathrm{Exp}^{\mathrm{MA}}$ if some \emph{fresh} instance
accepts (completes $\mathsf{Confirm}$) and has no honest partner instance. We write
$\mathrm{Adv}^{\mathrm{MA}}_{\mathrm{QSA}}(\mathcal{A}):=
\Pr[\mathcal{A}\text{ wins }\mathrm{Exp}^{\mathrm{MA}}]$.
Informally: \emph{false acceptance without a matching honest counterparty}.
\end{definition}
 
\paragraph*{Completeness is measured, not assumed.}
Separately from security, $\Pr[\text{honest session accepts}]\ge1-k\,p_{\mathrm{flip}}$,
where $p_{\mathrm{flip}}$ is the per-instance bucket error rate. Figure~\ref{fig:ldqpe-noise-sweep} and
Table~\ref{tab:ibm-fez-ldqpe} measure $p_{\mathrm{flip}}$ in simulation and on hardware. Neither is a
security experiment, and neither bounds
$\mathrm{Adv}^{\mathrm{MA}}$; we state this explicitly to separate the two.
 
\subsection*{The assumptions}
 
The advantage is bounded under exactly one non-standard assumption, stated in two
parts. Both are average-case conjectures tied to the specific compiler and
ensemble used here. We do \emph{not} reduce either to a canonical worst-case
problem, and we do not claim such a reduction is implied by the attack analyses of
the previous subsection.
 
\begin{assumption}[Label hiding, $\mathrm{LH}$]
\label{ass:lh}
Let $(b_i,\ket{\psi_i},V_i)\leftarrow\mathsf{Compile}$ for the symmetric compiler of
Methods, with $b_i$ uniform on $\{0,1\}^n$ and $\ket{\psi_i}=P_i^\dagger\ket{0^n}$
independent of $b_i$. Then for every QPT $\mathcal{A}$,
\begin{equation}
\big|\Pr[\mathcal{A}(U_i)=b_i]-2^{-n}\big|\ \le\ \varepsilon_{\mathrm{LH}}(\lambda),
\label{eq:lh}
\end{equation}
and more strongly the statistical distance between $(U_i,b_i)$ and $(U_i,b')$ for
independent uniform $b'$ is at most $\varepsilon_{\mathrm{LH}}$.
\end{assumption}
 
\begin{remark}[Status of $\mathrm{LH}$, and a construction that removes it]
\label{rem:lh}
Given the public $U_i$, the adversary knows the eigenbasis $\{V_i\ket{x}\}$ and the
full spectrum $\{\theta(x)\}$; the only secret is the index $b_i$. Blind guessing
succeeds with probability $2^{-n}$, and Eq.~\eqref{eq:lh} asserts that nothing better
is available. What is being assumed is therefore \emph{not} that the spectrum hides
the feature as Lemma~\ref{lem:unif} settles that unconditionally but that the
compiled circuit does not encode its own label. That is a property of the compilation
procedure rather than of quantum mechanics: $U_i$ is published as the parameter vector
$\vec\alpha^\star_i$ returned by an optimiser that was given $b_i$, and no counting
argument over the eigenbasis bears on whether $\vec\alpha^\star_i$ correlates with
$b_i$. The supporting intuition is a group action. With a leading layer of universal
single-qubit gates, $X^{b}$ can be absorbed into that layer: for every $b$ there is
an $\alpha'=\alpha'(\alpha,b)$ with $V(\alpha')=V(\alpha)X^{b}$ up to phase. Since
$X^{b}\ket{0^n}=\ket{b}$, the label-$b$ compilation objective is then the label-$0^n$
objective at reindexed parameters,
\begin{equation}
F_b(\alpha)\;=\;\bigl|\bra{\psi}V(\alpha)\ket{b}\bigr|^2
\;=\;\bigl|\bra{\psi}V(\alpha')\ket{0^n}\bigr|^2\;=\;F_{0^n}(\alpha'),
\label{eq:lh-reindex}
\end{equation}
so compiling for label $b$ is the task for $0^n$ under the bijection
$\alpha\leftrightarrow\alpha'$, and $\mathrm{LH}$ holds. It can
fail because descent from a fixed initialisation need not be equivariant under it.
Since $\mathrm{LH}$ states that $(U_i,b_i)$ and $(U_i,b')$ are statistically close for
independent uniform $b'$, any efficient predictor of $b_i$ from $\vec\alpha^\star_i$ is
by definition a distinguisher and refutes it.
 
$\mathrm{LH}$ carries one \emph{necessary} condition on the compiler of Methods:
\emph{the planted-state ensemble must be typical for the ansatz}. The compiler fits $V(\vec\alpha)$ so that
$V_i\ket{b_i}\approx\ket{\psi_i}$, hence index $b_i$ is the one eigenvector of $U_i$
that the fit was made to control, while $\{V_i\ket{x}\}_{x\neq b_i}$ are whatever the
ansatz returns. Suppose $\ket{\psi_i}$ is atypical for the ansatz: shallow, low-entanglement, or
otherwise structured relative to $V$'s reachable set. Then $b_i$ is the one anomalous
index, and an adversary reads it off by enumerating $x$ and ranking $V_i\ket{x}$ under
any statistic that separates the two ensembles into no training, no knowledge of $P_i$, and
no quantum computation required. We verify in Appendix~\ref{app:lh-tests} that this attack has no power when $\ket{\psi_i}$
is typical i.e the rank of $b_i$ under bipartite entanglement, participation ratio, and
Shannon entropy of the computational-basis distribution is uniform on $\{0,\dots,2^n-1\}$
to within sampling error and that it succeeds with probability approaching $1$ when
$\ket{\psi_i}$ is deliberately made atypical. $P_i$ must therefore be specified, not left
free: it is a requirement on the primitive, not a deployment note. We stress that
typicality is \emph{necessary and not sufficient}; ranking $V_i\ket{x}$ under a fixed
battery of statistics is one attack, and Appendix~\ref{app:lh-tests} also reports the parameter-level
predictor, which is another.

We ran both tests against the compiler of Methods with ensemble-matched planted states.
Across $N$ compiled instances at $n\le6$ (compilation overlap $\ge0.99$), the planted
index $b_i$ ranks uniformly under all ten structural statistics of the battery: at $n=6$,
$N=300$, the smallest $p$-value over the ten is $0.33$, against a Bonferroni threshold of
$0.005$, with the mean rank of $b_i$ equal to $31.5$ of $63$ as expected under
$\mathrm{LH}$. The supervised per-bit predictor $\hat b=f(\vec\alpha^\star)$ attains
held-out accuracy consistent with $1/2$ at every bit (maximum $0.57$ over six bits at
$n=6$, within one standard error of the baseline). As a positive control, replacing the
planted state by a deliberately atypical low-entanglement state makes the entanglement
statistic locate $b_i$ in the top rank for the majority of instances
($p<10^{-15}$), confirming that the null result is a property of ensemble matching and
not of a powerless test. The full protocol, statistic definitions, classifier family, and
per-$(n,m)$ tables are in Appendix~\ref{app:lh-tests}; the reference implementation is released with the
code ~\cite{kish_2026_21450694}.
 
$\mathrm{LH}$ is not an artefact of a poorly tuned optimiser; it is forced by a
tension between three things QSA would ideally want at once. First, the honest
prover must be able to \emph{regenerate} the planted state from the seed, so that
no physical register need be shipped to the endpoint (seed-derivability). Second,
the planted state must sit almost entirely on one eigenvector,
$|\bra{\psi_i}V_i\ket{b_i}|^2\ge1-\delta$, so that low-depth phase estimation
converges to its eigenphase at practical shot cost (high overlap). Third, the
public $U_i$ should reveal nothing about which eigenvector is planted, i.e.\
$\varepsilon_{\mathrm{LH}}=0$ (label independence). Any two of these can be had
together; all three cannot, and which one is sacrificed is exactly what
distinguishes the two provisioning modes.

Seed mode sacrifices label independence. The seed pins $\ket{\psi_i}$ first, so
$V_i$ must be \emph{fitted} to place that state on eigenvector $b_i$ at high
overlap. Because the fit is performed with knowledge of $b_i$, its output $V_i$
could in principle carry a trace of the label; $\mathrm{LH}$ is the assumption tested in Appendix~\ref{app:lh-tests} and there upheld for our compiler that it
does not. This is why seed mode rests on an assumption at all: it is the price of
regenerating the credential from a classical seed rather than delivering it.

State mode sacrifices seed-derivability instead, and thereby removes the
assumption. The order of generation reverses: sample $V_i$ and $b_i$ independently,
with no optimisation, and \emph{define} the planted state as
$\ket{\psi_i}:=V_i\ket{b_i}$, which the provisioning layer prepares in depth
$\mathrm{depth}(V)$ and delivers physically. The fit is gone, so no step ever sees
$b_i$ in a way that could bias $V_i$; the law of $U_i$ is independent of $b_i$ by
construction, $\varepsilon_{\mathrm{LH}}=0$ exactly, and Theorem~\ref{thm:ma} rests
on Assumption~\ref{ass:fu} alone. The expressive-unitary compiler of Methods is
thus not intrinsic to QSA but is the machinery of seed-derivability; a deployment
able to teleport its credential does not run it, and the compiler ceases to be an
attack surface.

The third corner is unavailable: one cannot keep both seed-derivability and label
independence. Sampling $V_i$ independently of the label while insisting on a
seed-derived $\ket{\psi_i}$ leaves the planted state aligned with its nearest
eigenvector only to $O(n\,2^{-n})$. The signal eigenphase still carries nonzero
weight, so a prover willing to expend exponentially many shots could in principle
recover it; but the shot cost defeats the completeness budget, so this corner is
ruled out on cost, not on principle. Label hiding is therefore precisely the price
of high-overlap seed-derivable operation, and delivering the state physically is
what buys it back. Empirically certifying LH enumerates the $2^n$-dimensional eigenbasis and is therefore limited to the small-$n$, exactly-enumerable regime; this is inherent to testing the assumption rather than a gap in the evidence, and is precisely why LH is stated as an assumption rather than proved. State mode removes it outright ($\varepsilon_{\mathrm{LH}}=0$ by construction), leaving it load-bearing only in seed mode.
\end{remark}
 
\begin{assumption}[Multi-session feature unpredictability, $\mathrm{FU}_{q_s}$]
\label{ass:fu}
For every QPT $\mathcal{A}$ given the public challenges of $q_s$ sessions derived from
a common $S_0$, together with the honest feature vectors $\Theta^{(1)},\dots,
\Theta^{(q_s-1)}$ of all but one,
\begin{equation}
\Pr\!\big[\mathcal{A}(\cdot)=\Theta^{(q_s)}\big]\ \le\ 2^{-mk}+\varepsilon_{\mathrm{FU}}(\lambda,q_s).
\label{eq:fu}
\end{equation}
\end{assumption}
 
\noindent Assumption~\ref{ass:fu} is where \emph{repeated sessions} are priced. It is
strictly stronger than a single-instance statement because all sessions share $S_0$.
For QSA-Q with the per-instance HKDF planted-state schedule of Eq.~(8), each session
uses independent $(P_i,b_i)$, so leakage does not accumulate against a single target
and $\varepsilon_{\mathrm{FU}}$ is not expected to grow with $q_s$; this is precisely
the design content of Attack~II. For QSA-M and QSA-C deployments with a persistent
$\ket{\psi}$, no such argument is available, and the renewal policy is therefore a
\emph{protocol requirement}: $q_s$ per planted state must be bounded by the renewal
interval, and Assumption~\ref{ass:fu} is only invoked for $q_s$ below that bound.
 
\paragraph*{An unconditional entropy lemma.}
For the symmetric compiler the second half of the security argument is not an
assumption but a theorem. Recall $D_i=\bigotimes_q R_z(\beta_{i,q})$ and
Eq.~\ref{eq:phase_closed_form}, $\theta(b)=\tfrac12\sum_{q=1}^n(2b_q-1)\beta_q \bmod 2\pi$.
 
\begin{lemma}[Feature uniformity]
\label{lem:unif}
Let $\beta_1,\dots,\beta_n$ be i.i.d.\ uniform on an interval of length $2\pi$ and
public, let $b$ be uniform on $\{0,1\}^n$ and independent of the published circuit,
and let $\mathsf{B}(\theta)\in\{0,1\}^m$ denote the $m$-bit bucket of $\theta$. Then
\begin{enumerate}
\item[(i)] the law $\mu_\beta$ of $\theta(b)$ has \emph{exact} Fourier coefficients
$\hat\mu_\beta(j)=\prod_{q=1}^n\cos(j\beta_q/2)$ for all $j\in\mathbb{Z}$;
\item[(ii)] for every $J\ge1$,
$\max_c\big|\Pr[\mathsf{B}=c]-2^{-m}\big|\le \tfrac{1}{J+1}
+C\,(1+\ln J)\max_{1\le j\le J}|\hat\mu_\beta(j)|$ for an absolute constant $C$;
\item[(iii)] with probability at least $1-2^{-m}$ over $\beta$,
$\max_{1\le j\le 2^{2m}}|\hat\mu_\beta(j)|\le 2^{-n+c_1\sqrt{nm}}$ for an absolute
constant $c_1$; consequently, whenever $n\ge 2m+c_1\sqrt{nm}+\log_2 m+O(1)$,
\begin{equation}
\max_c\Pr[\mathsf{B}=c]\ \le\ 2^{-m}\big(1+2^{-m+2}\big),
\qquad\text{i.e.}\qquad
H_\infty\!\left(\theta^\star_i\,\middle|\,U_i\right)\ \ge\ m-1 .
\label{eq:minent}
\end{equation}
\end{enumerate}
\end{lemma}
 
\begin{corollary}[Feature min-entropy]
\label{cor:minent}
Under Assumption~\ref{ass:lh} and the hypothesis of Lemma~\ref{lem:unif}(iii), and
with $(b_i,\beta_i)$ independent across $i$,
$H_\infty(\Theta\mid U_1,\dots,U_k)\ \ge\ k(m-1)-k\log_2(1+\varepsilon_{\mathrm{LH}}2^{n})$.
\end{corollary}
 
\begin{remark}[Regime of validity, and what we verify numerically]
\label{rem:mm}
Lemma~\ref{lem:unif}(iii) requires $n\gtrsim 3m$; the constants are not optimised.
The instances demonstrated in this paper use $n=m$, which violates that hypothesis.
There we do not claim the bound. Instead we compute $\max_c\Pr[\mathsf{B}=c]$
\emph{exactly}, by enumerating all $2^n$ labels, for $n=m\le18$, and report
$\kappa:=2^m\max_c\Pr[\mathsf{B}=c]$ (Appendix~\ref{app:kappa}). We find $\kappa$
grows \emph{linearly}, $\kappa\approx0.36\,n+1.1$ for $n\ge4$, so the per-instance
min-entropy loss $\log_2\kappa$ rises from about one bit at $n=m=4$ to under three
bits at $n=m=18$: a loss of a few bits rather than a collapse; Appendix~\ref{app:uniformity-highn} confirms the complementary regime $n\gtrsim3m$, where the proven bound applies and the loss falls below a fraction of a bit. The growth is slightly
faster than the independent balls-into-bins prediction because the bucket occupancies
are correlated, $\theta(b)$ being a signed sum of the \emph{same} $n$ angles across
all labels; we report this as a numerical finding, not a theorem. This distinction,
provable for $n\gtrsim3m$ and numerically verified at $n=m$, is exactly the kind of
separation we intend throughout.
\end{remark}
 
\subsection*{The theorem}
 
\begin{theorem}[Mutual authentication]
\label{thm:ma}
Model the KDF as a quantum-accessible random oracle. Let $\mathcal{A}$ be QPT,
making at most $q_s$ session queries and $q_H$ random-oracle queries against QSA
instantiated with parameters $(n,m,k)$, nonce length $2\lambda$, confirmation
response length $\ell$, and an AEAD $\mathcal{E}/\mathcal{D}$. Then, under
Assumption~\ref{ass:fu},
\begin{equation}
\mathrm{Adv}^{\mathrm{MA}}_{\mathrm{QSA}}(\mathcal{A})\ \le\
\underbrace{4q_s\sqrt{q_H\big(2^{-mk}+\varepsilon_{\mathrm{FU}}\big)}}_{\text{feature unpredictability}}
\;+\;\underbrace{q_s\big(\mathrm{Adv}^{\mathrm{auth}}_{\mathcal{E}}(\mathcal{B})+2^{-\ell}\big)}_{\text{confirmation forgery}}
\;+\;\underbrace{q_s^2\,2^{-2\lambda}}_{\text{nonce collision}} .
\label{eq:mabound}
\end{equation}
For the symmetric compiler, Assumption~\ref{ass:lh} together with
Corollary~\ref{cor:minent} gives
$2^{-mk}+\varepsilon_{\mathrm{FU}}\le 2^{-k(m-1)}+k\,\varepsilon_{\mathrm{LH}}$ at
$q_s=1$; for $q_s>1$, Assumption~\ref{ass:fu} is invoked directly.
\end{theorem}
 
\begin{corollary}[Key indistinguishability]
\label{cor:ki}
Under the same hypotheses,
$\mathrm{Adv}^{\mathrm{KI}}_{\mathrm{QSA}}(\mathcal{A})\le
4q_s\sqrt{q_H(2^{-mk}+\varepsilon_{\mathrm{FU}})}$.
Session-key indistinguishability is thus a step in the proof of
Theorem~\ref{thm:ma} rather than an independent target.
\end{corollary}
 
\noindent Proofs are in Appendix~\ref{app:proofs}. The structure is three game hops: (G1) abort on
nonce collision; (G2) replace each fresh session key by an independent uniform
string, at a cost bounded by one-way-to-hiding in the QROM with a
measure-and-extract reduction to Assumption~\ref{ass:fu}; (G3) with uniform
independent keys, acceptance without a partner is an AEAD forgery.
 
\paragraph*{Parameter guidance implied by Eq.~\eqref{eq:mabound}.}
Two consequences are worth stating because they change our recommended parameters.
First, the $\sqrt{\;\cdot\;}$ loss is not an artefact. \emph{Key confirmation
inherently supplies an offline verifier}: an eavesdropper who observes
$\mathcal{C}_1=\mathcal{E}(M_1,K)$ can test candidate $K$, hence candidate $\Theta$,
without interacting. Rate limiting bounds \emph{online} impersonation only; it does
not bound offline recovery. Generic recovery therefore costs $2^{mk}$ classical and
$\Theta(2^{mk/2})$ quantum queries, and $mk$ must be sized as an offline symmetric
key length: $mk\ge2\lambda$ for $\lambda$-bit post-quantum security, or $mk\ge\lambda$
under the same convention by which AES-256 is assigned 256-bit security despite
Grover; we size $mk$ as an offline key length rather than relying on the protocol
boundary to exclude oracle access. Second, $\ell\ge128$ and nonces of
$2\lambda$ bits suffice for the remaining terms.
 
\subsection*{Non-claims}
 
\begin{figure}[t!]
  \centering

  % --- Symmetric: U = V D V^\dagger (fast-power) ---
  \begin{subfigure}[t]{0.49\linewidth}
    \centering
    \includegraphics[width=\linewidth]{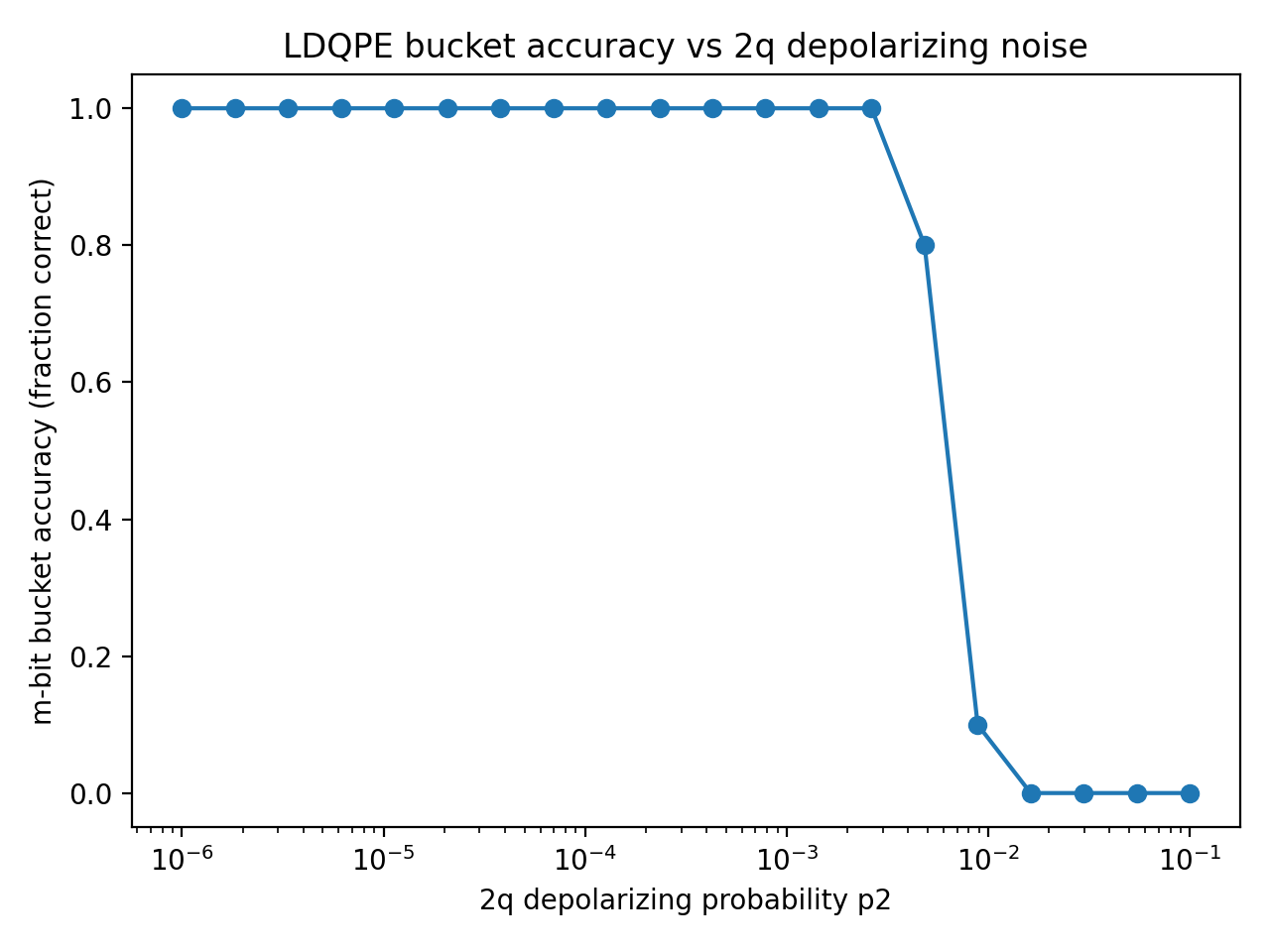}
    \caption{\textbf{Symmetric $U=VDV^\dagger$ (fast-power): bucket accuracy vs two-qubit depolarizing noise.}
    Fraction of trials (num\_reps $=20$ seeds) that returned the correct $m$-bit bucket for $m=8$ and $n=8$.
    Transpiled gate counts for the largest-moment real-part circuit are
    \{rz: 579, cx: 385, rx: 225, ry: 149, measure: 1\}.}
    \label{fig:ldqpe-sym-acc-vs-p2}
  \end{subfigure}
  \hfill
    % --- Asymmetric: U = V_L V_R^\dagger ---
  \begin{subfigure}[t]{0.49\linewidth}
    \centering
    \includegraphics[width=\linewidth]{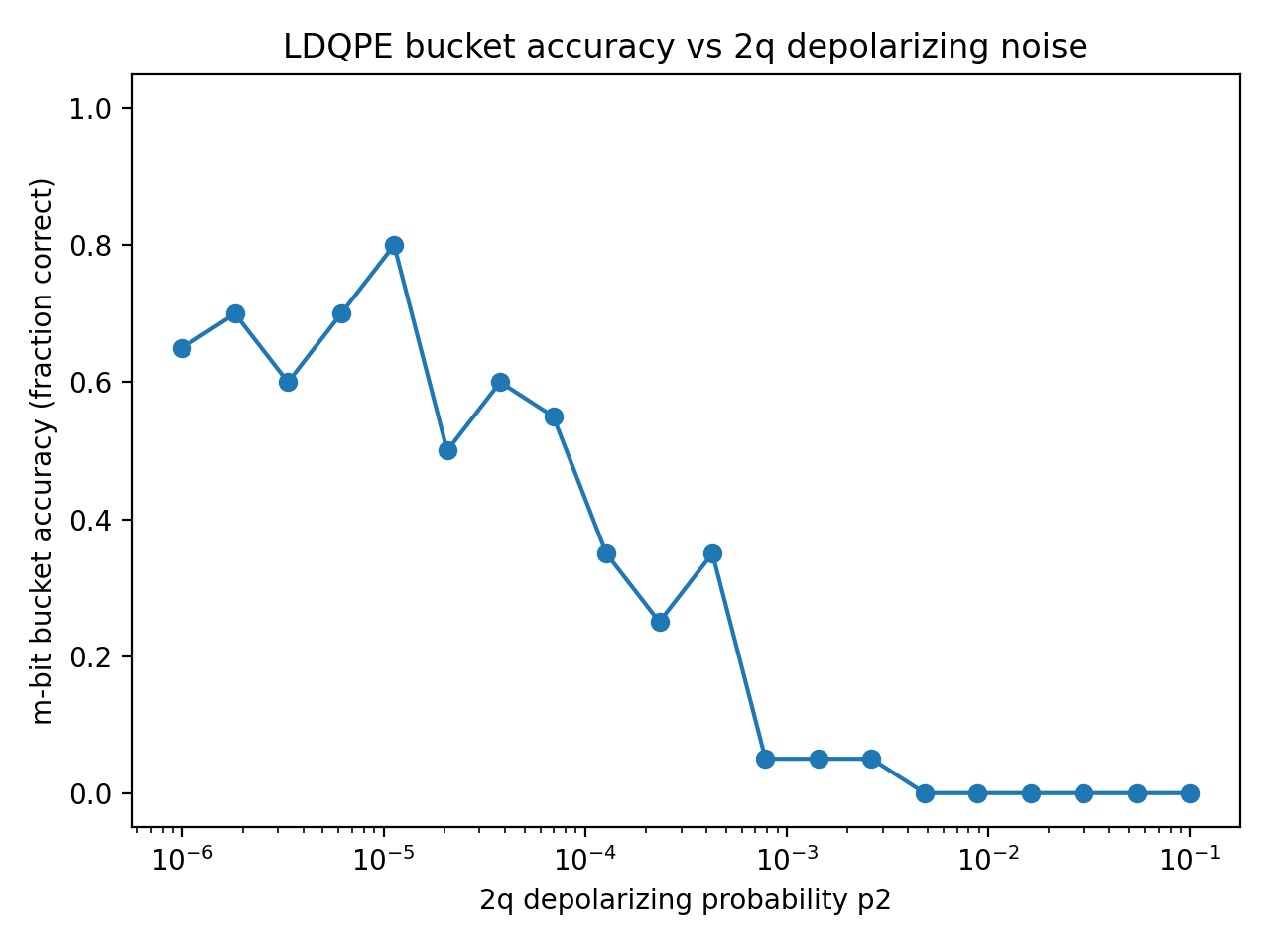}
    \caption{\textbf{Asymmetric $U=V_LV_R^\dagger$: bucket accuracy vs two-qubit depolarizing noise.}
    Transpiled gate counts for the largest-moment real-part circuit are
    \{rz: 51345, cx: 39047, rx: 25989, ry: 11285, measure: 1\}.}
    \label{fig:ldqpe-asym-acc-vs-p2}
  \end{subfigure}
  
  % \begin{subfigure}[t]{0.49\linewidth}
  %   \centering
  %   \includegraphics[width=\linewidth]{ldqpe_phaseerror_vs_p2_n8m8.png}
  %   \caption{\textbf{Symmetric $U=VDV^\dagger$ (fast-power): circular phase error vs two-qubit depolarizing noise.}
  %   Median circular error $d(\hat{\theta},\theta^\star)$ with interquartile range (IQR) over num\_reps $=20$ seeds, for $m=8$ and $n=8$.
  %   Here $d(\hat{\theta},\theta^\star)=\min_{k\in\mathbb{Z}}|\hat{\theta}-\theta^\star+2\pi k|$.}
  %   \label{fig:ldqpe-sym-phaseerr-vs-p2}
  % \end{subfigure}

  % \vspace{0.35em}

  % \hfill
  % \begin{subfigure}[t]{0.49\linewidth}
  %   \centering
  %   \includegraphics[width=\linewidth]{ldqpe_phaseerror_vs_p2_n8m8asymmetric20.png}
  %   \caption{\textbf{Asymmetric $U=V_LV_R^\dagger$: circular phase error vs two-qubit depolarizing noise.}}
  %   \label{fig:ldqpe-asym-phaseerr-vs-p2}
  % \end{subfigure}

  \caption{\textbf{Noise sensitivity of LDQPE on compiled QSA-Q instances: symmetric challenge compilers.}
  Depolarizing noise is applied with two-qubit error probability $p_2$ on all two-qubit gates, single-qubit rate $p_1=0.1p_2$, and fixed symmetric readout error probability $0.01$.
  Each point aggregates num\_reps $=20$ independent trials (distinct simulator seeds and shot noise), with shots $=4000$ per circuit.}
  \label{fig:ldqpe-noise-sweep}
\end{figure}

\begin{enumerate}
\item \textbf{No worst-case reduction.} Assumptions~\ref{ass:lh} and~\ref{ass:fu} are
average-case conjectures tied to the specific ensembles and compilers of Methods. We
do not reduce them to Local Hamiltonian, UniqueQMA, or any canonical problem, and
Appendix~\ref{app:complexity-perspective} is motivation, not a reduction.
\item \textbf{No forward secrecy against verifier compromise.} $\Theta$ is a
deterministic function of $S_0$ and the public schedule, so an adversary who obtains
$S_0$ and $\{b_i\}$ recomputes $\Theta$, hence $K$, for every recorded past session.
QSA is not a forward-secure key exchange in the Diffie-Hellman sense and should be
composed with one where per-session forward secrecy against \emph{both} endpoints is
required. Two qualifications sharpen this. First, \emph{state mode is forward-secure against prover
compromise}, and structurally so: the prover holds no $S_0$, only the physical
registers $\ket{\psi_i}$, and measurement consumes them. An adversary who captures the
endpoint at time $t$ obtains the states not yet spent; the states that produced past
$\Theta$ no longer exist, and by no-cloning no copy of them can have been retained.
Provided the endpoint erases $\Theta$ and $K$ after use, the same hygiene any
forward-secrecy claim requires past session keys survive endpoint capture. This
asymmetry matches the deployment threat: the endpoint is the component that is
physically seized, the verifier sits in a controlled facility. Seed mode has no such
property, since the prover holds $S_0$. Second, rotating $S_0$ with verifiable erasure
of the old value gives forward secrecy at epoch granularity, but the guarantee is then
supplied by the re-provisioning channel and not by QSA, and we do not count it here.
\item \textbf{No security under a state-copy oracle.} Since $V_i$ is public, a
\emph{single} copy of $\ket{\psi_i}$ breaks Assumption~\ref{ass:lh} at any $n$: the
adversary runs the honest fast path, applying $V_i^\dagger$ and measuring in the
computational basis to recover $b_i$ with probability at least $1-\delta$ (exactly, in
state mode), with no tomography and no need for polynomially many copies. Confidentiality
of the provisioning layer is therefore load-bearing, and lies outside the QSA boundary by
assumption.
\item \textbf{No claim that quantumness adds security in seed mode.} In seed mode the
adversary's task is unpredictability of a keyed public function, as it would be for a
classical PRF challenge-response. The distinguishing claims are state mode and
evaluability, as set out above.
\item \textbf{Attacks are cryptanalysis, not lower bounds.} Attacks I to III and the
appendix reference attacks bound the success probability of \emph{specific}
strategies. They are evidence for Assumptions~\ref{ass:lh} and~\ref{ass:fu}, not proofs
of them, and no bound in this paper depends on the claim that they are exhaustive.
\item \textbf{Hardware and noise results are completeness measurements.} Figures~\ref{fig:ldqpe-noise-sweep}
and~\ref{fig:ldqpe-hardware-fez} and Tables~\ref{tab:QSA-costs-cvp} and~\ref{tab:ibm-fez-ldqpe} quantify $p_{\mathrm{flip}}$ and gate cost. They do not enter
Eq.~\eqref{eq:mabound}.
\end{enumerate}
\subsection*{Compilation and verification of unitary challenges for QSA-Q}
\label{subsec:compilation-fast-eval-QSAq}

We report two compilation families for QSA-Q challenge unitaries that (i) embed a high-overlap planted eigenstate and (ii) yield robust low-precision phase features for honest evaluation on noisy devices. Full compilation details and optimisation settings are given in Methods and are available at \cite{kish_2026_21450694}; here we focus on the resulting planted spectral structure and on the evaluation consequences of each family.

%\subsubsection*{Two-party (single-prover) unitary challenges: symmetric vs asymmetric compilers}
\label{subsec:QSAq_compilers_two_party}

In the symmetric compiler we publish challenges of the form $U=VDV^\dagger$, where $V$ is learned so that $V\ket{b}\approx\ket{\psi}$ for a hidden computational basis state $\ket{b}$ and planted state $\ket{\psi}=P^\dagger\ket{0^n}$. The diagonal layer is
\[
D=\bigotimes_{q=1}^{n}R_z(\beta_q),\qquad \beta_q\sim\mathrm{Unif}[-\pi,\pi],
\]
and the corresponding ``signal'' eigenphase associated with the hidden label $b$ is available in closed form:
\[
\langle b|D|b\rangle=e^{i\theta(b)},\qquad
\theta(b)=\frac{1}{2}\sum_{q=1}^{n}(2b_q-1)\beta_q\;(\mathrm{mod}\;2\pi).
\tag{\ref{eq:phase_closed_form}}
\]
Thus, the verifier can compute the intended phase feature directly from the hidden bitstring $b$ and angles $\{\beta_q\}$, without diagonalisation or LDQPE evaluation. 

To remove this direct ``read-off'' structure, the asymmetric compiler publishes challenges of the form $U=V_LV_R^\dagger$, where $V_L$ and $V_R$ are independently learned expressive circuits. In this family, the verifier has no closed-form eigenphase prediction; the relevant phase feature must be extracted by the same LDQPE-based routine used by the prover (or by dense diagonalisation in small-n evaluation experiments). In our reference implementation at fixed $(n,m)$, the asymmetric routine required roughly $2\times$ the compilation time of the symmetric family, consistent with learning two maps instead of one.

To visualise the planted-eigenstate effect, we diagonalise each compiled symmetric unitary instance and compute overlap weights $|\langle v_i|\psi\rangle|^2$ with eigenvectors $\{|v_i\rangle\}$. We aggregate overlap mass into $M=2^6$ uniform eigenphase bins over $[0,2\pi)$:
\[
p_k \;=\; \sum_{i\,:\,\mathrm{bin}(\theta_i)=k} |\langle v_i|\psi\rangle|^2.
\]
Figure~\ref{fig:bin-mass-honest-vs-eve} shows strong localisation for the planted state, while a baseline ``Eve'' state (random and independent of the planted structure) remains broadly delocalised. This is the relevant effect for the scalable symmetric construction used in the main QSA-Q pathway. The asymmetric compiler showed the same qualitative trend in small reference experiments, but we do not emphasise it because its compilation and evaluation costs are substantially higher due to exponential gate depth, as will be shown in the next subsection.

% ============================================================
% Replace the entire existing LDQPE figure block with this one
% ============================================================

% ============================================================
% Replace the subsection text with this revised version
% ============================================================

\subsection*{Simulated LDQPE noise-sensitivity on compiled QSA-Q instances}
\label{subsec:ldqpe-noise-sensitivity}

To complement the calibration-style cost models above, we provide the prover demonstration of the low-depth QPE (LDQPE) primitive \cite{ni2023lowdepthqpe} that underlies the honest QPU evaluation pathway of QSA-Q. The honest evaluation estimates complex moments
\begin{equation}
Z_j \;=\;\langle\psi|U^{2^j}|\psi\rangle,\qquad j=0,1,\ldots,m-1,
\end{equation}
via Hadamard tests, and then performs a candidate-set unwrapping step to output an $m$-bit phase bucket. Concretely, for each $j$ we estimate $Z_j$ from the real and imaginary Hadamard-test circuits, and apply the standard unwrap rule
\[
S_j=\Bigl\{\tfrac{2\pi k+\arg(Z_j)}{2^j}\Bigr\}_{k=0}^{2^j-1},\qquad
\theta_j=\arg\min_{\theta\in S_j}|\theta-\theta_{j-1}|_{2\pi},
\]
followed by quantisation of $\theta_{m-1}$ into an $m$-bit bucket.

%\paragraph{Hadamard-test primitive.}
Each moment $Z_j$ is obtained from two circuits (real and imaginary parts) of the form:
\begin{equation}
\Re(Z_j)=\Pr(0)-\Pr(1)\ \text{on the ancilla in the $X$ basis},\qquad
\Im(Z_j)=\Pr(0)-\Pr(1)\ \text{with an $S^\dagger$ phase on the ancilla},
\end{equation}
where the ancilla controls the application of $U^{2^j}$ to the system register prepared in $\ket{\psi}$.
A compact circuit sketch is shown in Fig. \ref{fig:app_hadamard_moment1}:

%\paragraph{Symmetric vs asymmetric evaluation.}
For the asymmetric compiler $U=V_LV_R^\dagger$, implementing $U^{2^j}$ typically requires repeating (or compiling) controlled-$U$ blocks whose two-qubit gate count grows rapidly with $j$, so increasing $m$ incurs a sharp depth penalty.
In contrast, for the symmetric compiler $U=VDV^\dagger$, we exploit the fast-power identity
\begin{equation}
U^{2^j} \;=\; V D^{2^j} V^\dagger,
\end{equation}
and evaluate the moment circuits by composing controlled-$V$, controlled-$D^{2^j}$, and controlled-$V^\dagger$ once per $j$.
Crucially, $D$ is a diagonal $R_z$ layer,
\[
D=\bigotimes_{q=1}^{n}R_z(\beta_q),
\qquad
D^{2^j}=\bigotimes_{q=1}^{n}R_z(2^j\beta_q),
\]
so the structure (and in practice the gate count) of the diagonal controlled layer does not scale exponentially with $m$; only the rotation angles are rescaled. This is the main reason the symmetric $VDV^\dagger$ family can support larger $m$ on NISQ devices.

%\paragraph{Noise model and simulation settings.}
We emphasise that these experiments are sanity checks rather than hardware benchmarks. We apply depolarising noise with two-qubit error probability $p_2$ on all two-qubit gates, set the single-qubit rate proportionally as $p_1=0.1p_2$, and include a fixed symmetric readout error probability of $0.01$.
For each noise point, we run num\_reps $=20$ independent trials (distinct simulator seeds) and use shots $N_s=4000$ per circuit. For the \(m=8\) LDQPE simulations, we set \(\epsilon=2^{-8}\), corresponding to 8-bit bucket precision, and used \(\xi=1\), the standard full-depth choice in Algorithm~2 of Ref.~\cite{ni2023lowdepthqpe}. Here \(\xi\) is the parameter that trades off depth against robustness by shrinking the retained phase interval at each iteration, and it sets the maximal runtime scale as \(T_{\max}=O(\xi\epsilon^{-1})\). For the typical compiled-instance range \(\delta\in[0.1,0.25]\), Eq.~(16) in Ref. \cite{ni2023lowdepthqpe} therefore implies that the required shot count is still only \(O(10^2)\) per moment circuit for standard constant failure probability, so our use of \(N_s=4000\) shots for \(n=m=8\) is conservative.

Figure~\ref{fig:ldqpe-noise-sweep} summarises the $m$-bit bucket accuracy as a function of the two-qubit depolarizing probability $p_2$ for $n=8$ and $m=8$, and we include both the symmetric fast-power LDQPE and the asymmetric $U=V_LV_R^\dagger$ case. The symmetric setting remains highly accurate at low noise, with bucket recovery staying essentially perfect up to a few $\times 10^{-3}$, then degrading sharply: the accuracy is still around $0.8$ at $p_2\approx 5\times 10^{-3}$, falls to about $0.1$ near $10^{-2}$, and reaches zero for $p_2\gtrsim 2\times 10^{-2}$. By contrast, the asymmetric sweep is already error-prone at extremely small two-qubit noise, with bucket accuracy fluctuating around only $0.6$ to $0.8$ even for $p_2\sim 10^{-6}$ to $10^{-5}$, dropping below roughly $0.6$ by $p_2\sim 10^{-5}$ to $10^{-4}$, falling to around one third by $p_2\sim 10^{-4}$ to $10^{-3}$, and becoming essentially unusable beyond about $p_2\approx 10^{-3}$; by $p_2\gtrsim 5\times 10^{-3}$ the success rate is effectively zero. The corresponding transpiled largest-moment real-part circuits differ dramatically in size: the symmetric circuit has gate counts {rz: 579, cx: 385, rx: 225, ry: 149, measure: 1}, whereas the asymmetric circuit has {rz: 51345, cx: 39047, rx: 25989, ry: 11285, measure: 1}. Overall, these results show that two-qubit noise at the $10^{-3}$ level is already too large for reliable asymmetric LDQPE bucket recovery, while the symmetric fast-power construction remains viable until the noise approaches the $10^{-2}$ regime. 

For comparison, if one replaces LDQPE by conventional $m$-ancilla QPE in a regime where the planted state overlap with the dominant eigenvector is only moderate, say $p_0\sim 0.05$ to $0.1$, then the circuit cost becomes much less attractive even when the eigenphase gap is still favourable. The advantage of QPE is that, unlike LDQPE, it does not require the stringent LDQPE threshold $p_0>4-2\sqrt{3}\approx 0.536$ for stable bucket recovery; however, this comes at the price of coherently implementing all controlled powers $U^{2^0},U^{2^1},\ldots,U^{2^{m-1}}$ in a single circuit together with an inverse QFT on the $m$-qubit phase register. For the symmetric fast-power family $U=VDV^\dagger$, where each $U^{2^j}$ can still be compiled as $VD^{2^j}V^\dagger$ with essentially the same structure as the largest LDQPE moment circuit, a rough estimate for $m=n=8$ is therefore obtained by summing eight controlled-power blocks of approximately the same size as the $j=7$ LDQPE circuit. Using the observed largest-moment counts {rz: 579, cx: 385, rx: 225, ry: 149}, this gives a controlled-power subtotal of about {rz: 4632, cx: 3080, rx: 1800, ry: 1192}, before the inverse QFT is added. An 8-qubit inverse QFT contributes a further $8$ Hadamards and $28$ controlled-phase rotations, which, after standard decomposition, corresponds to roughly another $\sim 56$ two-qubit entangling gates plus a modest number of single-qubit phase gates. Thus, even in the optimistic fast-power setting, QPE at $m=8$ is already a few thousand two-qubit-gate procedure, roughly an order of magnitude larger than a single LDQPE moment circuit. Moreover, when $p_0\sim 0.05$ to $0.1$, one should also expect on the order of $\log(1/\varepsilon_{\text{fail}})/p_0$ repetitions before the dominant eigenphase is sampled with high confidence, further multiplying the effective experimental cost.

%\paragraph{Context: present-day gate error rates.}
On present-day devices, reported two-qubit gate error rates in the \(10^{-3}\) range are already achievable in best-case settings, but they are not yet generally at the \(10^{-4}\) to \(10^{-5}\) levels implied by the strictest ``single-shot'' heuristics for deep controlled blocks. For example, IBM has reported best two-qubit gate error rates on the order of \(8\times 10^{-4}\) on its Heron-generation superconducting processors~\cite{ibm_annual_letter_2024}, while Quantinuum has reported two-qubit gate fidelity of \(99.921\%\) across all qubit pairs for Helios, corresponding to an error rate of approximately \(7.9\times 10^{-4}\)~\cite{quantinuum_helios_blog,quantinuum_helios_press}. IonQ has also announced prototype two-qubit gate fidelities exceeding \(99.99\%\), corresponding to error rates on the order of \(10^{-4}\), for research prototypes~\cite{ionq9999}. Against this backdrop, the symmetric fast-power construction is attractive because it enables larger \(m\) without forcing exponentially larger controlled-\(U^{2^j}\) blocks, shifting the primary feasibility lever back to compilation efficiency and two-qubit gate count rather than to the exponential moment-depth growth typical of generic unitaries. This confirms that the asymmetric construction is effectively overwhelmed by realistic two-qubit noise, whereas the symmetric fast-power pathway is substantially more noise-tolerant and remains the only plausible scalable NISQ-oriented route. These simulation results are further supported by small-instance hardware executions on the Heron-class IBM \texttt{ibm\_fez}, discussed next.

\begin{figure}[ht!]
    \centering
    \begin{subfigure}[ht!]{0.49\linewidth}
        \centering
        \includegraphics[width=\linewidth]{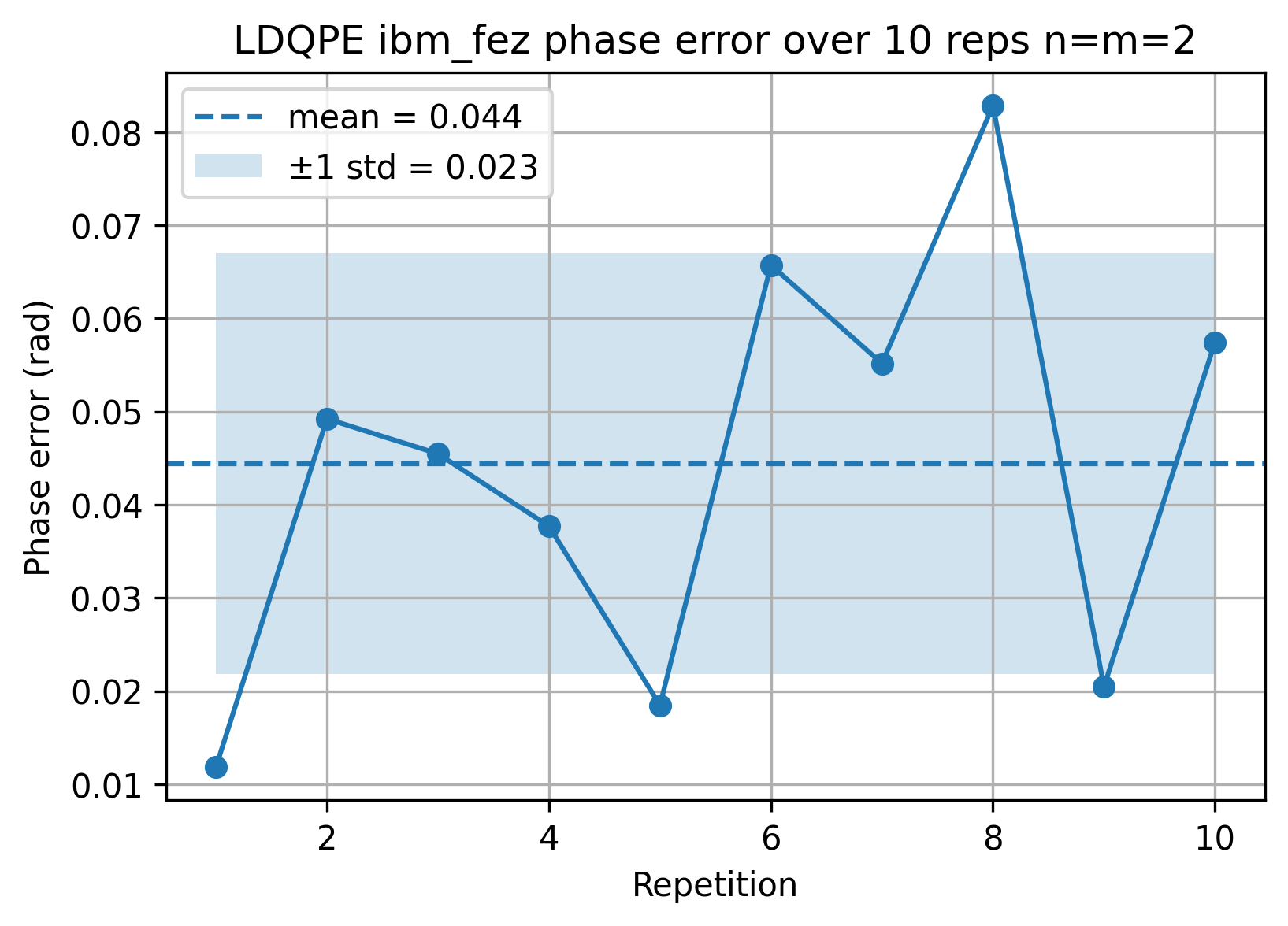}
        \caption{\textbf{Hardware phase error over $10$ repetitions for the symmetric LDQPE instance with $n=m=2$ on \texttt{ibm\_fez}.} All $10$ repetitions recovered the correct bucket. The mean phase error was $0.0444$ rad with standard deviation $0.0226$ rad.}
        \label{fig:ldqpe-hardware-n2m2}
    \end{subfigure}
    \hfill
    \begin{subfigure}[ht!]{0.49\linewidth}
        \centering
        \includegraphics[width=\linewidth]{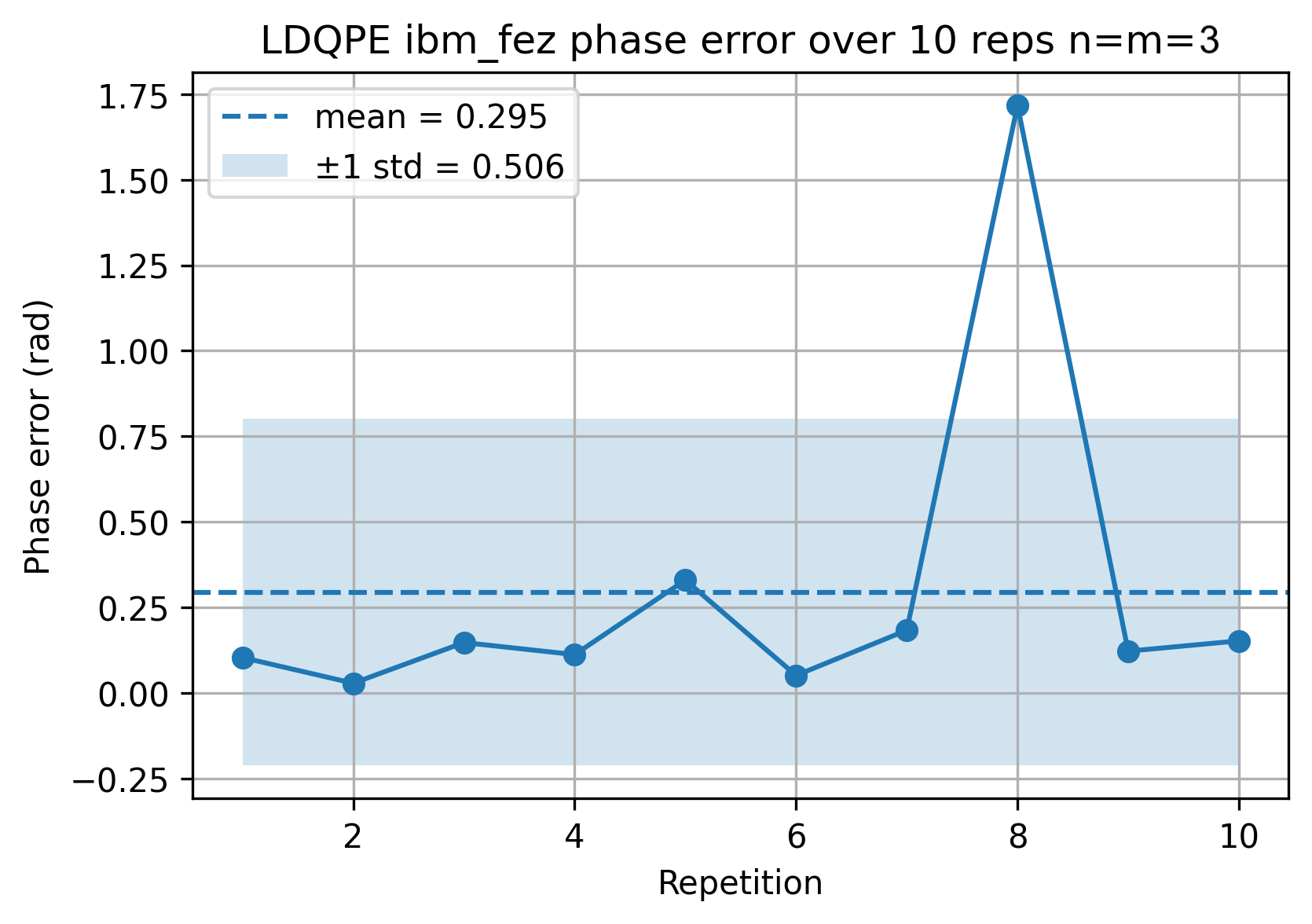}
        \caption{\textbf{Hardware phase error over $10$ repetitions for the symmetric LDQPE instance with $n=m=3$ on \texttt{ibm\_fez}.} Nine of ten repetitions recovered the correct bucket; one large-error outlier produced a bucket flip, yielding bucket accuracy $90\%$.}
        \label{fig:ldqpe-hardware-n3m3}
    \end{subfigure}

    \vspace{0.5em}

    \begin{subfigure}[ht!]{0.62\linewidth}
        \centering
        \includegraphics[width=\linewidth]{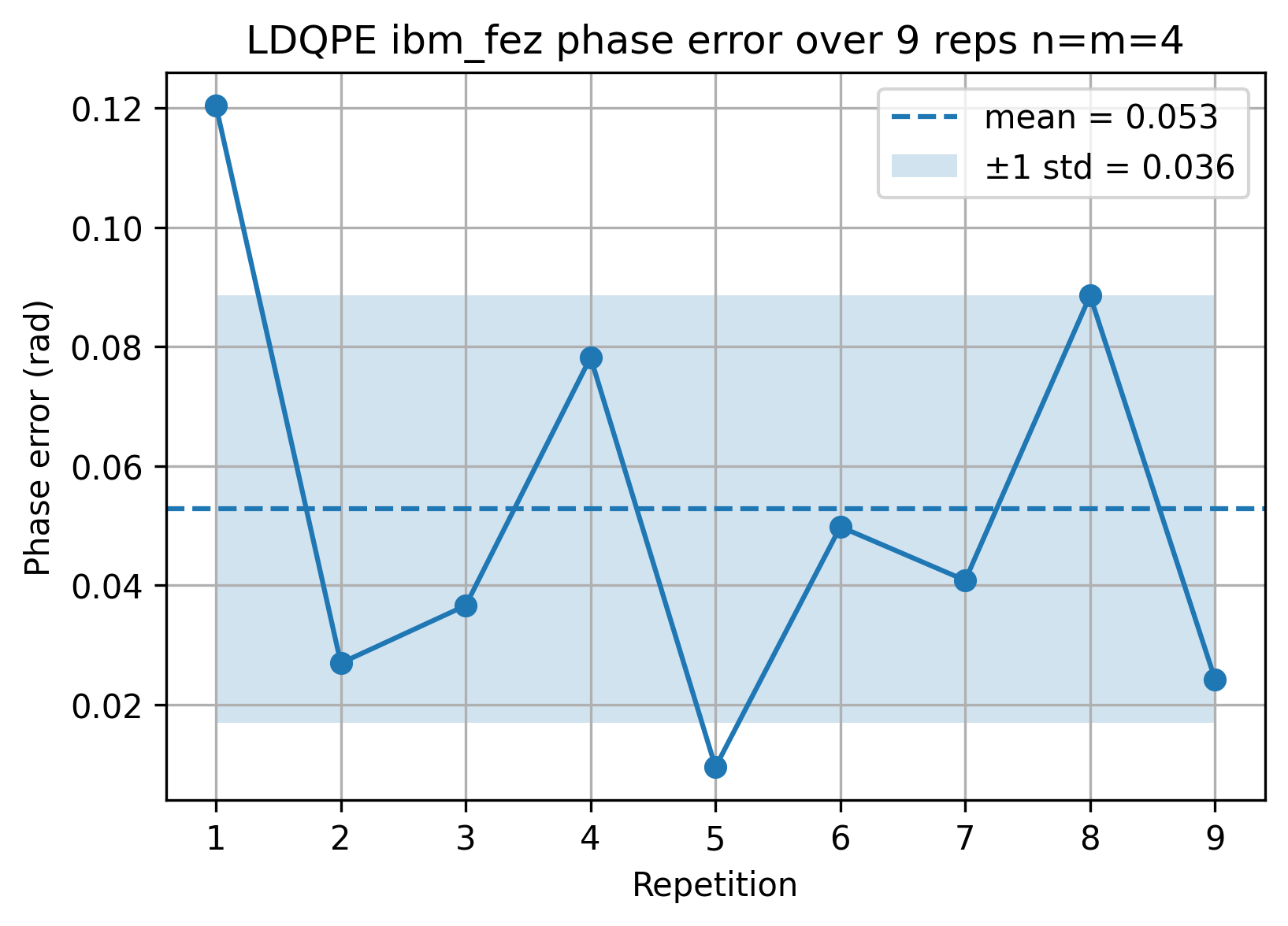}
        \caption{\textbf{Hardware phase error over $9$ completed repetitions for the symmetric LDQPE instance with $n=m=4$ on \texttt{ibm\_fez}.} Eight of the nine completed repetitions recovered the correct bucket. The mean phase error was $0.0528$ rad with standard deviation $0.0361$ rad.}
        \label{fig:ldqpe-hardware-n4m4}
    \end{subfigure}

    \caption{\textbf{Real-device LDQPE demonstrations on IBM \texttt{ibm\_fez}.} The $n=m=2$ instance is stably recovered on hardware, while the $n=m=3$ and partially completed $n=m=4$ instances remain mostly successful but exhibit occasional bucket failures once the transpiled controlled-Hadamard-test circuits reach the $10^2$-two-qubit-gate regime. We used \(512\) shots per real/imaginary Hadamard-test circuit for the \(m=n=2,3,4\) instances; using Eq.~(16) of Ref.~\cite{ni2023lowdepthqpe} with \(\epsilon=2^{-m}\), the standard full-depth choice \(\xi=1\), a representative compiled-instance range \(\delta\in[0.05,0.25]\), and a standard constant failure probability \(\eta=10^{-2}\), the corresponding theorem-level minimum shot counts are only \(124\text{ to }356\) for \(m=2\), \(128\text{ to }370\) for \(m=3\), and \(132\text{ to }382\) for \(m=4\), so \(512\) shots per circuit is comfortably conservative in all three cases.}
    \label{fig:ldqpe-hardware-fez}
\end{figure}
\subsection*{Hardware feasibility of the general LDQPE route on IBM \texttt{ibm\_fez}}
\label{subsec:ldqpe-hardware-fez}

Figure~\ref{fig:ldqpe-hardware-fez} shows the real-device phase-error distributions for the symmetric fast-power LDQPE pathway on IBM's \texttt{ibm\_fez}, while Table~\ref{tab:ibm-fez-ldqpe} summarises the corresponding transpiled gate counts and phase-recovery statistics. These runs complement the simulation model noise sweeps by providing a direct hardware sanity check of the structured verifier-driven route \(U=VDV^\dagger\) on a present-day superconducting backend. We benchmark LDQPE deliberately, since it is the general evaluator required for QSA-C, QSA-M, and the asymmetric compiler; the symmetric fast path of Methods is strictly easier on the same device (a single ancilla-free forward circuit $V^\dagger$ followed by a measurement, with no controlled powers or Hadamard test), so these runs are a conservative lower bound on what QSA-Q can execute rather than a performance ceiling. At the time of writing, IBM reports for \texttt{ibm\_fez} a median two-qubit error of \(2.77\times 10^{-3}\), layered two-qubit error of \(5.06\times 10^{-3}\), and median readout error of \(1.55\times 10^{-2}\), placing the device squarely in the noise regime discussed above.

We restricted attention to the symmetric compiler, since this is the practically relevant pathway identified by the simulation study. For each instance, LDQPE moments were estimated on hardware via Hadamard tests using \(512\) shots per circuit, and the full LDQPE evaluation was repeated multiple times. As visible in Fig.~\ref{fig:ldqpe-hardware-fez}, the \(n=m=2\) instance is highly stable, with all \(10\) repetitions recovering the correct bucket and only small phase spread. The \(n=m=3\) instance remains mostly successful but shows one clear outlier that produces a bucket flip, reducing bucket accuracy to \(90\%\). The partially completed \(n=m=4\) run likewise remains mostly successful, with \(8/9\) correct buckets and phase estimates still clustered fairly close to the target despite one failure.

\begin{table*}[ht!]
\centering
\caption{\textbf{Hardware LDQPE results on IBM \texttt{ibm\_fez}} for the symmetric fast-power pathway. Each Hadamard-test circuit used \(512\) shots. The \(n=m=4\) case includes the first \(9\) completed repetitions only, since the final repetition was not completed within the available runtime budget.}
\label{tab:ibm-fez-ldqpe}
\begin{tabular}{cccccccccc}
\toprule
\(n=m\) & reps & bucket accuracy & bucket histogram & \(\#\texttt{sx}\) & \(\#\texttt{rz}\) & \(\#\texttt{cz}\) & \(\#\texttt{x}\) & phase error mean \(\pm\) SEM (rad) & median [IQR] (rad) \\
\midrule
2 & 10 & \(100\%\) & \(\{0:10\}\) & 130 & 117 & 52  & 10 & \(0.0444 \pm 0.00714\) & \(0.0473\,[0.0248,\,0.0569]\) \\
3 & 10 & \(90\%\)  & \(\{3:9,\,5:1\}\) & 264 & 231 & 123 & 19 & \(0.2948 \pm 0.1601\) & \(0.1349\,[0.1057,\,0.1757]\) \\
4 & 9  & \(88.9\%\) & \(\{6:1,\,7:8\}\) & 393 & 349 & 177 & 24 & \(0.0528 \pm 0.0120\) & \(0.0408\,[0.0256,\,0.0782]\) \\
\bottomrule
\end{tabular}
\end{table*}

Taken together, Fig.~\ref{fig:ldqpe-hardware-fez} and Table~\ref{tab:ibm-fez-ldqpe} show a clear qualitative trend: the symmetric pathway remains executable at small instance size, but becomes progressively more fragile as the transpiled two-qubit-gate count grows from roughly \(52\) to \(123\) to \(177\) \texttt{cz} gates. This is consistent with the noise simulation model study in Fig.~\ref{fig:ldqpe-noise-sweep}, which already indicated that the symmetric fast-power pathway should remain viable into the low-\(10^{-3}\) to few-\(10^{-3}\) two-qubit-noise regime before accumulated entangling-gate noise begins to dominate. On \texttt{ibm\_fez}, whose reported median and layered two-qubit errors lie precisely in this range, the \(n=m=2\) instance is robust, while the \(n=m=3\) and \(n=m=4\) instances enter a visibly more marginal regime.

The practical conclusion is therefore unchanged but now supported directly by both the per-run phase-error plots and the compact summary statistics: the symmetric \(VDV^\dagger\) compiler provides a genuine NISQ-feasible LDQPE pathway at small instance size, and its current hardware limitation is governed primarily by transpiled two-qubit-gate count and backend calibration quality rather than by any exponential structural blow-up in the powered unitaries.

Although the present hardware demonstrations are limited to \(n=m\le 4\), the observed transpiled gate counts on `ibm\_fez` grow approximately linearly over the tested range. This makes it useful to give an indicative large-\(n\) extrapolation for the symmetric pathway, since in this regime the powered unitary retains the form \(U^{2^j}=V D^{2^j} V^\dagger\) and does not incur the exponential-in-\(2^j\) depth blow-up associated with naive repeated application. Table~\ref{tab:qsaq-gate-extrapolation} shows a simple linear extrapolation of the observed transpiled counts to \(n=m\in\{10,15,20\}\). Under this extrapolation, the dominant two-qubit-gate cost remains roughly linear in \(n\), with \(\#\texttt{cz}\approx 555,867,1180\) respectively. The same table also reports a crude two-qubit error budget obtained by requiring a \(95\%\) survival factor from CZ gates alone, namely \((1-p_{2q})^{N_{\texttt{cz}}}\gtrsim 0.95\). This gives rough target error rates in the \(10^{-4}\) to few\(\times 10^{-5}\) range. These values should not be over-interpreted, since they ignore routing overhead, coherent error accumulation, SPAM, and any mitigation strategy, but they suggest that larger symmetric QSA-Q instances are not ruled out by gate-count scaling alone. Rather, the main barrier is hardware fidelity, not an intrinsic exponential growth in transpiled circuit size over the currently tested range.
\section{Discussion}
\label{sec:discussion}

QSA is best understood as a post-provisioning authentication primitive for quantum control planes rather than as a standalone authenticated key exchange protocol. It does not solve the initial authenticated distribution of the credential; instead, it begins after a quantum credential has already been established by some upstream mechanism, such as teleportation, secure commissioning, entanglement-assisted distribution, or seed-based enrolment. Its purpose is to issue fresh public unitary challenges and convert consistent spectral responses of the hidden provision into transcript-bound session material for explicit confirmation. In this way, QSA turns possession of a previously installed quantum resource into an application-facing authentication token without disclosing the resource itself. The security picture is correspondingly shaped by repeated fresh challenges. An adversary may observe the public challenge family and any associated metadata, but does not receive the planted state, its seed, or the honest prover's witness copies. Forgery therefore reduces either to reproducing the accepted feature response for fresh challenges without the planted resource, or to constructing an alternative witness that induces the same accepted transcript. For independently generated expressive challenge instances, the associated eigenbases are expected to be effectively decorrelated, so information gained about one challenge should not transfer cheaply to the next. Operationally, this pushes the adversary away from simple reuse strategies and toward harder tasks such as planted-state search, spectrum reconstruction, or satisfying many independent spectral constraints at once. Two parameters govern this tradeoff most directly: the total extracted feature length \(mk\), which controls resistance to direct guessing once explicit confirmation is applied, and the planted-state dimension \(n\), which controls the difficulty of witness-recovery attacks in the \(2^n\)-dimensional eigenstructure.

From an implementation perspective, the central practical result is the verifier-driven symmetric compiler based on unitaries of the form \(U = V D V^\dagger\). Its key advantage is structural: powered blocks retain the form \(U^{2^j}=VD^{2^j}V^\dagger\), so the diagonal layer can be updated through simple per-qubit \(R_Z\) rotations rather than suffering uncontrolled growth with LDQPE precision. This makes low-depth phase extraction feasible in a way that generic or asymmetric constructions do not presently match. The asymmetric construction remains useful as a conceptual comparison because it removes direct diagonal read-off and represents a stricter extraction setting, but it is less attractive for near-term hardware because compilation and controlled evaluation are heavier. More broadly, the symmetric route supports a continuum of honest-evaluation options: LDQPE is the preferred near-term protocol, while full QPE remains available as a higher-cost fallback when lower planted overlap must be tolerated. Our numerical and hardware results support this interpretation. Simulations indicate that honest low-depth phase extraction remains viable in the low-noise regime relevant to current devices, and that performance is limited mainly by accumulated hardware noise and compiled circuit overhead rather than by any conceptual obstacle in the protocol itself. Small-instance executions on IBM \texttt{ibm\_fez} provide a real-device sanity check of this structured pathway. Taken together, these results suggest that the main bottlenecks are now device-level issues such as routing, two-qubit fidelity, and compilation quality, rather than the basic viability of QSA as a hardware-meaningful authentication interface.

This viewpoint also clarifies where QSA fits in quantum-network workflows. If a quantum credential is teleported or otherwise provisioned to a remote endpoint, QSA can be run afterwards to test whether that endpoint still holds the intended provision and can respond correctly to fresh public challenges. In this sense, QSA composes naturally with teleportation, transported memories, entanglement-enabled links, and related quantum-network models, including quantum sneakernets~\cite{devitt2016ship, srikara2024sneakernet}: the upstream mechanism delivers the quantum resource, while QSA authenticates possession and usability of that resource at the endpoint. The teleported-QSA variant, however, occupies a very different resource regime from QKD. As calculated in the Supplemental Information, estimating all LDQPE moments across multiple challenge instances can require Bell-pair budgets that are already large for moderate \(m\) and \(n\), reaching \(2N_snmk \simeq 500{,}000+\) Bell pairs under LDQPE evaluation for representative choices such as \(k=36\) and \(n=m=8\), although the symmetric fast path removes the moment and Hadamard-test factors, and because the teleported variant is a state-mode deployment (\(\delta=0\)) a single teleported copy per instance suffices, lowering this to \(\approx nk\sim 300\) (see Supplemental Information). Teleported QSA is therefore not best viewed as a competitor to QKD for classical key establishment, but rather as a specialised endpoint-validation layer for remotely provisioned quantum states or memories when the task is to confirm that a specific hidden quantum credential has arrived intact and remains operational. The same framework also extends beyond the point-to-point setting. As shown in Methods, the symmetric compiler can embed multiple hidden signal eigenvectors within a single broadcast unitary, each associated with a different party-specific planted state and secret label, suggesting that the same challenge-generation philosophy may scale from bilateral authentication to broadcast or multi-party control-plane settings.

Several limitations remain, and these define the most important directions for future work. Our attack catalogue is not exhaustive, and the security framing rests on a single average-case label-hiding assumption rather than a reduction to a canonical worst-case problem, though we prove the accompanying feature-uniformity property unconditionally and show that the state-mode construction removes the assumption entirely (Sec.~\ref{sec:secmodel}). Closing the gap to a worst-case reduction in seed mode also remains open. New algorithmic ideas, especially those exploiting leakage, collusion, adaptive challenge selection, or partial transcript reuse, could change the practical security picture. On the implementation side, although we have demonstrated the honest prover-side LDQPE pathway for small instances, we have not yet realised a full end-to-end workflow including provisioning, challenge generation, prover evaluation, and explicit confirmation in a single live deployment, nor have we experimentally demonstrated the teleported-state variant. A natural next step is to study QSA-Q at larger hidden-state dimension \(n\) with  extracted precision \(m\), for example \(n\sim 20\text{ to }30\) and \(m\sim 20\text{ to }30\), where the provisioned state is genuinely high dimensional while the symmetric compiler still permits practical powered evaluation without naive exponential depth growth. Benchmarking transpiled depth, two-qubit gate counts, tolerated error rates, and approximate classical surrogates in this regime would help identify when QSA-Q moves beyond proof-of-pathway and begins to access planted state sizes that are no longer comfortably classically tractable. Further work is also needed on challenger-side variational generation loops and architecture-specific ans\"atze, since practical performance will depend on hardware-aware optimisation and compilation choices. Even with these caveats, the main conclusion is clear: once a quantum credential has been established by an authenticated upstream mechanism, QSA provides a concrete way to re-authenticate possession of that credential under fresh public challenges and to derive conventional transcript-bound authentication material from it.

The two provisioning modes trace a single design axis: who delivers the planted state. In seed mode the prover regenerates it from a classical seed, which forces the verifier to solve an inverse problem: fitting an expressive $V(\vec{\alpha})$ so that $V\ket{b}\approx\ket{\psi}$, and it is precisely this variational step that introduces the label-hiding assumption and the ensemble-matching requirement. In state mode the planted register is delivered physically, so the verifier may instead sample $V$ and the label independently and \emph{define} the planted state as $V\ket{b}$; no optimisation is performed, the overlap is exact, and label hiding holds by construction. The expressive-unitary compiler is thus not intrinsic to QSA but is the price of seed-derivability, and a deployment able to teleport its credential does not pay it. This clarifies the near-term tradeoff. After a one-time provisioning of $S_0$ (which may itself be quantum, for example a QKD-delivered seed, but lies outside the QSA boundary), seed mode needs no per-session quantum communication and runs entirely classically, at the cost of the label-hiding assumption; state mode instead consumes a freshly delivered quantum register each session, but is assumption-free at the compiler and forward-secure against endpoint capture. The axis is thus one-time versus per-session quantum communication, not classical versus quantum.

Finally, although the demonstrations here are NISQ-scale and a fault-tolerant realisation is beyond our present scope, the construction is unusually well suited to logical hardware, for two concrete reasons. First, in the symmetric family the precision knob is the diagonal layer $D=\bigotimes_q R_z(\beta_q)$, and powering only rescales its angles, $D^{2^j}=\bigotimes_q R_z(2^j\beta_q)$. Under a fault-tolerant cost model, where non-Clifford $R_z$ rotations dominate through their magic-state or $T$ budget, the per-instance non-Clifford cost is therefore $n$ single-qubit rotations independent of the moment index, so extracting more phase bits does not multiply the magic-state budget; what is merely NISQ-convenient here becomes a bounded $T$-count statement under fault tolerance. Second, the honest fast path is nearly Clifford: the prover applies $V^\dagger$ and measures in the computational basis, and $\theta(b)$ is computed classically, so the only non-Clifford content is the $T$-count of $V^\dagger$ (and of the planting circuit), with no controlled-$U$ powers and no quantum Fourier transform. This makes the symmetric fast path a natural beyond-NISQ primitive. A logically encoded provisioning route is a natural complement rather than a security change: preparing a planted state on a logical register genuinely requires a fault-tolerant stack and makes the physical-layer attacks we scope out (cloning, tomography of $\ket{\psi_i}$) harder to mount, strengthening the provisioning-layer assumptions of Sec.~\ref{sec:results} without altering the label-hiding assumption on which the security bound rests. We leave a fault-tolerant resource analysis and a logically encoded QSA-Q demonstration to future work.
\section{Methods}\label{sec:methods-unitary-generation}

\subsection*{Symmetric unitary challenge generation: Quantum-native compiler for QSA-Q public unitaries}
\label{sec:methods-quantum-compiler}

This section describes a purely quantum compilation procedure used to generate the public unitary family in \textbf{QSA-Q}. The goal is to produce, for each index \(i\in\{1,\ldots,k\}\), a shallow circuit \(U_i\) that admits a \emph{hidden} eigenvector with large overlap with a planted secret state \(\ket{\psi_i}\), while keeping the associated eigenphase (and the identity of the eigenvector within the eigenbasis) unpredictable to an adversary. However, although this method gives the unitary eigenphases for free without needing a re-application of a low-depth QPE, it is not exponentially protected against Attack \Romannum{2}A.  However, it is still exponentially protected against eigenphase coupon-collecting Attack \Romannum{1}B.

%\paragraph*{Private planting map and secret state.}
Let \(P_i\) be a planted state (planting) circuit shared by the honest parties, and define the planted state
\begin{equation}
  \ket{\psi_i} \;:=\; P_i^\dagger \ket{0^n}.
\end{equation}
The circuit \(P_i\) (or its seed) is secret; it is \emph{not} published.

%\paragraph*{Hidden computational basis label.}
We sample a uniformly random bit-string \(b_i \in \{0,1\}^n\), and define the corresponding computational basis state prepared from \(\ket{0^n}\) by a layer of \(X\) gates:
\begin{equation}
  \ket{b} \;=\; X^{b_0}\otimes \cdots \otimes X^{b_{n-1}} \ket{0^n}.
\end{equation}
The string \(b\) is kept secret by the challenger; it determines \emph{which} eigenvector of \(U_i\) carries the planted overlap signal.

%\paragraph*{Variational alignment subroutine (quantum compilation loop).}
Let \(V(\vec{\alpha})\) be an expressive parameterised circuit (ansatz) on \(n\) qubits with real parameters \(\vec{\alpha} \in \mathbb{R}^p\).
We choose \(\vec{\alpha}\) so that \(V(\vec{\alpha})\ket{b}\) aligns with \(\ket{\psi_i}\). Concretely, we maximise the overlap
\begin{equation}
  F_i(\vec{\alpha})
  \;:=\;
  \bigl|\langle\psi_i \,|\, V(\vec{\alpha})\,|\, b\rangle\bigr|^2
  \;=\;
  \bigl|\bra{0^n} P_i^\dagger V(\vec{\alpha}) \ket{b}\bigr|^2.
  \label{eq:Fi}
\end{equation}
This objective is estimated \emph{directly on quantum hardware} by appending \(P_i^\dagger\) and measuring in the computational basis:
\begin{equation}
  F_i(\vec{\alpha})
  \;=\;
  \Pr\!\Big[\,0^n \text{ upon measuring } P_i^\dagger V(\vec{\alpha})\ket{b}\,\Big].
  \label{eq:prob0}
\end{equation}
Equivalently, one may minimise the loss \( \mathcal{L}_i(\vec{\alpha}) := 1 - F_i(\vec{\alpha})\).
Optimisation can be performed with gradient-free methods (e.g.\ SPSA) or with parameter-shift gradients, depending on the ansatz and hardware constraints. Let \(\vec{\alpha}_i^\star\) denote the final parameters, and define
\begin{equation}
  V_i \;:=\; V(\vec{\alpha}_i^\star).
\end{equation}
By construction, the compiler enforces
\begin{equation}
  \bigl|\langle \psi_i \,|\, V_i\,|\, b \rangle \bigr|^2 \;\ge\; 1-\delta,
  \label{eq:overlap-target}
\end{equation}
for a chosen target \(\delta\) (empirically tuned).

%\paragraph*{Public unitary construction \(U_i = V_i D_i V_i^\dagger\).}
After learning \(V_i\), we define the published unitary \(U_i\) in diagonalised form:
\begin{equation}
  U_i \;:=\; V_i \, D_i \, V_i^\dagger,
  \label{eq:UVDVdag}
\end{equation}
where \(D_i\) is a diagonal unitary in the computational basis implemented as a layer of \(R_z\) rotations,
\begin{equation}
  D_i
  \;:=\;
  \bigotimes_{j=0}^{n-1} R_z(\beta_{i,j}),
  \qquad
  \beta_{i,j} \in [0,2\pi).
  \label{eq:Ddef}
\end{equation}
The angles \(\{\beta_{i,j}\}\) are public and define the eigenphases of \(U_i\) in the \(V_i\)-rotated eigenbasis.

%\paragraph*{Eigenstructure and the hidden ``signal'' eigenphase.}
Since \(D_i\) is diagonal in the computational basis, its eigenvectors are \(\{\ket{x} : x\in\{0,1\}^n\}\). Therefore the eigenvectors of \(U_i\) are \(\{V_i\ket{x}\}\) and
\begin{equation}
  U_i \, V_i\ket{b} \;=\; V_i D_i \ket{b} \;=\; e^{i\theta(b)}\, V_i\ket{b},
\end{equation}
where the phase \(\theta(b)\) is determined by the diagonal action of \(D_i\).
Using \(R_z(\theta)=\mathrm{diag}(e^{-i\theta/2},e^{+i\theta/2})\), one may write,
\begin{equation}
\theta(b)=\frac{1}{2}\sum_{q=1}^{n}(2b_q-1)\,\beta_q \; (\mathrm{mod}\; 2\pi).
\label{eq:phase_closed_form}
\end{equation}
In particular, the state
\begin{equation}
  \ket{u_i^\star} \;:=\; V_i \ket{b}
\end{equation}
is an eigenvector of \(U_i\) with eigenphase \(\theta_i(b)\), but the identity of this eigenphase is hidden because \(b\) is secret and \(V_i\) is only revealed through the composite public circuit for \(U_i\).

%\paragraph*{Planted overlap guarantee.}
Combining \eqref{eq:overlap-target} with \(\ket{u_i^\star}=V_i\ket{b}\) yields
\begin{equation}
  \bigl|\langle \psi_i \,|\, u_i^\star\rangle\bigr|^2
  \;=\;
  \bigl|\langle \psi_i \,|\, V_i\,|\,b\rangle \bigr|^2
  \;\ge\; 1-\delta.
  \label{eq:planted-overlap}
\end{equation}
Thus, the honest planted state has high overlap with a \emph{single} eigenvector of the published \(U_i\), while an adversary (lacking \(P_i\) and \(b\)) does not know which eigenphase \(\theta(x)\) corresponds to the planted eigenvector \(x=b\).

%\paragraph*{Use in QSA-Q evaluation.}
Given the public circuit description of \(U_i\), honest parties prepare \(\ket{\psi_i}\) and use a low-depth phase-estimation primitive (LDQPE) to extract an \(m\)-bit approximation of the signal eigenphase \(\theta(b)\). Repeating over \(i=1,\ldots,k\) yields the eigenphase feature vector \(\boldsymbol{\Theta}\), which is then passed to a classical KDF to derive session keys.

\paragraph*{Honest fast path (symmetric compiler).}
For the symmetric family \(U_i=V_iD_iV_i^\dagger\) the prover need not run LDQPE at all. Since \(V_i\) is public and \(|\langle\psi_i|V_i|b\rangle|^2\ge 1-\delta\), applying \(V_i^\dagger\) to the planted state gives
\[
V_i^\dagger\ket{\psi_i}=\ket{b}+\sqrt{\delta}\,\ket{\perp},\qquad
\bigl|\langle b|V_i^\dagger|\psi_i\rangle\bigr|^2=\bigl|\langle\psi_i|V_i|b\rangle\bigr|^2\ge 1-\delta,
\]
a state supported almost entirely on the single computational-basis string \(b\). A computational-basis measurement then returns \(b\), realised by the ancilla-free circuit
\begin{center}
\begin{quantikz}[row sep=0.25cm, column sep=0.35cm]
\lstick{$\ket{0^n}$} &
\gate[wires=4]{P_i^\dagger} &
\gate[wires=4]{V_i^\dagger} &
\meter{} \\
\end{quantikz}
\end{center}
whose outcome is the \(n\)-bit string \(b\). Because \(\delta<\tfrac12\) in every compiled regime of interest, \(b\) is the unique modal outcome, so majority voting over
\[
N_{\mathrm{shot}}\ \ge\ \frac{\ln(1/\varepsilon_{\mathrm{fail}})}{2\left(\tfrac12-\delta\right)^2}
\]
independent shots, re-preparing \(\ket{\psi_i}=P_i^\dagger\ket{0^n}\) from the seed each time, returns \(b\) except with probability \(\varepsilon_{\mathrm{fail}}\); for the typical compiled range \(\delta\in[0.05,0.25]\) this is a few dozen shots (for example \(N_{\mathrm{shot}}\le 37\) at \(\varepsilon_{\mathrm{fail}}=10^{-2}\)). In state mode \(\ket{\psi_i}=V_i\ket{b}\) exactly (\(\delta=0\)), so a single shot recovers \(b\) deterministically. Given \(b\), the prover computes the feature classically from the public angles by Eq.~\eqref{eq:phase_closed_form}, a signed sum \(\theta(b)=\tfrac12\sum_{q=1}^{n}(2b_q-1)\beta_{i,q}\ \mathrm{mod}\ 2\pi\) evaluated in \(O(n)\) time and quantised to \(m\) bits: this is the same \(O(n)\) read-off the verifier performs, now available to the prover because the measurement supplies \(b\). The fast-path circuit has depth \(\mathrm{depth}(P_i)+\mathrm{depth}(V_i)\), uses no ancilla and no controlled operations, and is strictly cheaper than the LDQPE moment circuits. It does not weaken security: the adversary lacks \(\ket{\psi_i}\) and so cannot execute \(V_i^\dagger\) on it, and the concentration it exploits is exactly the localisation visualised in Fig.~\ref{fig:bin-mass-honest-vs-eve}. Evaluability is preserved, since simulating \(V_i^\dagger\ket{\psi_i}\) classically remains exponential in \(n\). LDQPE remains the general-purpose route, required for QSA-C, QSA-M, and the asymmetric compiler \(U=V_LV_R^\dagger\), which is designed precisely to remove this shortcut.

\begin{figure}[ht!]
\centering
\begin{quantikz}[row sep=0.25cm, column sep=0.35cm]
\lstick{$\ket{0^n}$} &
\gate[wires=4]{X^{b_i}} &
\gate[wires=4]{V(\vec{\alpha})} &
\gate[wires=4]{P_i^\dagger} &
\meter{} \\
\end{quantikz}
\caption{Quantum-native compilation loop objective. The hidden bitstring \(b_i\in\{0,1\}^n\) prepares \(\ket{b}=X^{b_i}\ket{0^n}\). The ansatz \(V(\vec{\alpha})\) is optimised to maximise the probability of measuring \(0^n\) after applying the private planting inverse \(P_i^\dagger\), i.e.\ \(F_i(\vec{\alpha})=\Pr[0^n]=|\langle 0^n|P_i^\dagger V(\vec{\alpha})|b\rangle|^2\).}
\label{fig:QSAq_compile_objective}
\end{figure}

\subsection*{Multi-prover extension: broadcast symmetric challenges for many parties}
\label{sec:methods-quantum-compiler-multiparty}

The symmetric compiler above extends naturally from a two-party setting to a \emph{broadcast} (one-to-many) setting in which a single published unitary instance $U_i$ is used to authenticate \emph{multiple} provers (e.g.\ Bob, Charlie, and Daniel) against the same verifier-side challenge transcript. The key idea is to learn a \emph{single} expressive map $V_i$ that simultaneously aligns multiple secret planted states to multiple hidden computational-basis labels, so that each party’s planted state has high overlap with its own hidden signal eigenvector of the \emph{same} public unitary.

%\paragraph*{Per-party planted states and hidden labels.}
Fix a set of parties $\mathcal{P}=\{\mathrm{B},\mathrm{C},\mathrm{D}\}$ (for Bob, Charlie, Daniel).
For each party $P\in\mathcal{P}$ and index $i$, let $P_{i,P}$ denote a private planting circuit (or planting seed) shared between the verifier and party $P$, and define the planted state
\begin{equation}
  \ket{\psi_{i,P}} \;:=\; P_{i,P}^\dagger\ket{0^n}.
\end{equation}
Independently sample secret bitstrings $b_{i,P}\in\{0,1\}^n$ and define the corresponding hidden basis states
\begin{equation}
  \ket{b_{i,P}} \;=\; X^{(b_{i,P})_0}\otimes \cdots \otimes X^{(b_{i,P})_{n-1}}\ket{0^n}.
\end{equation}
Each $b_{i,P}$ is kept secret by the challenger; it determines which eigenvector of the eventual public unitary carries party $P$'s planted overlap signal.

%\paragraph*{Multi-objective variational alignment.}
We now learn a \emph{single} parameterised ansatz $V(\vec{\alpha})$ that aligns \emph{all} party labels and planted states simultaneously. For each party $P$, define the fidelity objective
\begin{equation}
  F_{i,P}(\vec{\alpha})
  \;:=\;
  \bigl|\langle \psi_{i,P}\,|\,V(\vec{\alpha})\,|\,b_{i,P}\rangle\bigr|^2
  \;=\;
  \bigl|\bra{0^n}P_{i,P}^\dagger V(\vec{\alpha})\ket{b_{i,P}}\bigr|^2.
  \label{eq:Fip}
\end{equation}
Each $F_{i,P}(\vec{\alpha})$ is estimated on quantum hardware in the same way as in \eqref{eq:prob0}, by appending $P_{i,P}^\dagger$ and measuring $\ket{0^n}$.
We then optimise a \emph{multi-party loss} that aggregates per-party misalignment:
\begin{equation}
  \mathcal{L}_i(\vec{\alpha})
  \;:=\;
  \sum_{P\in\mathcal{P}}\bigl(1-F_{i,P}(\vec{\alpha})\bigr),
  \label{eq:multiparty-loss}
\end{equation}
using the same optimiser class as in the two-party compiler (e.g.\ SPSA with restarts).
Let $\vec{\alpha}_i^\star$ be the final parameters and define
\begin{equation}
  V_i \;:=\; V(\vec{\alpha}_i^\star).
\end{equation}
Empirically, feasibility depends on ansatz expressivity and on how ``orthogonal'' the targets $\{(\ket{b_{i,P}},\ket{\psi_{i,P}})\}_P$ are; in practice, one can trade compilation effort for accuracy by increasing ansatz depth or optimiser budget.

%\paragraph*{Broadcast unitary construction and per-party signal eigenvectors.}
Given the learned $V_i$, we publish a \emph{single} symmetric challenge
\begin{equation}
  U_i \;:=\; V_i D_i V_i^\dagger,
\end{equation}
with $D_i$ defined as in \eqref{eq:Ddef} and public angles $\{\beta_{i,j}\}$.
As before, eigenvectors of $U_i$ are $\{V_i\ket{x}\}$, and for each party $P$ the hidden signal eigenvector is
\begin{equation}
  \ket{u_{i,P}^\star} \;:=\; V_i\ket{b_{i,P}},
\end{equation}
with corresponding eigenphase
\begin{equation}
  \theta_i(b_{i,P})
  \;=\;
  \frac{1}{2}\sum_{q=1}^{n}\bigl(2(b_{i,P})_q-1\bigr)\,\beta_{i,q}
  \quad (\mathrm{mod}\;2\pi),
\end{equation}
by the same closed-form argument as \eqref{eq:phase_closed_form}.

%\paragraph*{Multi-party planted-overlap guarantee.}
If optimisation succeeds to tolerance, the learned $V_i$ enforces a per-party overlap target,
\begin{equation}
  \bigl|\langle \psi_{i,P}\,|\,u_{i,P}^\star \rangle \bigr|^2
  \;=\;
  \bigl|\langle \psi_{i,P}\,|\,V_i\,|\,b_{i,P}\rangle \bigr|^2
  \;\ge\; 1-\delta_P,
  \qquad \forall P\in\mathcal{P},
  \label{eq:multiparty-overlap}
\end{equation}
where $\delta_P$ can be tracked per party (or replaced by a uniform $\delta$ if one uses a worst-case bound).
Thus the same public $U_i$ simultaneously contains multiple hidden ``signal'' eigenvectors, one per authenticated party, each keyed by its own secret $b_{i,P}$ and planting circuit $P_{i,P}$.

%\paragraph*{Protocol use and practical notes.}
In a broadcast authentication setting, the verifier publishes the same challenge instance $U_i$ to all parties, and each party $P$ runs the standard QSA-Q evaluation against its own planted state $\ket{\psi_{i,P}}$, extracting an $m$-bit feature corresponding to $\theta_i(b_{i,P})$.
These per-party features are transcript-bound and compressed by a KDF/MAC/AEAD layer as usual, yielding independent acceptance tokens per party under a single broadcast challenge.

The main engineering consideration is \emph{capacity}: a single $V_i$ can only simultaneously realise a finite number of high-fidelity alignments at fixed depth and optimiser budget. In practice this can be managed by (i) limiting the broadcast fan-out $|\mathcal{P}|$ per compiled instance, (ii) using deeper ans\"atze when compiling broadcast instances, or (iii) rotating the hidden labels $\{b_{i,P}\}$ and planting circuits across $i$ so that any residual correlations do not accumulate across epochs. In Appendix \ref{subsec:QSAq_compilers_multi_party}, we evaluate this for an example $n=m=8$ qubits system. However, we find that this converges very slowly to a solution that minimizes the loss function for all parties.

From a security viewpoint, the broadcast extension preserves the primary hiding mechanism: an adversary still sees only the public circuit for $U_i$ and the public diagonal angles, while the mapping from eigenphases to parties remains hidden behind the private $(P_{i,P}, b_{i,P})$ pairs.

\begin{figure}[ht!]
\centering
\begin{quantikz}[row sep=0.25cm, column sep=0.40cm]
\lstick{$\ket{0^n}$} &
\gate[wires=4]{X^{b_{i,P}}} &
\gate[wires=4]{V(\vec{\alpha})} &
\gate[wires=4]{P_{i,P}^\dagger} &
\meter{} \\
\end{quantikz}
\caption{Multi-party compilation objective for a broadcast symmetric challenge. For each party $P\in\mathcal{P}$, the compiler estimates $F_{i,P}(\vec{\alpha})=\Pr[0^n]=|\langle 0^n|P_{i,P}^\dagger V(\vec{\alpha})|b_{i,P}\rangle|^2$ and minimises the aggregated loss $\mathcal{L}_i(\vec{\alpha})=\sum_{P\in\mathcal{P}}(1-F_{i,P}(\vec{\alpha}))$. A single learned $V_i$ is then used to form the broadcast public unitary $U_i=V_iD_iV_i^\dagger$, which embeds a distinct hidden signal eigenvector $V_i|b_{i,P}\rangle$ for each party.}
\label{fig:QSAq_compile_objective_multiparty}
\end{figure}
\subsection*{Asymmetric challenge generation: dual-compiler construction without an exposed diagonal layer}
\label{sec:methods-asymmetric-compiler}

The compiler above produces challenges of the form \(U_i = V_i D_i V_i^\dagger\), which has an explicit diagonal layer \(D_i\) and therefore admits efficient \emph{verifier-side} prediction of the planted eigenphase via \(\arg\langle b|D|b\rangle\) once the hidden label is known. In some deployments, however, it is preferable that \emph{even the verifier} does not obtain an algebraic ``phase-for-free'' handle, and instead evaluates the challenge using the same LDQPE pipeline as the prover. This yields an asymmetric trade-off: it removes exposure of a terminal diagonal structure (mitigating the direct eigenphase read-off of Appendix Attack A.2(a) in the \(VDV^\dagger\) family), at the cost of requiring the verifier to perform LDQPE online.

%\paragraph*{Dual variational compilers.}
Fix the planted secret state \(\ket{\psi_i}=P_i^\dagger\ket{0^n}\) as before, with \(P_i\) kept secret.
We sample two independent hidden computational labels
\(
b_{L,i},b_{R,i}\leftarrow\{0,1\}^n
\)
and define \(\ket{b_{L,i}}\) and \(\ket{b_{R,i}}\) via \(X\)-layers.
We then run \emph{two} independent alignment loops, each starting from a distinct random seed / initial parameters:
\begin{align}
F_{L,i}(\vec{\alpha_L})
&:=\bigl|\langle \psi_i\,|\,V_L(\vec{\alpha})\,|\,b_{L,i}\rangle \bigr|^2
=\Pr\!\big[\,0^n\text{ upon measuring }P_i^\dagger V_L(\vec{\alpha_L})\ket{b_{L,i}}\,\big],\\
F_{R,i}(\vec{\alpha_R})
&:=\bigl|\langle \psi_i\,|\,V_R(\vec{\alpha})\,|\,b_{R,i}\rangle \bigr|^2
=\Pr\!\big[\,0^n\text{ upon measuring }P_i^\dagger V_R(\vec{\alpha_R})\ket{b_{R,i}}\,\big].
\end{align}
Let \(\vec{\alpha}_i^\star,\beta_i^\star\) denote the optimised parameters and define
\(
V_{L,i}:=V_L(\vec{\alpha}_i^\star)
\)
and
\(
V_{R,i}:=V_R(\beta_i^\star).
\)
The compiler targets overlap guarantees
\begin{equation}
\bigl|\langle \psi_i\,|\,V_{L,i}\,|\,b_{L,i}\rangle \bigr|^2\ge 1-\delta,
\qquad
\bigl|\langle \psi_i\,|\,V_{R,i}\,|\,b_{R,i}\rangle \bigr|^2\ge 1-\delta,
\label{eq:dual-overlap-target}
\end{equation}
for a chosen tolerance \(\delta\).

%\paragraph*{Published unitary challenge.}
We publish the unitary challenge
\begin{equation}
U_i \;:=\; V_{L,i}\,V_{R,i}^\dagger.
\label{eq:U-asym}
\end{equation}
In contrast to \eqref{eq:UVDVdag}, \eqref{eq:U-asym} does not expose a terminal diagonal layer whose eigenphases can be read out directly.
Both \(V_{L,i}\) and \(V_{R,i}\) are public only through the composite circuit for \(U_i\); the hidden labels \(b_{L,i},b_{R,i}\) and the planting circuit \(P_i\) remain secret.
\begin{figure}[ht!]
\centering
\begin{quantikz}[row sep=0.25cm, column sep=0.35cm]
\lstick{$\ket{0^n}$} &
\gate[wires=4]{X^{b_{L,i}}} &
\gate[wires=4]{V_L(\vec{\alpha_L})} &
\gate[wires=4]{P_i^\dagger} &
\meter{} \\
\end{quantikz}
\hspace{0.8cm}
\begin{quantikz}[row sep=0.25cm, column sep=0.35cm]
\lstick{$\ket{0^n}$} &
\gate[wires=4]{X^{b_{R,i}}} &
\gate[wires=4]{V_R(\vec{\alpha_R})} &
\gate[wires=4]{P_i^\dagger} &
\meter{} \\
\end{quantikz}
\caption{Dual alignment objectives for the asymmetric compiler. Two independently initialised ans\"atze \(V_L(\vec{\alpha})\) and \(V_R(\vec{\alpha_R})\) are optimised (with different hidden labels \(b_{L,i},b_{R,i}\)) to maximise \(\Pr[0^n]\) after \(P_i^\dagger\), i.e.\ to align \(V_L\ket{b_{L,i}}\) and \(V_R\ket{b_{R,i}}\) with the same planted \(\ket{\psi_i}=P_i^\dagger\ket{0^n}\). The published challenge is then \(U_i=V_{L,i}V_{R,i}^\dagger\), which does not expose a terminal diagonal layer and therefore requires LDQPE evaluation by the verifier as well as the prover.}
\label{fig:QSAq_asym_compile_objective}
\end{figure}
%\paragraph*{Why a planted spectral signal exists.}
Hence, we can define the (approximately) aligned states
\(
\ket{u_{L,i}}:=V_{L,i}\ket{b_{L,i}}
\)
and
\(
\ket{u_{R,i}}:=V_{R,i}\ket{b_{R,i}}.
\)
By \eqref{eq:dual-overlap-target}, both satisfy
\(
|\langle \psi_i|u_{L,i}\rangle|^2\ge 1-\delta
\)
and
\(
|\langle \psi_i|u_{R,i}\rangle |^2\ge 1-\delta,
\)
so \(\ket{\psi_i}\) has most of its weight supported on the two-dimensional subspace spanned by \(\{\ket{u_{L,i}},\ket{u_{R,i}}\}\) (up to leakage controlled by \(\delta\)).
Within this dominant subspace, the operator \(U_i=V_{L,i}V_{R,i}^\dagger\) acts as a relative basis change between the two aligned frames.
Heuristically, when \(\delta\) is small, this induces a stable spectral signature on \(\ket{\psi_i}\) that LDQPE can extract from the power moments
\(
Z_t^{(i)}=\langle\psi_i|U_i^t|\psi_i\rangle
\)
at modest precision.
(We empirically validate that the overlap mass concentrates into a phase bin for the honest \(\ket{\psi_i}\), while a random adversarial state yields a flatter bin mass distribution; see Sec.~\ref{sec:results}.) 

%\paragraph*{Security and evaluation consequences.}
Because \eqref{eq:U-asym} does not contain an explicit diagonal layer, the verifier cannot compute the intended phase response by a closed-form expression such as \eqref{eq:phase_closed_form}.
Instead, both verifier and prover evaluate \(\theta_i^\star\) using the same LDQPE routine (Algorithm~2 of~\cite{ni2023lowdepthqpe}) applied to \(\ket{\psi_i}\) and the public circuit for \(U_i\).
This removes the ``eigenphases-for-free'' shortcut that underlies the \(VDV^\dagger\) construction, while preserving the exponential cost barrier against eigenphase coupon-collecting attacks that attempt to cover many phase bins without the planted state.
An adversary who wishes to predict the correct response without the secret state-preparation circuit or provisioning seed is pushed toward spectrum-level strategies for $m\ge n$ (see Appendix \ref{app:reference-attacks}), which require running substantially deeper QPE-style procedures and/or searching for eigencomponents that correlate with the planted state across independent instances.

\section*{Funding}
This work did not receive dedicated project grant funding. The lead author was supported by the Commonwealth Scientific and Industrial Research Organisation (CSIRO) and by the Office of the Chief Scientist of CSIRO through the Impossible Without You program.
\section*{Acknowledgements}
We thank Gavin Brennen and the BTQ team for helpful discussions and feedback during the early stages of this work. During the preparation of this manuscript, the author(s) used a generative AI assistant (Claude, Anthropic) to help draft and revise portions of the text, with LaTeX preparation and formatting, and to structure the response to reviewers. All AI-assisted output was reviewed, verified, and edited by the author(s), who take full responsibility for the content, analysis, and conclusions of this work.

\section*{Author contributions}
S.\ P.\ Kish conceived the project, developed the QSA constructions, implemented the software, performed the experiments, generated all figures and tables, and wrote the manuscript. H. J. \ Vallury co-conceived and developed the core ideas and methods, contributed expertise on quantum phase estimation and quantum hardware considerations, and devised the chained-QPE attack as well as the efficient unitary challenge method. J.\ Pieprzyk contributed to the classical cryptographic security analysis, including security against known classical attacks, and formulated the key indistinguishability game. C. Thapa and S.\ Camtepe contributed to critical comments, manuscript editing, and proofreading. All authors reviewed and approved the final manuscript.

\section*{Competing interests}
The authors declare the following competing interests: an Australian patent application related to this work has been filed. The applicant is the authors’ institution. The inventors are S. P. Kish and H. J. Vallury. The original application number is AU 2025903524 and the addendum application number is AU 2026902415. The application is currently pending. The application covers aspects of the manuscript relating to the QSA primitive, including planted state and basis seeded spectral feature extraction from public unitaries, the associated unitary generation and compilation methods, and the use of the resulting eigenphase feature vectors as input to a conventional key derivation schedule for deriving cryptographic keys.

\section*{Code availability}
The source code used in this study is available at the repository \cite{kish_2026_21450694}. %During peer review, the code is retained privately to support the review process and can be provided to editors or reviewers upon request.

\section*{Data availability}
The data generated and analysed during this study will be released upon request from the corresponding lead author.

\bibliographystyle{naturemag-doi}

\bibliography{sample}
\appendix
\counterwithin{figure}{section}
\counterwithin{table}{section}
\renewcommand{\thefigure}{\Alph{section}\arabic{figure}}
\renewcommand{\thetable}{\Alph{section}\arabic{table}}

\begin{appendix}

\section*{Appendix}

\section{Supplementary figures and tables}
\label{app:si-floats}
% Floats relocated from the main text (Intro/Results/Discussion).
% When splitting the SI into its own file these renumber as S1, S2, ...

% ---- moved: tab:QSA-costs-cvp ----
\begin{table}[ht!]
\centering
\caption{Asymptotic cost summary for QSA regimes. Here $n$ is the number of system qubits, $m$ is the phase precision (bits) per unitary, and $D(n)$ denotes the depth of a typical public challenge circuit. For QSA-Q, we distinguish the symmetric $U=VDV^\dagger$ compiler (fast-power structure) from the asymmetric $U=V_LV_R^\dagger$ compiler (no read-off and no fast powering). We write $F_V(n)$ for the compilation cost of one expressive map $V$ (including optimisation and transpilation) and treat it as implementation-dependent. The symmetric compiler additionally admits an ancilla-free honest \emph{fast path}: the prover measures $V^\dagger\ket{\psi}$ to recover the label $b$ and evaluates $\theta(b)$ by Eq.~\eqref{eq:phase_closed_form} in $O(n)$ time, with LDQPE and QPE as the general-purpose routes.}
\label{tab:QSA-costs-cvp}
\begin{tabular}{llll}
\toprule
Regime & Challenge setup (per $U_i$) & Verifier eval (per $U_i$) & Prover eval (per $U_i$) \\
\midrule

QSA-M (dense)
& \parbox[t]{0.20\linewidth}{$O(2^{3n})$ (dense Haar/QR sampling)}
& \parbox[t]{0.20\linewidth}{$O(2^{3n})$ (eigendecomp.\ + overlap scan)}
& \parbox[t]{0.20\linewidth}{$O(2^{3n})$ (eigendecomp.\ + overlap scan)} \\[0.5em]

QSA-C (classical spectral eval)
& \parbox[t]{0.20\linewidth}{$O(\mathrm{poly}(n)\,D(n))$ (seeded random circuit)}
& \parbox[t]{0.20\linewidth}{$O(2^m\cdot 2^n\,\mathrm{poly}(n)\,D(n))$ (autocorr.\ moments + FFT/peak pick)}
& \parbox[t]{0.20\linewidth}{$O(2^m\cdot 2^n\,\mathrm{poly}(n)\,D(n))$ (same as verifier)} \\[0.5em]

QSA-Q (i: symmetric, $U=VDV^\dagger$)
& \parbox[t]{0.20\linewidth}{$O(F_V(n))$ (compile $V$ once; choose diagonal $D$ and form $U=VDV^\dagger$)}
& \parbox[t]{0.20\linewidth}{$O(n)$ (closed-form $\arg\langle b|D|b\rangle$; Eq.~\eqref{eq:phase_closed_form})}
& \parbox[t]{0.20\linewidth}{Fast path: $O(\mathrm{depth}(V){+}n)$, no ancilla ($O(\log(1/\varepsilon_{\mathrm{fail}})/(\tfrac12-\delta)^2)$ shots); \newline or LDQPE: $O(\mathrm{poly}(n)\,m)$ (structured); \newline or QPE: $O\!\bigl(N_{\mathrm{rep}}\mathrm{poly}(n,m)\bigr)$} \\[0.5em]

QSA-Q (ii: asymmetric, $U=V_LV_R^\dagger$)
& \parbox[t]{0.20\linewidth}{$O(F_{V_L}(n)+F_{V_R}(n)) \approx O(2F_V(n))$ (compile $V_L$ and $V_R$)}
& \parbox[t]{0.20\linewidth}{$O(\mathrm{poly}(n)\cdot (2^m-1))$ (LDQPE on $(U_i,\ket{\psi_i})$)}
& \parbox[t]{0.20\linewidth}{$O(\mathrm{poly}(n)\cdot (2^m-1))$ (LDQPE, Alg.~2~\cite{ni2023lowdepthqpe})} \\

\bottomrule
\end{tabular}
\end{table}

% ---- moved: fig:chained-qpe-attack ----
\begin{figure}[b!]
\centering
\resizebox{\linewidth}{!}{%
\begin{tikzpicture}[x=1cm,y=1cm,>=latex]

  \tikzset{
    block/.style={draw, rounded corners, thick, align=center, inner sep=4pt},
    smallblock/.style={draw, rounded corners, thin, align=center, inner sep=2pt},
    line/.style={->, thick},
    cline/.style={->, dashed, thin}
  }

  \node[block] (guess) at (0,0) {Adversary guesses\\input state $\ket{\phi_E}$};
  \node[block] (qpe1)  at (3,0) {QPE on $U_1$};
  \node[smallblock] (phase1) at (6,1.0) {Measured phase\\$\widetilde{\theta}_1$};
  \node[smallblock] (state1) at (6,-1.0) {Post-measurement\\eigenstate $\ket{u_1^*}$};

  \node[align=left] (noteprob) at (1.5,-1.9)
    {\footnotesize Success prob.\ that $\ket{u_1^*}$\\\footnotesize matches honest eigenstate\\\footnotesize $\approx 2^{-n}$};

  \node[block] (qpe2)  at (9,0) {QPE on $U_2$};
  \node[smallblock] (phase2) at (12,1.0) {Measured phase\\$\widetilde{\theta}_2$};
  \node[smallblock] (state2) at (12,-1.0) {Post-measurement\\eigenstate $\ket{u_2^*}$};

  \node at (14.5,0) {$\cdots$};

  \node[block] (qpek)  at (17,0) {QPE on $U_k$};
  \node[smallblock] (phasek) at (20,1.0) {Measured phase\\$\widetilde{\theta}_k$};
  \node[smallblock] (statek) at (20,-1.0) {Post-measurement\\eigenstate $\ket{u_k^*}$};

  \node[block, align=center] (phases) at (20,3.0)
    {Collected eigenphase vector\\$\bigl(\widetilde{\theta}_1,\widetilde{\theta}_2,\ldots,\widetilde{\theta}_k\bigr)$};

  \draw[line] (guess) -- (qpe1);
  \draw[line] (qpe1) -- (state1);
  \draw[line] (state1) -- (qpe2);
  \draw[line] (qpe2) -- (state2);
  \draw[line] (state2) -- (qpek);
  \draw[line] (qpek) -- (statek);

  \draw[cline] (qpe1) -- (phase1);
  \draw[cline] (qpe2) -- (phase2);
  \draw[cline] (qpek) -- (phasek);

  \draw[cline] (phase1) |- (phases);
  \draw[cline] (phase2) |- (phases);
  \draw[cline] (phasek) -- (phases);

  \node[align=center] at (3,-2.9) {\footnotesize First QPE stage:\\\footnotesize search over eigenstates of $U_1$};
  \node[align=center] at (11,-2.9) {\footnotesize Chained QPE:\\\footnotesize reuse $\ket{u_i^*}$ as input to $U_{i+1}$};
  \node[align=center] at (19,-2.9) {\footnotesize Goal: reconstruct honest\\\footnotesize eigenphase vector $\boldsymbol{\Theta}$};

\end{tikzpicture}%
}
\caption{\textbf{Attack \Romannum{1} (Chained-QPE / eigenstate propagation) schematic.}
The adversary first guesses an input state \(\ket{\phi_E}\) and runs QPE on \(U_1\), obtaining a measured phase \(\widetilde{\theta}_1\) and a post-measurement eigenstate \(\ket{u_1^*}\). With probability about \(2^{-n}\), \(\ket{u_1^*}\) coincides with the honest signal eigenstate. The adversary then feeds \(\ket{u_1^*}\) into QPE for \(U_2\) to obtain \(\widetilde{\theta}_2\) and \(\ket{u_2^*}\), and so on along the chain \(U_1,\ldots,U_k\), hoping to accumulate an eigenphase vector \((\widetilde{\theta}_1,\ldots,\widetilde{\theta}_k)\) that matches the honest vector \(\boldsymbol{\Theta}\).}
\label{fig:chained-qpe-attack}
\end{figure}

% ---- moved: fig:topeigenvector ----
\begin{figure}[ht!]
  \centering
  \begin{subfigure}[t]{0.49\linewidth}
    \centering
    \includegraphics[width=\linewidth]{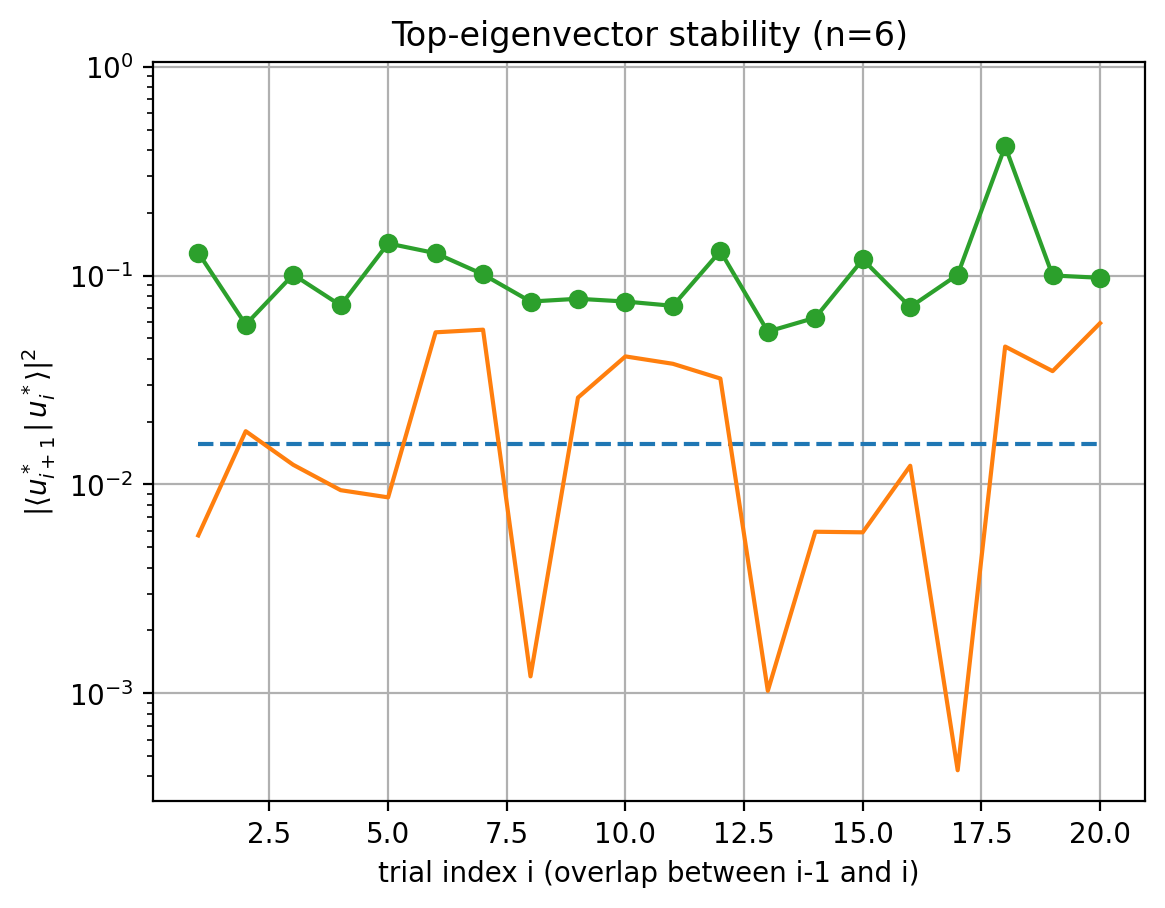}
    \subcaption{QSA-C top-eigenvector stability for \(n=6\) (representative). Dashed line is \(2^{-n}\).}
    \label{fig:topeigenvector:a}
  \end{subfigure}\hfill
  \begin{subfigure}[t]{0.49\linewidth}
    \centering
    \includegraphics[width=\linewidth]{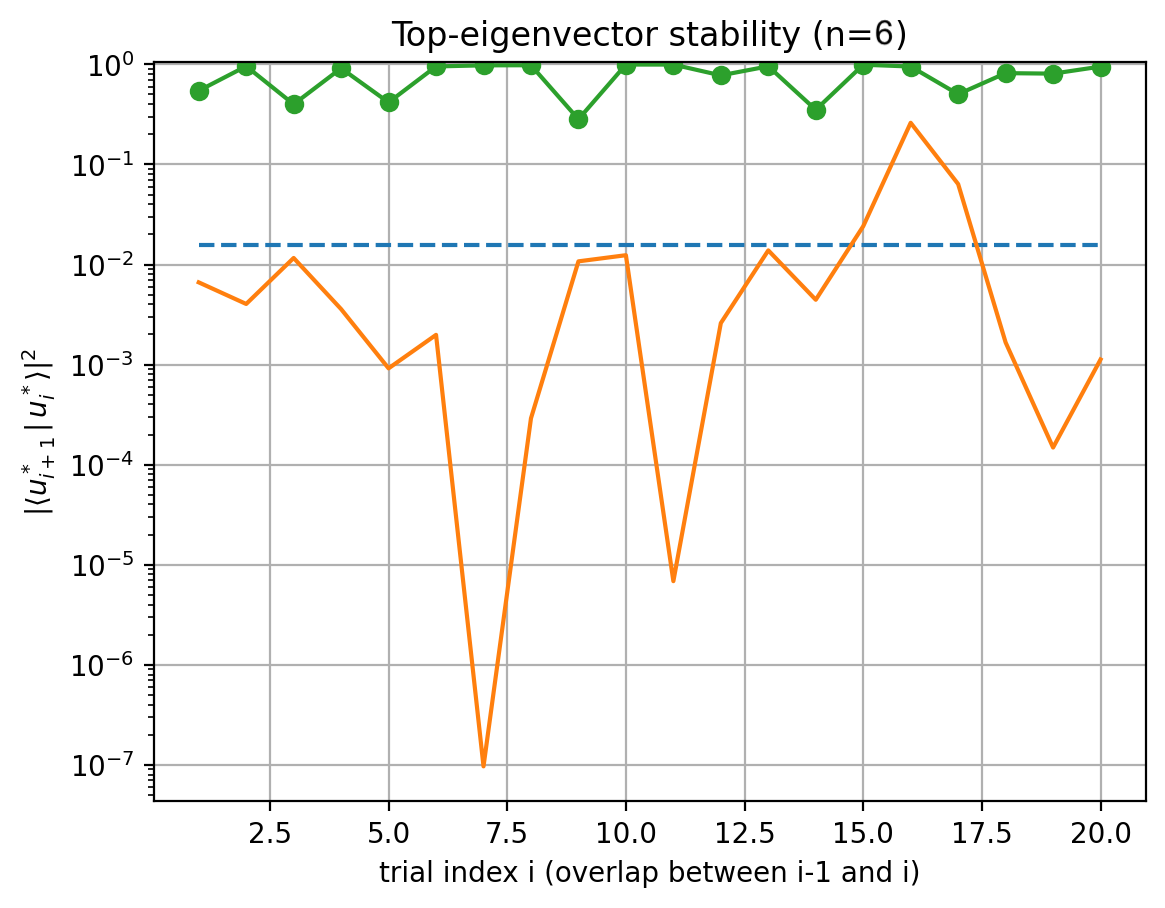}
    \subcaption{QSA-Q top-eigenvector stability for \(n=6\) (representative). Dashed line is \(2^{-n}\).}
    \label{fig:topeigenvector:b}
  \end{subfigure}
 \caption{\textbf{Attack \Romannum{1} decorrelation proxies.}
Representative cross-instance overlaps for independently randomised challenge instances at \(n=6\). In both panels, the orange curve shows the squared overlap \(|\langle u_{i+1}^\ast|u_i^\ast\rangle|^2\) between successive signal (or top) eigenvectors, which is the quantity relevant to chained-QPE reuse across instances. The dashed horizontal line marks the Haar benchmark \(2^{-n}\). In panel~(a) (QSA-C), the public circuit instances are independently randomised and the signal eigenvector is defined \emph{a posteriori} by the classical spectral extractor, so no high planted-state overlap is enforced. In panel~(b) (QSA-Q), each compiled instance is built around its own planted state \(\ket{\psi_i}\), so the green curve shows the planted-state overlap \(|\langle \psi_i|u_i^\ast\rangle|^2\) for \(i\ge 2\), while the orange curve again shows successive signal-eigenvector overlaps. The key observation is that the cross-instance overlaps remain near the Haar scale \(2^{-n}\), indicating that successive signal eigenvectors are effectively decorrelated even in the compiled high-overlap QSA-Q setting.}
  \label{fig:topeigenvector}
\end{figure}

% ---- moved: tab:notation ----
\begin{table}[t]
\caption{Notation used in the security model and Appendix~\ref{app:proofs}.}
\label{tab:notation-sec}
\begin{ruledtabular}
\begin{tabular}{ll}
Symbol & Meaning \\
\hline
$\lambda$ & security parameter \\
$S_0$ & long-term root secret held by the verifier \\
$n$ & qubits per challenge instance \\
$m$ & bits of eigenphase read out per instance (bucket width) \\
$k$ & instances per session \\
$U_i$ & $i$-th public challenge unitary, $U_i=V_iD_iV_i^\dagger$ \\
$V_i$ & compiled eigenbasis circuit (public) \\
$D_i=\bigotimes_q R_z(\beta_{i,q})$ & diagonal factor (public) \\
$\beta_{i,q}$ & diagonal angles of $D_i$ (public) \\
$P_i$ & planting circuit, $\ket{\psi_i}=P_i^\dagger\ket{0^n}$ (verifier-private) \\
$b_i\in\{0,1\}^n$ & hidden label; the planted eigenvector index (secret) \\
$\ket{\psi_i}$ & planted state held by the prover \\
$\theta(x)$ & eigenphase of $U_i$ at eigenvector index $x$, Eq.~\ref{eq:phase_closed_form} \\
$\theta^\star_i=\theta(b_i)$ & honest eigenphase feature of instance $i$ \\
$\mathsf{B}(\theta)\in\{0,1\}^m$ & $m$-bit bucket (rounding) of $\theta$ \\
$\Theta\in\{0,1\}^{mk}$ & feature vector, the concatenation of $k$ buckets \\
$\textit{sid}$ & session identifier, $(N_A,N_B,\mathrm{digest}(U_1\|\cdots\|U_k))$ \\
$N_A,N_B$ & per-session nonces, $2\lambda$ bits each \\
$H$ & key-derivation function, modelled as a (quantum) random oracle \\
$K$ & session key, $K=H(\Theta,\textit{sid})$ \\
$\mathcal{E},\mathcal{D}$ & AEAD encryption / decryption \\
$\ell$ & length of the confirmation response \\
$\pi_P^s$ & $s$-th instance (session) of party $P$ \\
$q_s$ & number of sessions the adversary may run \\
$q_H$ & number of random-oracle queries the adversary may make \\
$\mathcal{A}$ & adversary, quantum polynomial time (QPT) \\
$\varepsilon_{\mathrm{LH}},\varepsilon_{\mathrm{FU}}$ & slack terms in Assumptions~\ref{ass:lh} and~\ref{ass:fu} \\
$p_{\mathrm{flip}}$ & per-instance bucket error rate (completeness, not security) \\
$\kappa$ & $2^m\max_c\Pr[\mathsf{B}=c]$, measured bucket non-uniformity \\
\end{tabular}
\end{ruledtabular}
\end{table}

% ---- moved: fig:bin-mass-honest-vs-eve ----
\begin{figure}[ht!]
    \centering
    \begin{subfigure}[t]{0.49\linewidth}
        \centering
        \includegraphics[width=\linewidth]{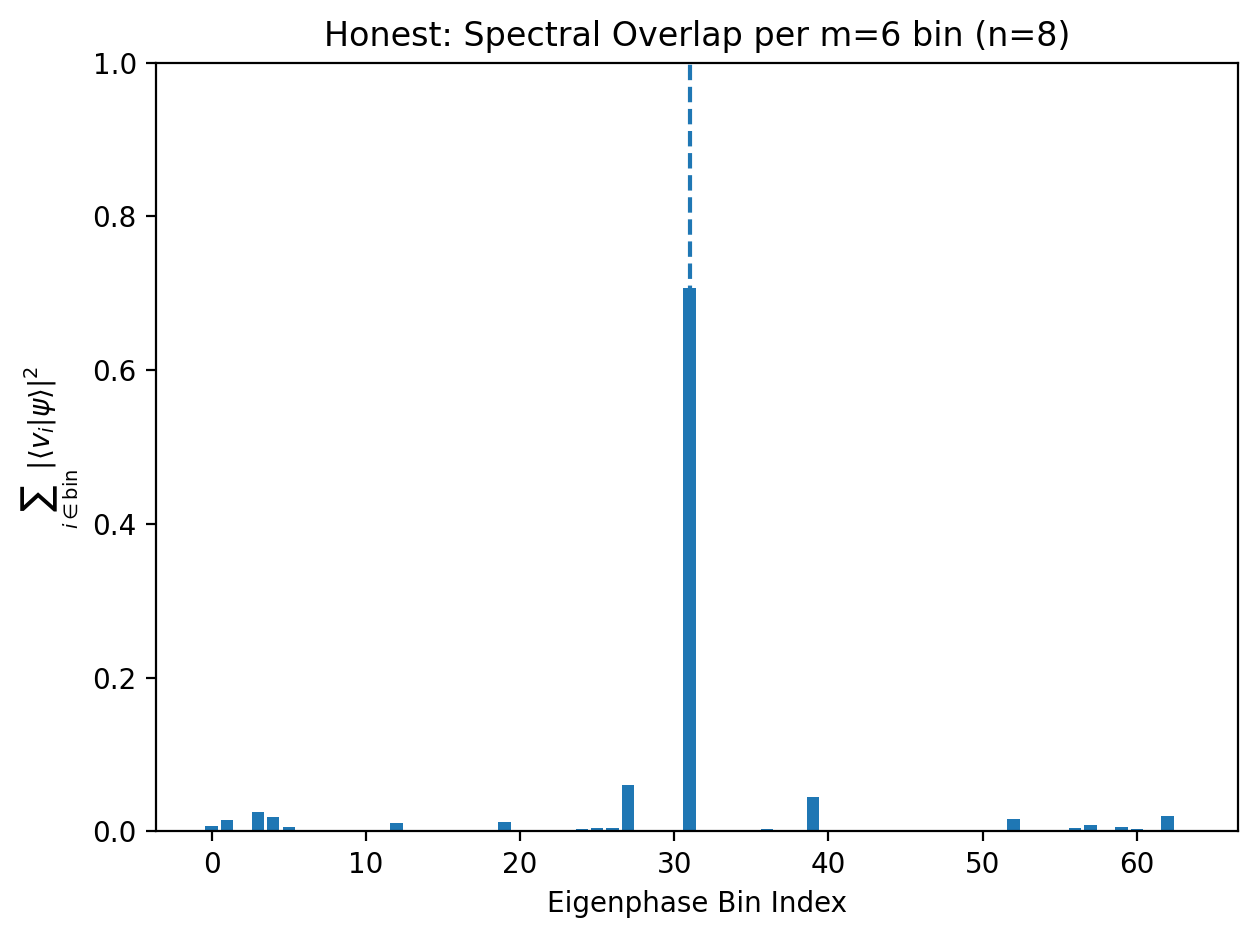}
        \caption{Symmetric ($U=VDV^\dagger$), honest planted state.}
        \label{fig:honest-bin-mass}
    \end{subfigure}
    \hfill
    \begin{subfigure}[t]{0.49\linewidth}
        \centering
        \includegraphics[width=\linewidth]{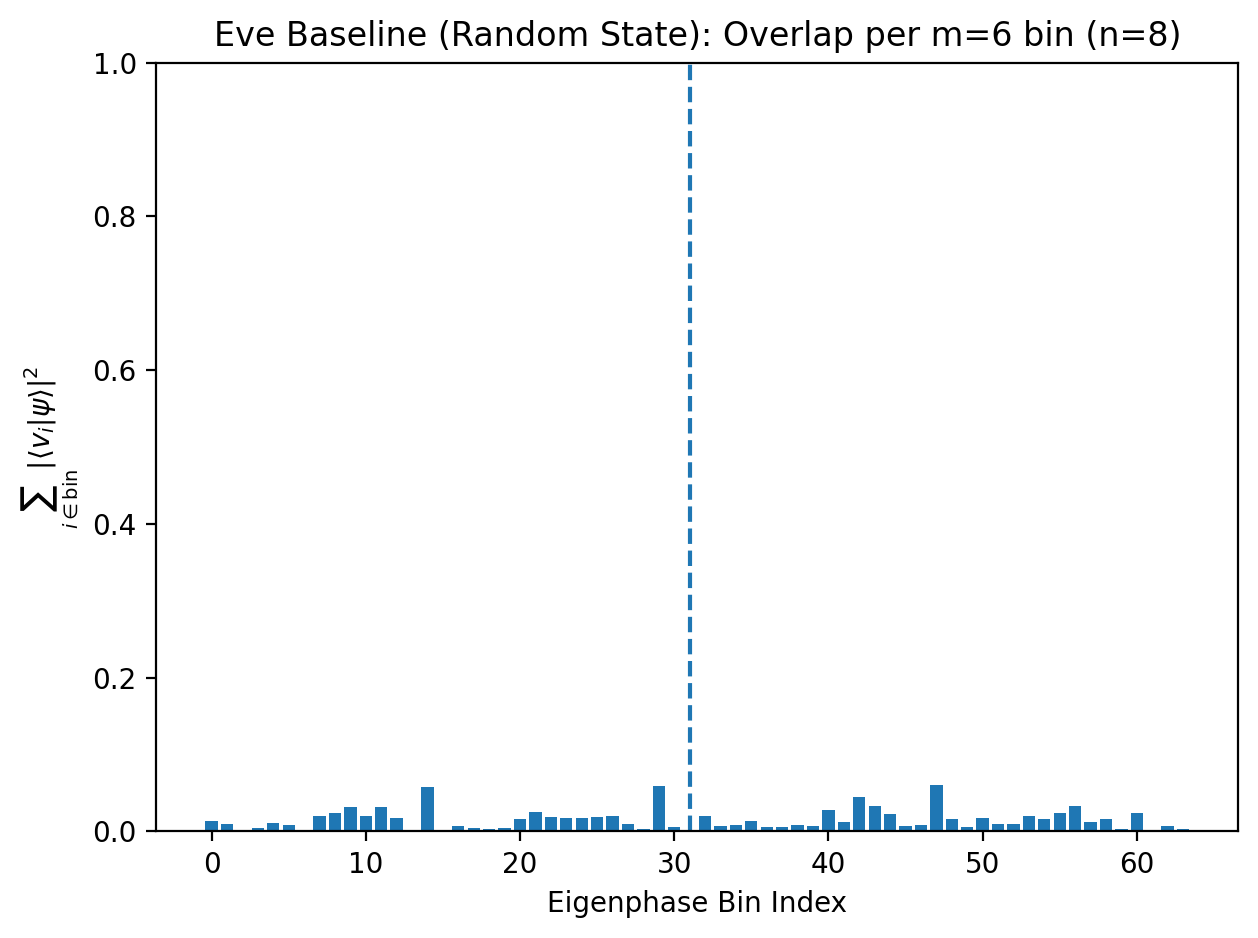}
        \caption{Symmetric ($U=VDV^\dagger$), Eve baseline (random state).}
        \label{fig:eve-bin-mass}
    \end{subfigure}

    \caption{Spectral overlap mass aggregated into $M=2^6$ eigenphase bins ($n=8$, $m=6$) for the symmetric compiler.
    For each compiled instance, we diagonalise $U$ and form
    $p_k=\sum_{i:\mathrm{bin}(\theta_i)=k}|\langle v_i|\cdot\rangle|^2$.
    The planted state $\ket{\psi}$ produces a highly non-uniform binned spectrum (localisation), whereas a random baseline state is broadly spread across bins (delocalisation).}
    \label{fig:bin-mass-honest-vs-eve}
\end{figure}

% ---- moved: fig:app_hadamard_moment1 ----
\begin{figure}[ht!]
\centering
\begin{quantikz}
\lstick{$\ket{0}$} & \gate{H} & \ctrl{1} & \gate{\text{I/$S^\dagger$}} & \gate{H} & \meter{} \\
\lstick{$\ket{\psi_i}$} &  & \gate{U(\vec{\alpha},\vec{\beta})^{t}} & & &
\end{quantikz}
\caption{\textbf{Hadamard-test estimation of $Z(\vec{\alpha},\vec{\beta})_t^{(i)}=\langle\psi_i|U(\vec{\alpha},\vec{\beta})^t|\psi_i\rangle$.}
Measuring the ancilla in the $X$ basis yields $\mathbb{E}[(-1)^x]=\Re [Z(\vec{\alpha},\vec{\beta})_t^{(i)}]$.
To obtain $\Im[Z(\vec{\alpha},\vec{\beta})_t^{(i)}]$, insert $S^\dagger$ before the final Hadamard (equivalently measure in the $Y$ basis).
In practice, $U(\vec{\alpha},\vec{\beta})^t$ is implemented as $t$ sequential applications of $U(\vec{\alpha},\vec{\beta})$ (and controlled-$U(\vec{\alpha},\vec{\beta})$ within the Hadamard test), avoiding synthesis blow-ups for explicit controlled powers.}
\label{fig:app_hadamard_moment1}
\end{figure}

% ---- moved: tab:qsaq-gate-extrapolation ----
\begin{table}[ht!]
\centering
\caption{\textbf{Illustrative scaling extrapolation for symmetric QSA-Q from the observed \(n=m\in\{2,3,4\}\) `ibm\_fez` runs.}
Gate counts are obtained from a simple linear extrapolation of the transpiled circuits and should be interpreted only as indicative scaling estimates, not as calibrated forecasts for larger-\(n\) hardware. The final column gives a rough two-qubit error budget obtained by requiring \((1-p_{2q})^{N_{\texttt{cz}}}\gtrsim 0.95\), i.e. a \(95\%\) survival factor from CZ gates alone.}
\label{tab:qsaq-gate-extrapolation}
\begin{tabular}{cccccc}
\toprule
\(n=m\) & \(\#\texttt{sx}\) & \(\#\texttt{rz}\) & \(\#\texttt{cz}\) & \(\#\texttt{x}\) & rough \(p_{2q}^{\max}\) for \(95\%\) survival \\
\midrule
10 & \(\approx 1183\) & \(\approx 1044\) & \(\approx 555\)  & \(\approx 67\)  & \(\approx 9.2\times 10^{-5}\) \\
15 & \(\approx 1840\) & \(\approx 1624\) & \(\approx 867\)  & \(\approx 102\) & \(\approx 5.9\times 10^{-5}\) \\
20 & \(\approx 2498\) & \(\approx 2204\) & \(\approx 1180\) & \(\approx 137\) & \(\approx 4.3\times 10^{-5}\) \\
\bottomrule
\end{tabular}
\end{table}

\section{Proofs}
\label{app:proofs}

\subsection{Proof of Lemma~\ref{lem:unif}}

\emph{(i)} Write $\theta(b)=\tfrac12\sum_q s_q\beta_q$ with $s_q:=2b_q-1$ uniform on
$\{\pm1\}$ and independent. Then
$\hat\mu_\beta(j)=\mathbb{E}_b[e^{ij\theta(b)}]
=\prod_q\mathbb{E}_{s_q}[e^{ijs_q\beta_q/2}]=\prod_q\cos(j\beta_q/2)$, exactly.

\emph{(ii)} This is the Erd\H{o}s-Tur\'an inequality applied to $\mu_\beta$ on the
circle: for any probability measure $\mu$ and any $J\ge1$, the discrepancy
$\sup_{I}|\mu(I)-|I|/2\pi|\le \tfrac{1}{J+1}+C\sum_{j=1}^{J}|\hat\mu(j)|/j$, and
$\sum_{j\le J}1/j\le1+\ln J$. Buckets are arcs of length $2\pi/2^m$.

\emph{(iii)} Fix $j\ge1$. Since $\beta_q$ is uniform on an interval of length $2\pi$,
the variable $j\beta_q/2$ ranges over an interval of length $j\pi$ and is therefore
\emph{exactly} uniform modulo $\pi$, for every integer $j\ge1$. Hence
$X_q:=\ln|\cos(j\beta_q/2)|$ are i.i.d.\ with
$\mathbb{E}[X]=\tfrac1\pi\int_0^\pi\ln|\cos u|\,du=-\ln2$ and
$\mathrm{Var}[X]=\pi^2/12$; $X\le0$, and the moment generating function
$\mathbb{E}[e^{tX}]=\Gamma(\tfrac{t+1}{2})/(\sqrt\pi\,\Gamma(\tfrac t2+1))$ is finite
for $t>0$, so a Chernoff bound applies to the upper tail of $\sum_q X_q$. Thus
$\Pr\big[\prod_q|\cos(j\beta_q/2)|\ge2^{-n}e^{t}\big]\le e^{-ct^2/n}$. A union bound
over $1\le j\le J=2^{2m}$ with $t=\Theta(\sqrt{n\ln J})$ gives failure probability at
most $1/J\le2^{-m}$ and
$\max_{j\le J}|\hat\mu_\beta(j)|\le2^{-n+c_1\sqrt{nm}}$. Substituting into (ii) with
$J=2^{2m}$ yields
$\max_c|\Pr[\mathsf{B}=c]-2^{-m}|\le2^{-2m}+O(m)\,2^{-n+c_1\sqrt{nm}}$, and the stated
hypothesis on $n$ makes the second term at most $2^{-2m}$, giving
Eq.~\eqref{eq:minent}. The constants $C,c,c_1$ are absolute and unoptimised. \qed

\subsection{Proof of Corollary~\ref{cor:minent}}
Under Assumption~\ref{ass:lh}, the conditional law of $b_i$ given $U_i$ is within
statistical distance $\varepsilon_{\mathrm{LH}}$ of uniform, so
Lemma~\ref{lem:unif} applies to each instance up to that distance. The $(b_i,\beta_i)$
are independent across $i$ by the HKDF schedule of Eq.~(8), and min-entropy is
additive over independent blocks. \qed

\subsection{Proof of Theorem~\ref{thm:ma}}

\textbf{Game $\mathrm{G}_0$.} The real experiment of Definition~\ref{def:ma}.

\textbf{Game $\mathrm{G}_1$.} Abort if two sessions share a nonce pair. Since nonces
are $2\lambda$ bits, $|\Pr[\mathrm{G}_0]-\Pr[\mathrm{G}_1]|\le q_s^2 2^{-2\lambda}$.
After $\mathrm{G}_1$, distinct sessions have distinct $\textit{sid}$, so partnering is
well defined and every fresh session queries $H$ at a distinct point
$(\Theta^{(s)},\textit{sid}^{(s)})$.

\textbf{Game $\mathrm{G}_2$.} For each fresh session $s$, replace
$K^{(s)}=H(\Theta^{(s)},\textit{sid}^{(s)})$ by an independent uniform string.
Proceed by a hybrid over the $q_s$ sessions. In hybrid $s$, the only way
$\mathcal{A}$ distinguishes is by querying $H$ at $(\Theta^{(s)},\textit{sid}^{(s)})$.
By the one-way-to-hiding lemma with semi-classical oracles, the distinguishing
advantage is at most $2\sqrt{\Pr[\mathrm{Find}]}$, and by the measure-and-extract
argument, $\Pr[\mathrm{Find}]\le4q_H\cdot\varepsilon$ where $\varepsilon$ bounds the
probability that a QPT algorithm outputs $\Theta^{(s)}$ given the adversary's view.
That view consists of the public challenges of all $q_s$ sessions and, in the leakage
variant, the features of the other sessions; it is exactly the view of
Assumption~\ref{ass:fu}, so $\varepsilon\le2^{-mk}+\varepsilon_{\mathrm{FU}}$.
Summing the hybrids,
$|\Pr[\mathrm{G}_1]-\Pr[\mathrm{G}_2]|\le
4q_s\sqrt{q_H(2^{-mk}+\varepsilon_{\mathrm{FU}})}$. This step is also the proof of
Corollary~\ref{cor:ki}.

\textbf{Game $\mathrm{G}_3$.} In $\mathrm{G}_2$ every fresh session key is uniform and
independent of the adversary's view. For a fresh instance to accept without a partner,
$\mathcal{A}$ must produce a confirmation ciphertext that decrypts and echoes
correctly under a key she does not hold. Per session this is bounded by
$\mathrm{Adv}^{\mathrm{auth}}_{\mathcal{E}}(\mathcal{B})+2^{-\ell}$, the second term
covering the case of a correct blind guess of the $\ell$-bit response. Summing over
sessions gives the second term of Eq.~\eqref{eq:mabound}, and collecting the hops
gives the theorem. \qed

\subsection{Remarks on tightness}
The $\sqrt{\;\cdot\;}$ loss in $\mathrm{G}_2$ is inherited from one-way-to-hiding and
is matched by an attack: an eavesdropper holding
$\mathcal{C}_1=\mathcal{E}(M_1,K)$ possesses an offline predicate for candidate
$\Theta$, and Grover search over $\{0,1\}^{mk}$ costs $\Theta(2^{mk/2})$. The bound is
therefore tight in its dependence on $mk$ up to constants, and the parameter
recommendation $mk\ge2\lambda$ follows. The factor $q_s$ is the usual multi-instance
degradation and can be removed only by per-session independent long-term secrets,
which QSA does not have by construction.

\section{Notation and conventions}
In table \ref{tab:notation}, we collect the symbols and acronyms used throughout. Unless stated otherwise, $n$ denotes the number of qubits, $d=2^n$ the Hilbert-space dimension, and $k$ the number of public unitaries.

\begin{table}[ht!]
\centering
\renewcommand{\arraystretch}{1.12}
\begin{tabular}{ll}
\toprule
\textbf{Object} & \textbf{Meaning / role} \\
\midrule
\multicolumn{2}{l}{\textbf{Acronyms / named objects}} \\
QSA & Quantum Spectral Authentication (this primitive). \\
QSA-M / QSA-C / QSA-Q & Dense-matrix / classically evaluated circuit / quantum-evaluated circuit instantiations. \\
KDF, HKDF & Key-derivation function; HMAC-based KDF (RFC 5869) used as extract-expand interface. \\
PRF & Pseudorandom function used to expand seeds into circuit parameters/schedules. \\
QPE, LDQPE & (Low-depth) Quantum phase estimation used for eigenphase extraction. \\
EVD & Eigenvalue/eigenvector decomposition (dense diagonalisation). \\
FFT & Fast Fourier transform (periodogram computation). \\
AEAD & Authenticated encryption with associated data (key confirmation wrapper). \\
\midrule
\multicolumn{2}{l}{\textbf{Dimensions, keys, and schedules}} \\
$\lambda$ & Security parameter (with $n=n(\lambda)$, $k=k(\lambda)$ at most polynomial in $\lambda$). \\
$n$, $d=2^n$ & Number of qubits; Hilbert-space dimension. \\
$k$ & Number of public unitaries in the instance $\{U_i\}_{i=1}^k$. \\
$m$ & Extracted phase bits per unitary (target phase resolution). \\
$\ell$ & Output key length. \\
$K$ & Derived session key (e.g. $K=\HKDF(\Theta,\mathrm{aux})$). \\
$\mathrm{aux}$ & Public KDF context (salt/labels/info). \\
$N_{\mathrm{ep}}$ & Number of key-derivation epochs, typically $N_{\mathrm{ep}}=\lceil 256/m\rceil$. \\
$N_s$ & Repetitions/samples per phase point in timing models. \\
\midrule
\multicolumn{2}{l}{\textbf{Public unitaries, planted state, and spectral features}} \\
$U_i$ & $i$-th public $n$-qubit unitary (matrix in QSA-M; circuit in QSA-C/Q). \\
$P$ (or $P_i$) & Planted state preparation circuit defining $|\psi\rangle=P^\dagger|0^n\rangle$ (or per-unitary $P_i$). \\
$|\psi\rangle$ & Planted state used by honest parties. \\
$\Theta=(\theta_1,\ldots,\theta_k)$ & Dominant-eigenphase feature vector extracted across $\{U_i\}$. \\
$Z_t$ & Autocorrelation: $Z_t=\langle\psi|U_i^t|\psi\rangle$, $t=0,\ldots,T-1$. \\
$S(\omega)$ & Periodogram/matched filter over an FFT grid: $S(\omega)=\left|\sum_{t=0}^{T-1} Z_t e^{-i\omega t}\right|$. \\
$\mu$ & Eigenvalue variable in characteristic equations (reserved so $\lambda$ can denote security parameter). \\
$\Lambda_i$ & Diagonal eigenvalue matrix in $U_i=V_i\Lambda_iV_i^\dagger$. \\
$\tau$ & Hamiltonian-simulation time in the mapping $U=e^{i(H-c\mathbb{I})\tau}$. \\
$G$ & Circuit gate count used in dense-conversion cost estimates. \\
\midrule
\multicolumn{2}{l}{\textbf{Promise / success-probability parameters}} \\
$\delta$ & High-overlap design knob (in QSA-Q). \\
$p_E^{\mathrm{tot}}$ & Total success probability for a composite adversarial event (as defined in the attacks). \\
$p_{\rm succ}$ & Generic success probability (e.g. state-guessing success). \\
\bottomrule
\end{tabular}
\caption{Notation and acronyms. Acronyms are expanded at first use in the text.}
\label{tab:notation}
\end{table}

\paragraph{Conventions.}
Vectors are columns; $\mathrm{Arg}(\cdot)\in(-\pi,\pi]$; overlaps use the standard inner product; and $\|\cdot\|$ denotes the operator norm for operators. Unless stated otherwise, depth counts two-qubit layers.
% ============================================================
% Appendix: Reference attacks (moved out of main taxonomy)
% ============================================================
%\appendix

\section{Reference spectrum and state-guessing attacks}
\label{app:reference-attacks}
This appendix collects \emph{reference} attack models that are useful for calibration and for the $m\gtrsim n$ spectrum-covering regime, but which are not the dominant threats under the intended online-forgery deployment assumptions (fresh transcript binding, rate-limited key confirmation, and no access to the secret state-preparation circuit/seed or copies of the planted state schedule). These attacks are referenced from Sec.~\ref{sec:impl-security} and are included here for completeness.

% ------------------------------------------------------------
\subsection{Appendix Attack A.1: Spectrum-oracle diagonalisation and candidate-response search}
\label{app:attack-spectrum-oracle}

\noindent\textbf{Appendix Attack A.1 (spectrum computation).}
In \textbf{Attack A.1(a)}, an adversary uses classical algorithms to compute the full eigenspectrum of each public challenge unitary.

For \textbf{QSA-M}, where $U_i$ is given explicitly as a dense $2^n\times 2^n$ matrix, the dominant cost is dense eigendecomposition, scaling as
\[
\mathrm{cost}(\text{Attack A.1(a)}) \;\approx\; k \times O(2^{3n}).
\]

For \textbf{QSA-C} and \textbf{QSA-Q}, the public description is a circuit. Converting an $n$-qubit circuit with $G$ gates into its explicit dense $2^n\times 2^n$ matrix requires simulating the circuit on all $2^n$ computational basis states, costing $O(G\,2^{2n})$ time and $O(2^{2n})$ memory, followed by diagonalisation (again $O(2^{3n})$ in the worst case). This remains exponentially hard in $n$ and is best viewed as a baseline \emph{spectrum access} cost.

\medskip
\noindent\textbf{Appendix Attack A.1 (candidate-response search and verification).}
In \textbf{Attack A.1(b)}, the adversary assumes spectra are known and attempts to identify the \emph{selected} phase-feature vector $\boldsymbol{\Theta}$ produced by the honest parties. Crucially, the spectrum alone does not reveal which eigenphase bucket is selected, because selection depends on the hidden planted state (or state-preparation seed) through overlaps such as $|\langle v_j|\psi\rangle|^2$. Thus, without additional leakage, Attack A.1(b) reduces to \emph{guessing} the response token produced by the hidden selector.

If the extracted phase precision is $m$ bits per unitary, the number of possible response vectors is at most $2^{mk}$, so any generic attacker that aims to \emph{forge a valid transcript-bound response in a single session} has success probability at most $2^{-mk}$ per attempt. In this online-forgery setting, the natural work factor is therefore $2^{mk}$ \emph{in the sense of one-shot guessing}, and the attack is dominated by directly guessing the derived session key/material $K$ (or its confirmation tag), rather than by diagonalising $U_i$.

If, on the other hand, the adversary is granted an unrealistically strong \emph{verification oracle} that can be queried many times (e.g., unlimited adaptive key-confirmation attempts on the same transcript), then the attacker can test candidate $\boldsymbol{\Theta}$ values until acceptance. In this oracle model, the query complexity is $O(2^{mk})$ classically, and $O(2^{mk/2})$ under Grover search. Such oracle access is not available in the intended deployment: key confirmation is transcript-bound (fresh challenges per session) and rate-limited, so the adversary obtains at most a small number of accept/reject outcomes across sessions, preventing scalable search.

Finally, in the extreme $m>n$ regime, one may also express the candidate space as $2^{nk}$ by counting eigenvectors rather than bins; however, this does not change the practical conclusion: without leakage that links public spectra to the hidden selector, the attacker does not gain advantage from enumerating eigenphases, and online impersonation remains governed by the $mk$-bit response length and confirmation.

\medskip
\noindent\textbf{Practicality and combined cost.}
For \textbf{QSA-M}, one may write $\text{Attack A.1}=\text{Attack A.1(a)}+\text{Attack A.1(b)}$ as an upper bound on offline work \emph{given a verification oracle}. For \textbf{QSA-C/Q}, the same spectrum-only reasoning applies only after paying the exponential spectrum-access cost in Attack A.1(a). In the intended authenticated-and-rate-limited setting, however, Attack A.1 is not the dominant online threat: the optimal generic strategy is to guess the confirmation key/tag directly, giving per-session success probability $\approx 2^{-mk}$ (or $\approx 2^{-|K|}$ after KDF compression), while spectrum computation primarily speaks to long-horizon credential-extraction attempts under additional leakage assumptions.

% ------------------------------------------------------------
\subsection{Appendix Attack A.2: QPE-based coupon-collecting over eigenphases}
\label{app:attack-qpe-coupon}

\noindent\textbf{Appendix Attack A.2: QPE-based coupon-collecting over eigenphases.}
In \textbf{Attack A.2(a)}, the adversary aims to learn the full spectrum of each public unitary, but instead of classical diagonalisation uses a quantum computer and quantum phase estimation (QPE). For a fixed $n$-qubit unitary $U$, the adversary prepares some input state
\[
\ket{\phi} = \sum_{j=1}^{2^n} c_j \ket{v_j},
\]
where $\{\ket{v_j}\}$ is an eigenbasis of $U$ with eigenphases $\{\theta_j\}$. A standard QPE routine on input $\ket{\phi}$ outputs an estimate of one eigenphase $\theta_j$ with probability $|c_j|^2$; repeating QPE i.i.d.\ therefore samples eigenphases according to the induced distribution $\{|c_j|^2\}$.

If $\ket{\phi}$ is a generic state (e.g., Haar-random, or a sufficiently expressive ansatz), then with high probability the weights satisfy $|c_j|^2 \approx 1/2^n$ for all $j$, and QPE outputs are close to uniformly random over the spectrum. In this regime, the problem of learning \emph{all} eigenphases of $U$ reduces to a coupon-collector problem over a set of size $2^n$: the expected number of QPE runs needed to see every eigenphase at least once is $\Theta\!\big(2^n \log(2^n)\big)$.
Each QPE run has cost $\mathrm{poly}(n)$ gates plus the cost of implementing controlled powers $U^{2^j}$, so the total cost to reconstruct the full spectrum of one unitary by QPE sampling is
\[
\tilde{O}\!\big(2^n \,\mathrm{poly}(n)\big),
\]
where $\tilde{O}$ hides polylogarithmic factors.

An adversary might try to bias the distribution $\{|c_j|^2\}$ in her favour by carefully choosing $\ket{\phi}$, for example via a variational routine that amplifies overlap with some subset of eigenstates. This can reduce the time to learn a \emph{few} high-weight eigenphases, but it does not help to discover the entire spectrum unless the number of significantly weighted eigenstates is itself small. In the QSA regime, the public unitaries are compiled to be expressive and to have non-degenerate spectra, so we do not expect such heavy concentration on a small subset of eigenstates.

QPE-based spectral recovery, therefore, has asymptotic cost
\[
\text{cost}(\text{Attack A.2}) \;\approx\; k \times \tilde{O}\!\big(2^n \,\mathrm{poly}(n)\big)+\text{Attack A.2(b)},
\]
where
\[
\begin{split}
\mathrm{cost}(\text{Attack A.2(b)})
\;\approx\;
O\!\left(2^{k\,\min\{m,n\}}\right)
\;=\;
\begin{cases}
O\!\left(2^{mk}\right), & m \le n,\\[4pt]
O\!\left(2^{nk}\right), & m > n.
\end{cases}
\end{split}
\]
This remains exponential in $n\times k$ or $m\times k$: QPE can reduce the barrier to \emph{collecting} eigenphases relative to dense diagonalisation, but it does not remove the exponential barrier associated with phase attribution across $k$ independent instances.

Once the adversary has collected an approximation of the full spectra, she still faces the combinatorial phase-assignment problem \textbf{Attack A.2(b)}: guessing which eigenphase per unitary corresponds to the honest planted state. Thus, QPE coupon-collecting is best viewed as a \emph{quantum variant of the classical baseline Attack A.1}: it changes constant factors and replaces dense diagonalisation by QPE sampling, but it does not remove the exponential barrier.

\begin{figure}[ht!]
  \centering
  \includegraphics[width=0.7\linewidth]{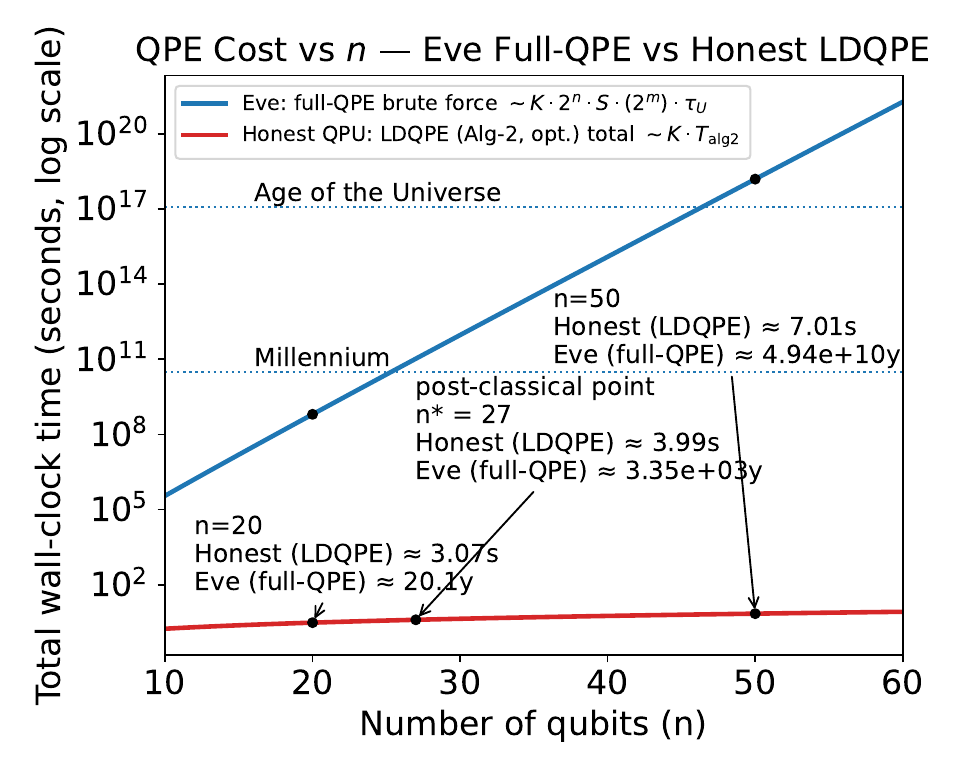}
  \caption{\textbf{Prototype scaling: honest LDQPE vs quantum Eve (Attack A.2, full-QPE brute force).}
  Total wall-clock time (log scale) versus (logical) qubit count~$n$ for $n\in[10,60]$.
  Honest evaluation uses LDQPE Algorithm~2 (Ni-Li-Ying), with $m=2$, key length $\ell_K=256$ bits and $k=\lceil 256/m\rceil=128$ unitaries, and the same depth model $\mathrm{depth}(U)=100+30n$.
  The hardware time per circuit layer is $T_{\mathrm{layer}}=5~\mu$s with control multiplier $c_{\mathrm{qpu}}=1.5$, so one apply-$U$ time is
  $T_{\mathrm{apply}}^{\mathrm{qpu}}=\mathrm{depth}(U)\cdot c_{\mathrm{qpu}}\cdot T_{\mathrm{layer}}$.
  Eve runs a full $m$-bit QPE circuit per trial with controlled-$U$ slowdown factor $c_{cU}=3$, $N_s=50$ repetitions per eigenphase, per-trial overhead factor $f_{\mathrm{oh}}=10$, and a measurement/reset overhead $T_{\mathrm{meas}}=5~\mu$s; the full-QPE circuit cost is modelled as
  $T_{\mathrm{QPE}} \approx (2^m-1)\cdot c_{cU}\cdot T_{\mathrm{apply}}^{\mathrm{qpu}} + T_{\mathrm{meas}}$ (plus a small $O(m^2)$ QFT term),
  and Eve's total cost scales as $T_{\mathrm{eve}} = N_{\mathrm{ep}}\cdot 2^n \cdot N_s \cdot f_{\mathrm{oh}} \cdot T_{\mathrm{QPE}}$.
  Annotated points show $n=27$ (honest $\approx 3.99$\,s; Eve $\approx 3,350$\,years) and $n=50$ (honest $\approx 7.01$\,s; Eve $\approx 4.94\times 10^{10}$\,years).}
  \label{fig:QSA-quantum-eve}
\end{figure}

\begin{figure}[ht!]
  \centering
  \includegraphics[width=0.7\linewidth]{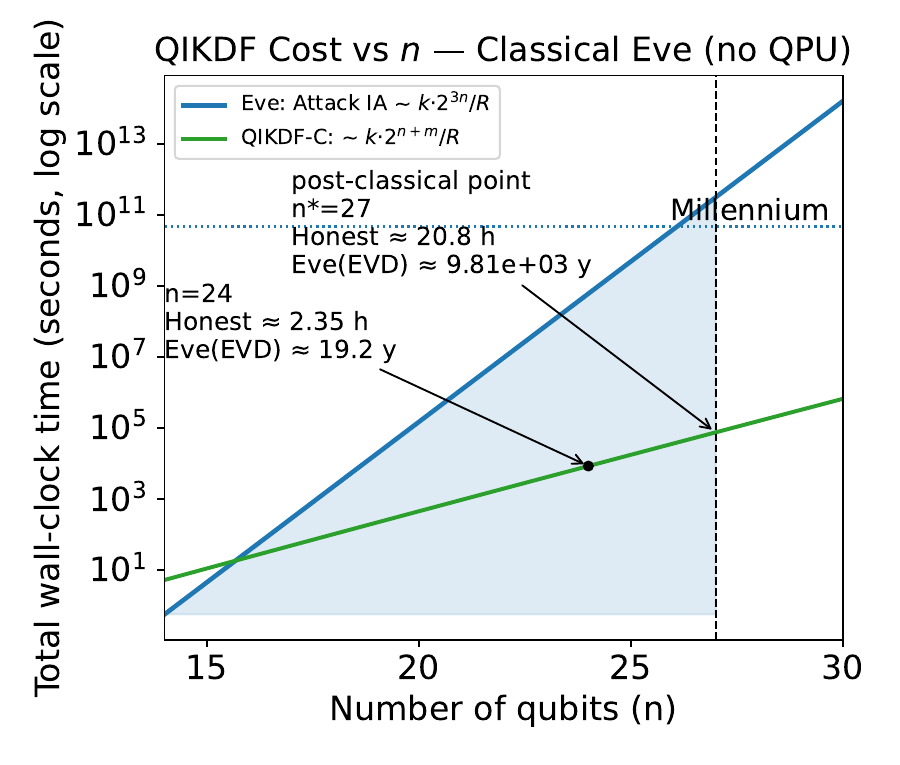}
  \caption{\textbf{Prototype scaling: QSA-C vs classical Eve (Attack A.1, dense EVD).}
  Total wall-clock time (log scale) versus (logical) qubit count~$n$ for $n\in[10,30]$.
  Parameters: $m=2$ phase bits; $N_{\mathrm{ep}}=\lceil 256/m\rceil=128$ epochs; circuit depth model $\mathrm{depth}(U)=d_0+d_1 n$ with $d_0=100$, $d_1=30$.
  Honest (QSA-C) classical evaluation uses $N_s=200$ samples per phase point and a classical apply-$U$ model
  $T_{\mathrm{apply}}^{\mathrm{class}}=\mathrm{depth}(U)\cdot c_{\mathrm{class}}\cdot 2^{n+m}/R_{\mathrm{class}}$
  with $c_{\mathrm{class}}=3$, $R_{\mathrm{classical}}=10^{12}$, giving
  $T_{\mathrm{honest}}=N_{\mathrm{ep}}\cdot m\cdot N_s\cdot T_{\mathrm{apply}}^{\mathrm{class}}$.
  Eve is modelled as dense EVD with arithmetic scaling
  $T_{\mathrm{EVD}} = N_{\mathrm{ep}}\cdot 2^{3n}/R_{\mathrm{SC}}$ using supercomputer values $R_{\mathrm{SC}}=10^{15}$.
  The annotated post-classical point is $n^\star=27$ (honest $\approx 7.5\times 10^4$\,s; Eve(EVD) $\approx 9.81\times 10^3$\,years).}
  \label{fig:QSA-classical-eve}
\end{figure}

\subsection*{Adversary cost scaling with the number of qubits}
\label{sec:prototype-performance}

To calibrate the constant factors implicit in Table~\ref{tab:QSA-costs-cvp}, we evaluated wall-clock time as a function of the system size~$n$. Figures~\ref{fig:QSA-quantum-eve} and~\ref{fig:QSA-classical-eve} summarise the resulting order-of-magnitude scaling for (i) an honest evaluator and (ii) representative adversarial attacks, with total runtime shown on a logarithmic scale. These plots should be interpreted as \emph{calibration models} rather than performance benchmarks; the assumptions and parameter values used in each cost model are stated in the figure captions. 

Figure~\ref{fig:QSA-quantum-eve} compares an honest QPU evaluator running low-depth QPE (LDQPE; Algorithm~2 of Ni-Li-Ying, using the per-epoch optimal scaling) against a quantum adversary executing Appendix Attack A.2(a) under a full-QPE brute-force model. The honest runtime is dominated by repeated applications of a shallow unitary~$U$ within LDQPE across $k=\lceil 256/m\rceil$ epochs. In the plotted model, the apply-$U$ time is set by the depth scaling $\mathrm{depth}(U)=100+30n$, a layer time $T_{\mathrm{layer}}=5\,\mu$s, and a control multiplier $c_{\mathrm{qpu}}=1.5$, giving $T_{\mathrm{apply}}^{\mathrm{qpu}}=\mathrm{depth}(U)\,c_{\mathrm{qpu}}\,T_{\mathrm{layer}}$. Eve is modelled as running an $m$-bit full-QPE circuit per trial with controlled-$U$ slowdown factor $c_{cU}=3$, $N_s=50$ repetitions per eigenphase, per-trial overhead factor $f_{\mathrm{oh}}=10$, and measurement/reset overhead $T_{\mathrm{meas}}=5\,\mu$s, with per-circuit cost $T_{\mathrm{QPE}}\approx (2^m-1)\,c_{cU}\,T_{\mathrm{apply}}^{\mathrm{qpu}} + T_{\mathrm{meas}}$ (plus a small $O(m^2)$ QFT term). Under this model, Attack A.2 scales as
$T_{\mathrm{eve}} = N_{\mathrm{ep}}\cdot 2^n \cdot N_s \cdot f_{\mathrm{oh}} \cdot T_{\mathrm{QPE}}$,
reflecting the need to ``cover'' an exponentially large set of eigenphases. The annotated points illustrate the resulting separation: at $n=27$ the honest runtime is $\approx 3.99$\,s while Eve requires $\approx 3{,}350$\,years; by $n=50$ the honest runtime remains $\approx 7.01$\,s while Eve rises to $\approx 4.94\times 10^{10}$\,years. Thus, for moderate~$n$, the full-spectrum quantum attack becomes infeasible long before honest evaluation.

Figure~\ref{fig:QSA-classical-eve} gives the analogous comparison when the attacker has no QPU and is restricted to classical computation. The honest evaluator runs QSA-C (classical simulation of the evaluation subroutine), with runtime dominated by repeated applications of a shallow circuit model for~$U$. In the plotted arithmetic model, honest evaluation uses $m=2$ phase bits and $N_{\mathrm{ep}}=\lceil 256/m\rceil=128$ epochs, with $\mathrm{depth}(U)=d_0+d_1 n$ where $d_0=100$ and $d_1=30$, and $N_s=200$ samples per phase point. The classical apply-$U$ time is modelled as
$T_{\mathrm{apply}}^{\mathrm{class}}=\mathrm{depth}(U)\,c_{\mathrm{class}}\,2^{n+m}/R_{\mathrm{class}}$
with $c_{\mathrm{class}}=3$ and $R_{\mathrm{class}}=10^{12}$, giving
$T_{\mathrm{honest}} = N_{\mathrm{ep}}\cdot m\cdot N_s\cdot T_{\mathrm{apply}}^{\mathrm{class}}$.
Eve is modelled as running Appendix Attack A.1(a): dense eigendecomposition of an effective $2^n\times 2^n$ operator with arithmetic scaling
$T_{\mathrm{EVD}} = N_{\mathrm{ep}}\cdot 2^{3n}/R_{\mathrm{SC}}$,
using a supercomputer throughput proxy $R_{\mathrm{SC}}=10^{15}$. The annotated post-classical point is $n^\star=27$: at this point the honest runtime is $\approx 7.5\times 10^4$\,s while Eve's dense EVD time is already $\approx 9.81\times 10^3$\, years.

Dense EVD is not limited only by FLOP/s: it requires storing at least one dense complex matrix (and typically additional work arrays), which imposes a hard feasibility cutoff in RAM. Table~\ref{tab:memory-cutoffs} summarises these memory cutoffs for representative top-tier systems: dense-matrix EVD becomes infeasible at roughly $n\approx 24$ to $25$ even under optimistic assumptions about usable aggregate memory, whereas state-vector simulation remains feasible to substantially larger $n$ (for the same $m$) because it scales as $O(2^{n+m})$ memory rather than $O(2^{2n})$.

% Preamble (if not already included):
% \usepackage{graphicx}
% \usepackage{subcaption}

These prototypes are not intended as performance benchmarks; rather, they provide a concrete calibration of the asymptotic scalings used in Table~\ref{tab:QSA-costs-cvp}. The main design takeaway is that QSA is most attractive in the high-dimensional regime and a prover must hold a QPU to evaluate: choose~$n$ large enough that attacks are either (i) astronomically slow under conservative arithmetic models (Attack A.2 strategies), or (ii) outright infeasible due to memory constraints for dense linear algebra (Attack A.1), while keeping $\mathrm{depth}(U)$ modest so that honest low-depth QPE (or its classical simulation) remains practical.

\begin{table}[t!]
\centering
\caption{Memory feasibility cutoffs for two classical approaches to spectral extraction on a $2^n$-dimensional space. Dense-matrix EVD requires storing at least one dense complex matrix ($\approx 16\cdot 2^{2n}$ bytes in complex64/complex128 conventions; here we use complex128 at 16 bytes/entry). State-vector circuit simulation requires storing a $2^{n+m}$-dimensional complex state ($\approx 16\cdot 2^{n+m}$ bytes), where $m$ is the phase register size (here $m=2$ as in the prototype plots). Values are theoretical upper bounds assuming the full system RAM were available to a single job; practical limits are typically 1 to 2 qubits lower due to distributed layout overheads, replication, checkpointing, and work buffers.}
\label{tab:memory-cutoffs}
\begin{tabular}{lccc}
\hline
System (aggregate RAM) & RAM & Dense EVD max $n$ & State-vector max $n$ (with $m{=}2$) \\
\hline
Fugaku & 4.85\,PiB & 24 & 46 \\
El Capitan & 5.4375\,PB & 24 & 46 \\
Frontier & 9.2\,PB & 24 & 47 \\
Aurora (DDR5+HBM) & 20.42\,PB & 25 & 48 \\
\hline
\end{tabular}
\end{table}

% ------------------------------------------------------------
\subsection{Appendix Attack A.3: Hilbert-space search and state guessing}
\label{app:attack-hilbert-guess}

\noindent\textbf{Appendix Attack A.3: Hilbert-space search/state guessing for the planted eigenphase features.}
In \textbf{Attack A.3}, Eve does not try to learn the spectrum of the public unitaries (as in Attacks A.1 and A.2), but instead tries to recover the secret QSA output by guessing the private planted state and using it to reproduce the same dominant eigenphase features that honest parties obtain. Operationally, this is a \emph{state-guessing} attack: Eve proposes a candidate state $\ket{\Phi}$, runs the public feature-extraction pipeline against each public unitary $U_i$, and checks whether the resulting $m$-bit eigenphase string matches the key-dependent features used by the protocol.

\smallskip
\noindent\emph{Attack procedure (one trial).}
Eve samples a candidate $\ket{\Phi}$ and for each $i\in\{1,\dots,k\}$ computes the dominant eigenphase feature associated with $(U_i,\ket{\Phi})$ using the cheapest available evaluation method for the implementation:
(i) dense EVD / spectrum extraction for QSA-M,
(ii) autocorrelation / classical circuit evaluation for QSA-C,
or (iii) low-depth QPE-style phase extraction for QSA-Q.
She succeeds on a trial only if \emph{all} $k$ extracted $m$-bit eigenphases coincide with the honest parties' $k$ eigenphases; otherwise, the derived key is wrong and the trial provides essentially no reliable ``gradient'' indicating how to modify $\ket{\Phi}$. This lack of a useful local improvement signal is the core reason this attack reduces to Hilbert-space search rather than a progressive optimisation routine.

\smallskip
\noindent\emph{High-overlap versus non-overlap planting.}
The distinguishing feature is whether the implementation enforces a high-overlap planted eigenstate.

\begin{itemize}
\item \textbf{QSA-M / QSA-C (no enforced high overlap).}
For expressive near-Haar unitaries, a typical fixed state $\ket{\Psi}$ has overlaps $|\langle \Psi|U_i|\Psi\rangle|^2$ on the order of $2^{-n}$, so there is no large ``signal'' amplitude that would single out a distinguished eigenphase from the perspective of a wrong guess $\ket{\Phi}$.
Consequently, Eve's extracted eigenphases behave (to a good approximation) like independent uniform $m$-bit strings, and a trial only succeeds by matching the entire $mk$-bit feature string by chance.
Thus, Attack A.3 in QSA-M/C is essentially a blind search with success probability exponentially small in $mk$ (and in particular in $nk$ when $m=\Theta(n)$).

\item \textbf{QSA-Q (compiler-enforced high overlap).}
In QSA-Q, the compilation is engineered so that the planted state $\ket{\psi_i}$ is close to a signal eigenstate for each $U_i$, i.e.\ $|\langle \psi_i|U_i|\psi_i\rangle|^2 \ge 1-\delta$.
This shrinks the effective search region: Eve can only hope to reproduce the correct dominant eigenphase for $U_i$ if her guess $\ket{\phi_i}$ has sufficiently high fidelity $F = |\langle \psi_i|\phi_i\rangle|^2$ with the true planted state.
However, even in this favourable regime, Eve must get \emph{all} $k$ eigenphases correct simultaneously; if even one unitary produces the wrong dominant eigenphase, the trial fails and (because the $U_i$ are designed to be expressive and decorrelated) the failure does not reliably reveal which direction in Hilbert space moves $\ket{\Phi}$ closer to $\ket{\psi_i}$.
Moreover, in our QSA-Q instantiation, we deliberately enforce this condition across $k$ independently randomised unitaries and planted states, so Eve cannot make progress by matching only a subset of eigenphases.
\end{itemize}
\begin{figure}[t!]
  \centering
  \begin{subfigure}{0.5\textwidth}
    \centering
    \includegraphics[width=.85\linewidth]{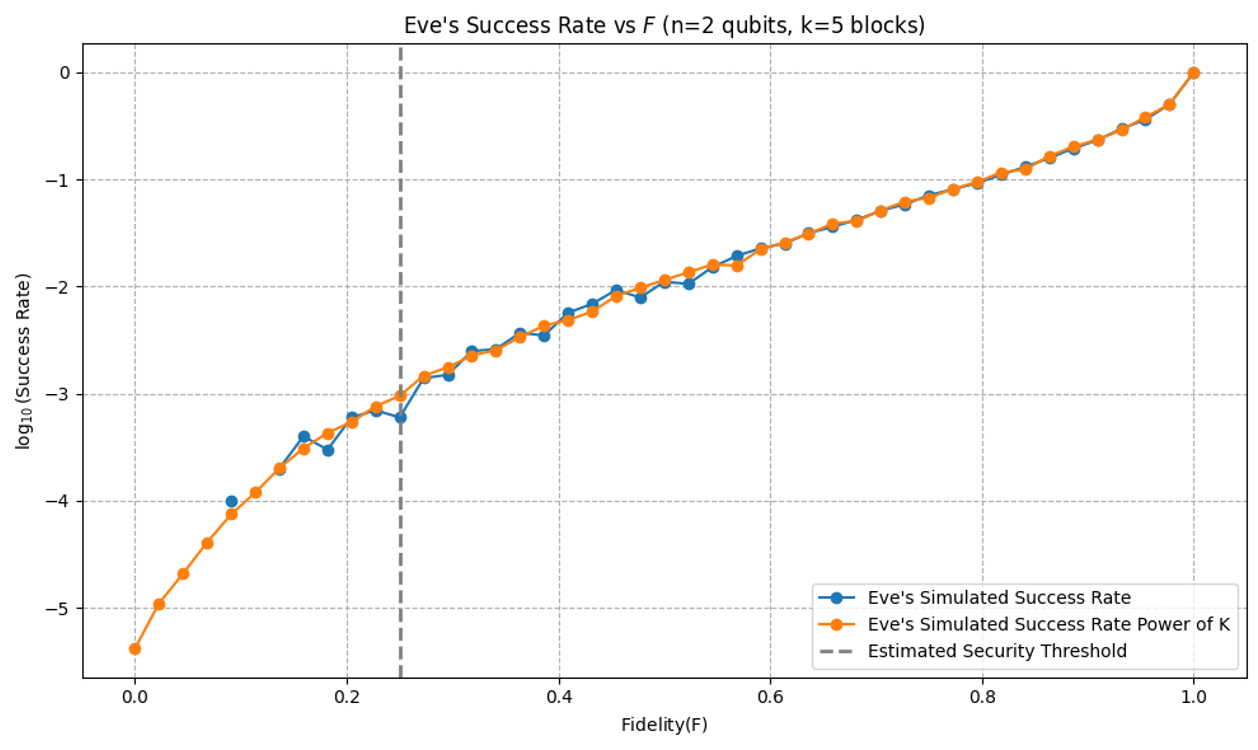}
    \caption{$P(F,m,k)$ and $p_\text{U}^k$ versus $F$ for $n=2$ qubits.}
    \label{prot}
  \end{subfigure}
  \begin{subfigure}{0.5\textwidth}
    \centering
    \includegraphics[width=0.85\linewidth]{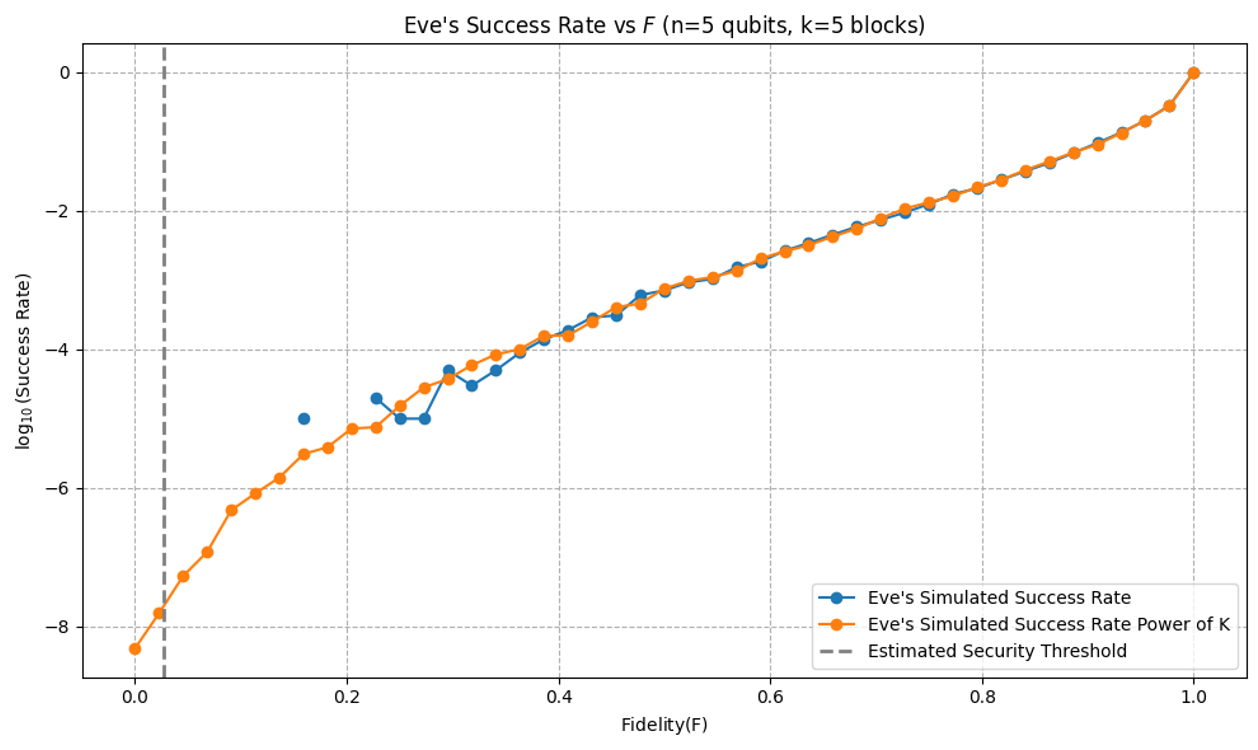}
    \caption{$P(F,m,k)$ and $p_\text{U}^k$ versus $F$ for $n=5$ qubits.}
    \label{fig:prob1_6_3}
  \end{subfigure}
  \caption{Eve's success rate vs fidelity $F$.}
\end{figure}

\begin{table*}[ht!]
\centering
\caption{Parameters $n$, $m$ and $k$ for $256$-bit level success probability and min-entropy.}
\begin{tabular}{|c|c|c|c|c|}
\hline
\textbf{Qubits} \( n \) & \textbf{Unitaries} \( k \) for $m=n$ & $\ell_k$ for $m=n$ & \textbf{Unitaries} \( k \) for $m=2$ & $\ell_k$ for $m=2$ \\
\hline
6  & 96  & 576 bits & 152  & 304 bits\\
7  & 48  & 336 bits & 130  & 260 bits\\
8  & 38  & 304 bits & 128  & 256 bits\\
9  & 31  & 279 bits & 128  & 256 bits\\
10 & 27  & 270 bits & 128  & 256 bits\\
11 & 24  & 264 bits & 128  & 256 bits\\
12 & 22  & 256 bits & 128  & 256 bits\\
15 & $\lceil{256/m}\rceil$ & 256 bits & 128  & 256 bits \\
\hline
\end{tabular}
\label{table2}
\end{table*}
\smallskip
\noindent\emph{Success probability model and min-entropy bound.}
Let the true planted state be $\ket{\Psi}$ in dimension $d=2^n$, and let Eve's guess $\ket{\Phi}$ have fidelity $F = |\langle \Psi|\Phi \rangle|^2$. The trace distance satisfies $D_{\rm trace}(\Psi,\Phi)=\sqrt{1-F}$, which upper bounds the distinguishing advantage between their measurement outcome distributions under any basis, including the eigenbasis of a random $U$.

Let $p_U(m,F)$ denote Eve's probability of outputting the correct $m$-bit dominant eigenphase for a \emph{single} unitary given fidelity $F$ (this is determined by the feature-extraction rule and is easily estimated by Monte Carlo). Under the decorrelation assumption implicit in using expressive independently randomised $U_i$, the probability of matching the entire $k$-unitary eigenphase sequence at fidelity $F$ is well-approximated by
\[
P(F,m,k)\;\approx\;p_U(m,F)^k,
\]
which we verify empirically in Fig.~\ref{prot} and~\ref{fig:prob1_6_3} (Appendix~\ref{hilbertspace}).

Finally, if Eve's guesses are sampled uniformly at random (Haar measure), then $F$ follows the Beta distribution $\mathrm{Beta}(1,d-1)$ with density $(d-1)(1-F)^{d-2}$ (see \cite{averagefid, MKus_1988}).
Therefore, the overall per-trial success probability of Attack A.3 can be upper bounded by
\[
p_{\rm succ}
\;=\;
\int_{F=0}^{1}
(d-1)(1-F)^{d-2}\,p_U(m,F)^k\,dF,
\]
and the corresponding guessing min-entropy is $H_\infty \approx -\log_2 p_{\rm succ}$.
Appendix~\ref{hilbertspace} evaluates this integral numerically (using simulated $p_U(m,F)$) and gives concrete $(n,k)$ choices achieving $\ge 256$-bit guessing security (Table~\ref{table2}), which only matter for low-depth $n\le 11$ where state guessing is an easy attack.

\smallskip
Attack A.3 is not ``recover the planted state by optimisation''; it is a \emph{verification-limited state-guessing attack}.
In QSA-M/C, it reduces to effectively uniform guessing over an exponentially large feature space; in QSA-Q, the enforced overlap imposes a fidelity threshold, but the requirement to match \emph{all} $k$ eigenphases (and our design choice to decorrelate signal eigenvectors across unitaries) keeps the overall success probability exponentially small in $nk$ (or $mk$), yielding the min-entropy bounds reported in this appendix. Beyond $n\ge 10 \gg m$, this attack is typically less likely to succeed than chained QPE unless the planting ansatz is structurally weak.

\label{hilbertspace}

To evaluate the guessing bound numerically it suffices to simulate the per-unitary success $p_U(m,F)$. In subfigures~\ref{prot} and~\ref{fig:prob1_6_3} we plot $P(F,m,k)$ and $p_U(m,F)^k$ versus fidelity for $N=100{,}000$ trials at each fidelity; the two agree, confirming the approximation $P(F,m,k)\approx p_U(m,F)^k$ and making the integral $p_{\rm succ}$ tractable. In Table~\ref{table2} we evaluate it and report the minimum key length needed for $256$-bit min-entropy against state guessing when $n=m$ and Eve guesses over $F\in[0,1]$.

\label{circuits}

% Preamble (once):
% \usepackage{graphicx}
% \usepackage{subcaption}
\section{Multi-party broadcast unitary challenges (one-to-many)}
\label{subsec:QSAq_compilers_multi_party}

The symmetric compiler can be extended to a broadcast (one-to-many) setting in which a single published unitary instance $U$ simultaneously embeds \emph{multiple} hidden signal eigenvectors, one for each party $P\in\{\mathrm{B},\mathrm{C},\mathrm{D}\}$. Concretely, each party has its own planted state $\ket{\psi_P}$ (derived from a private planting circuit or seed shared with the verifier) and its own hidden computational basis label $\ket{b_P}$, while the compiler learns a \emph{single} expressive map $V(\vec{\alpha})$ that aligns all targets at once. One convenient choice is the aggregate loss
\[
\mathcal{L}(\vec{\alpha}) \;=\; (1-F_{\mathrm{B}}(\vec{\alpha})) + (1-F_{\mathrm{C}}(\vec{\alpha})) + (1-F_{\mathrm{D}}(\vec{\alpha})),
\qquad
F_P(\vec{\alpha})=\bigl|\langle \psi_P\,|\,V(\vec{\alpha})\,|\,b_P\rangle \bigr|^2,
\]
so that the resulting $V$ yields high overlap between each $\ket{\psi_P}$ and its corresponding hidden eigenvector $V\ket{b_P}$ within the shared public $U=VDV^\dagger$ (see Methods for full details and discussion of optimisation feasibility as the fan-out increases).

\begin{figure}[t!]
  \centering
  \includegraphics[width=0.80\linewidth]{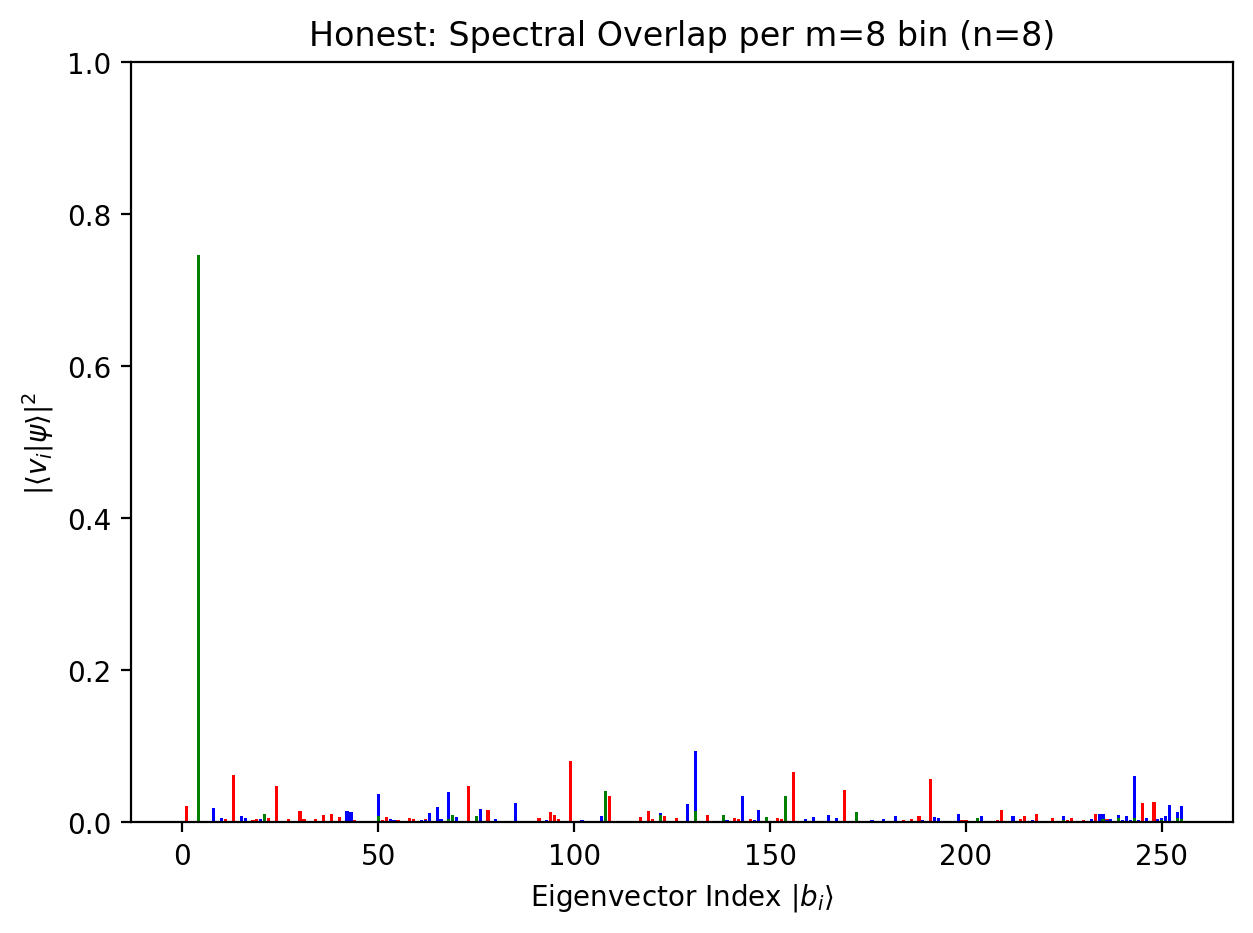}
  \caption{\textbf{One-to-three-party broadcast symmetric challenge ($U=VDV^\dagger$) for $n=m=8$.} 
  Three hidden basis labels $\ket{b_{\mathrm{B}}}  (\text{green}),\ket{b_{\mathrm{C}}} (\text{blue}),\ket{b_{\mathrm{D}}} (\text{red})$ are chosen for three planted states $\ket{\psi_{\mathrm{B}}},\ket{\psi_{\mathrm{C}}},\ket{\psi_{\mathrm{D}}}$, and a single $V(\vec{\alpha})$ is learned by minimising the aggregate loss $(1-F_{\mathrm{B}})+(1-F_{\mathrm{C}})+(1-F_{\mathrm{D}})$ (Methods).
  The plot shows per-party overlap weights $|\langle v_i|\psi_P\rangle|^2$ over the eigenvectors $\{\ket{v_i}\}$ of the same compiled $U$, with a different colour per party.}
  \label{fig:multiparty-overlap-placeholder}
\end{figure}

Figure~\ref{fig:multiparty-overlap-placeholder} provides a visualisation of a one-to-three-party compiled instance. The intended plot shows, for each party $P\in\{\mathrm{B},\mathrm{C},\mathrm{D}\}$, the overlap distribution $|\langle v_i|\psi_P\rangle|^2$ over eigenvectors $\{\ket{v_i}\}$ of the same broadcast unitary $U$. In a successful compilation, each party’s planted state concentrates its overlap mass on a distinct hidden signal eigenvector (or a narrow set of eigenvectors), with different parties peaking at different eigen-indices. This provides an operational picture of how a single public challenge can authenticate multiple provers under independent planted states, while keeping the signal eigenvector identities hidden behind the private $(P_P,b_P)$ pairs.

\paragraph*{{\bf Attack A.3(b) (ansatz-aware state guessing without copies)}.}
A sharper variant of Attack A.3 arises if Eve is given (or can accurately infer) the circuit \emph{template} used to generate the planted state $|\psi\rangle=P^\dagger|0^n\rangle$, even though she does \emph{not} obtain copies of $|\psi\rangle$ (so tomography and variational state learning are unavailable).
In this setting, the relevant question is not enumeration over the full Hilbert space but over the \emph{parameter manifold} induced by the chosen ansatz: if $P^\dagger(\theta)$ contains only single-qubit rotations (or is otherwise structurally restricted), then $|\psi\rangle$ is confined to a low-dimensional family (e.g., product states), which can enable structural shortcuts for predicting moments $\langle\psi|U^t|\psi\rangle$ or for mounting targeted searches over candidate states consistent with the ansatz.
Concretely, for an ansatz with $p$ continuous parameters and target angular precision $\varepsilon$, a naive $\varepsilon$-net cover has size on the order of $(2\pi/\varepsilon)^p$, so keeping $p$ small (or using commuting/diagonal-only layers) drastically shrinks Eve's effective search space compared to generic $n$-qubit states.

We therefore impose an \emph{expressivity requirement} on the planting circuit $P^\dagger$: it must include noncommuting single-qubit rotations and a nontrivial density of entangling gates on a connected interaction graph (e.g., alternating nearest-neighbour entanglers interleaved with full SU(2) layers), so that the induced state family is not confined to separable or efficiently simulable subclasses.
Under this requirement, and without copies of $|\psi\rangle$, Attack A.3(b) does not give Eve a practical advantage over the spectrum-based attacks analysed above; in particular, any remaining parameter-manifold search still requires an exponentially large effort in either precision or circuit depth to reach the fidelity needed to predict the LDQPE-derived phase features reliably.

% ============================================================
% Your existing detailed Hilbert-space analysis section (with
% figures + table) can remain as a dedicated appendix section.
% ============================================================

% =====================================================================
%  APPENDICES E and F for the QSA revision.
%  Insert after the existing appendices, inside the \appendix environment
%  (revtex4-1 letters them automatically; the \label commands are what the
%  main text \ref's resolve to).
%
%  Requires \usepackage{booktabs} (already in your preamble).
% =====================================================================

\section{Empirical tests of label hiding}
\label{app:lh-tests}

This appendix reports the tests that support Assumption~\ref{ass:lh} (label
hiding) for the symmetric compiler of Methods. The assumption is that the public
challenge $U_i=V_iD_iV_i^\dagger$ does not reveal the hidden label $b_i$ beyond the
$2^{-n}$ blind baseline. Because the spectrum is proved uniform unconditionally
(Lemma~\ref{lem:unif}) and the eigenbasis is useless to an adversary without a copy
of $\ket{\psi_i}$, the label can only leak through one of two channels: the
\emph{identity of the planted eigenvector} among the $2^n$ published eigenvectors,
or the \emph{published circuit parameters} $\vec\alpha^\star_i$ that the optimiser
returned while it had access to $b_i$. We test both. These are cryptanalytic
checks in the sense of the main text: a null result is evidence for
Assumption~\ref{ass:lh}, not a proof of it.

Throughout, planted states are ensemble-matched as required by the admissibility
condition of Methods: $\ket{\psi_i}=V(\vec\alpha_i^0)\ket{0^n}$ with
$\vec\alpha_i^0$ drawn from the same ansatz used for compilation. The ansatz is the
one used in the main text: $L$ layers of a universal single-qubit rotation
$R_z R_x R_z$ per qubit followed by a nearest-neighbour CZ brickwork. Compilation
uses the same variational objective $\max_{\vec\alpha}|\langle\psi_i|V(\vec\alpha)
\ket{b_i}|^2$ as Methods.

\subsection{Structural test: is the planted eigenvector distinguishable?}
\label{app:lh-structural}

The variational fit drives $V_i\ket{b_i}$ to $\ket{\psi_i}$, so $b_i$ is the one
index whose eigenvector is controlled; the remaining $\{V_i\ket{x}\}_{x\neq b_i}$
are whatever the ansatz returns. If $\ket{\psi_i}$ were atypical for the ansatz,
$b_i$ would be the anomalous index and an adversary would recover it by inspection.
For each published $U_i$ we therefore compute five scalar statistics of every
eigenvector $V_i\ket{x}$: bipartite (half-cut) entanglement entropy,
single-qubit entanglement entropy, inverse participation ratio, maximum
computational-basis amplitude, and Shannon entropy of the computational-basis
distribution. We record the \emph{rank} of the true index $b_i$ under each,
taken in both directions (highest and lowest), giving ten ranking statistics. Under
Assumption~\ref{ass:lh} each ranks $b_i$ uniformly on $\{0,\dots,2^n-1\}$, so the
top-rank frequency is $2^{-n}$ and the mean rank is $(2^n-1)/2$.

Table~\ref{tab:lh-null} reports the result for ensemble-matched planted states.
Across all statistics and all $n$ tested, the observed mean rank of $b_i$ matches
the uniform prediction, and the smallest $p$-value over the ten statistics (against
a one-sided binomial null, Bonferroni threshold $0.05/10=0.005$) is far above
threshold. The planted index is not distinguishable from the other $2^n-1$
eigenvectors by any statistic in the battery.

\begin{table}[h]
\centering
\caption{Structural test under ensemble matching (null expected). ``Mean rank''
is averaged over the ten ranking statistics; the uniform prediction is
$(2^n-1)/2$. ``$\min p$'' is the smallest $p$-value over the ten statistics; the
Bonferroni threshold is $0.005$. ``$\max$ acc.'' is the largest held-out per-bit
accuracy of the parameter distinguisher of App.~\ref{app:lh-param} (baseline
$0.5$). All runs use compilation overlap $\ge0.99$.}
\label{tab:lh-null}
\begin{tabular}{cccccc}
\toprule
$n$ & $N$ & overlap & mean rank (pred.) & $\min p$ & $\max$ acc. \\
\midrule
4 & 400 & 1.000 & 7.5\ \ (7.5)   & 0.13 & 0.62 \\
5 & 250 & 0.997 & 15.5\ (15.5)    & 0.11 & 0.70 \\
6 & 300 & 0.991 & 31.5\ (31.5)    & 0.33 & 0.57 \\
\bottomrule
\end{tabular}
\end{table}

As a positive control we repeat the structural test with a deliberately
\emph{atypical} planted state (a shallow product-like state that the ansatz would
not generically produce). The test then locates $b_i$ decisively: at $n=6$,
$N=150$, the half-cut entanglement statistic places $b_i$ in the top rank for
$77/150$ instances with mean rank $3.6$ of $63$, a $p$-value below $10^{-15}$. The
attack is therefore real, and the null of Table~\ref{tab:lh-null} reflects the
admissibility condition doing its job, not a test without power. This is the
empirical content of the statement in Methods that ensemble matching is
\emph{necessary and sufficient against the structural attack}.

\subsection{Parameter test: does $\vec\alpha^\star$ leak the label?}
\label{app:lh-param}

Even with the eigenvectors indistinguishable, the label could survive in the
published parameter vector, since the optimiser saw $b_i$ while converging. We test
this directly. For each $(n,L)$ we compile $N$ instances, retain
$(\vec\alpha^\star_i,b_i)$, and train a supervised classifier $\hat b=f(\vec
\alpha^\star)$ to predict each label bit, using an $80/20$ train/test split and two
classifier families (regularised logistic regression and gradient-boosted trees).
The angles are represented as $(\cos\vec\alpha^\star,\sin\vec\alpha^\star)$ so that
the $2\pi$-periodicity is respected. Under Assumption~\ref{ass:lh} the held-out
per-bit accuracy is $1/2$ up to sampling error; a reproducible excess would be a
distinguisher and would falsify Eq.~\eqref{eq:lh} for this ansatz. The largest
per-bit accuracy observed is reported in the final column of
Table~\ref{tab:lh-null}; in every case it lies within a few standard errors of the
baseline (one standard error is $\approx0.06$ to $0.11$ at these held-out set sizes),
consistent with no leakage.

\subsection{Scope}
\label{app:lh-scope}

These tests are at $n\le6$, where the eigenvector statistics can be computed by
exact enumeration of all $2^n$ indices. The structural test's top-rank statistic
requires $N\gg2^n$ to have power, so the tables above should be read as excluding
gross leakage rather than as bounding a small bias; the parameter test is limited
by the held-out set size in the same way. Both tests are provided as a reference
implementation \cite{kish_2026_21450694} and scale to larger $n$ on hardware faster
than the enumeration environment used here; we recommend $N\gtrsim10^4$ for
production parameter sweeps. The state-mode construction of Remark~\ref{rem:lh}
renders both tests vacuous, since there $\varepsilon_{\mathrm{LH}}=0$ holds by
construction and no fit is performed.

\section{Exact bucket non-uniformity at \texorpdfstring{$n=m$}{n=m}}
\label{app:kappa}

Lemma~\ref{lem:unif}(iii) proves feature uniformity in the regime $n\gtrsim3m$.
The instances demonstrated in this paper use $n=m$, which violates that hypothesis,
so the lemma does not apply to them and we verify uniformity there directly rather
than claiming it. For $n=m$ the eigenphase of each label is the exact signed sum
$\theta(b)=\tfrac12\sum_{q=1}^n(2b_q-1)\beta_q\bmod2\pi$, so the full distribution
over the $2^n$ labels can be enumerated with no sampling. For each of $300$ random
public angle vectors $\{\beta_q\}$ we compute the exact worst-case bucket
probability and report
\begin{equation}
\kappa\ :=\ 2^m\max_c\Pr[\mathsf{B}=c],
\end{equation}
the factor by which the most probable bucket exceeds the uniform value $2^{-m}$.
A value $\kappa=1$ is perfect uniformity; $\log_2\kappa$ is the number of min-entropy
bits lost relative to the ideal $m$ bits per instance.

Table~\ref{tab:kappa} reports the enumeration for $n=m$ up to $18$.

\begin{table}[h]
\centering
\caption{Exact bucket non-uniformity $\kappa=2^m\max_c\Pr[\mathsf{B}=c]$ at
$n=m$, over $300$ random public angle vectors per row. $\log_2\kappa$ is the
min-entropy loss per instance in bits.}
\label{tab:kappa}
\begin{tabular}{cccccc}
\toprule
$n=m$ & mean $\kappa$ & std & max $\kappa$ & $\log_2(\text{mean }\kappa)$ \\
\midrule
2  & 1.49 & 0.50 & 2.0  & 0.57 \\
4  & 2.48 & 0.79 & 7.0  & 1.31 \\
6  & 3.18 & 0.94 & 9.0  & 1.67 \\
8  & 3.86 & 0.85 & 8.0  & 1.95 \\
10 & 4.64 & 0.91 & 8.0  & 2.21 \\
12 & 5.38 & 0.97 & 9.0  & 2.43 \\
14 & 6.07 & 0.95 & 11.0 & 2.60 \\
16 & 6.76 & 0.85 & 10.0 & 2.76 \\
18 & 7.41 & 0.95 & 12.0 & 2.89 \\
\bottomrule
\end{tabular}
\end{table}

The growth is linear in $n$, $\kappa\approx0.36\,n+1.1$ for $n\ge4$, equivalently
$\kappa\sim n^{0.74}$ on a log fit; the min-entropy loss $\log_2\kappa$ grows from
about one bit at $n=m=4$ to under three bits at $n=m=18$. We note that this is
slower than balls-into-bins intuition would suggest only in that the buckets are
correlated ($\theta(b)$ is a signed sum of the \emph{same} $n$ angles across all
labels), but the practical conclusion is the one that matters: at $n=m$ the
feature loses a few bits of min-entropy per instance rather than collapsing, so the
$k$-instance response retains close to $k(m-\log_2\kappa)$ bits. This is the
numerical counterpart, in the demonstrated regime, of the asymptotic guarantee
$H_\infty\ge m-1$ that Lemma~\ref{lem:unif}(iii) provides for $n\gtrsim3m$. The
enumeration is deterministic and is included in the reference implementation \cite{kish_2026_21450694}.

\section{Feature uniformity in the \texorpdfstring{$n\gtrsim3m$}{n>=3m} regime}
\label{app:uniformity-highn}

Appendix~\ref{app:kappa} enumerates the bucket non-uniformity $\kappa$ at $n=m$,
the regime of the demonstrated instances, where Lemma~\ref{lem:unif}(iii) does not
apply. Here we complement that with the regime where the lemma \emph{does} apply,
$n\gtrsim3m$, to confirm that the proven guarantee and the numerics agree. For
$n>m$ the $2^n$ labels can no longer be enumerated exactly, so we estimate
$\kappa=2^m\max_c\Pr[\mathsf{B}=c]$ by sampling $1.2\times10^5$ uniform labels per
angle vector, over $150$ random public angle vectors $\{\beta_q\}$ per row.

\begin{table}[h]
\centering
\caption{Bucket non-uniformity $\kappa=2^m\max_c\Pr[\mathsf{B}=c]$ in the
lemma regime $n\gtrsim3m$, estimated by label sampling. $\log_2\kappa$ is the
min-entropy loss per instance in bits. Compare Appendix~\ref{app:kappa}, where
$n=m$ gives $\kappa$ growing to $\approx7$ at $n=m=18$.}
\label{tab:kappa-highn}
\begin{tabular}{ccccc}
\toprule
$m$ & $n$ & $n/m$ & mean $\kappa$ & $\log_2(\text{mean }\kappa)$ \\
\midrule
2 & 6  & 3 & 1.094 & 0.129 \\
2 & 8  & 4 & 1.055 & 0.077 \\
3 & 9  & 3 & 1.116 & 0.158 \\
3 & 12 & 4 & 1.035 & 0.050 \\
4 & 12 & 3 & 1.076 & 0.106 \\
4 & 16 & 4 & 1.026 & 0.037 \\
\bottomrule
\end{tabular}
\end{table}

Where the lemma applies, the buckets are essentially uniform: at $n=3m$ the
min-entropy loss is already below $0.16$ bit per instance, and at $n=4m$ below
$0.08$ bit, tightening as $n/m$ grows. This is the numerical face of
Lemma~\ref{lem:unif}(iii), and it contrasts sharply with the $n=m$ regime of
Appendix~\ref{app:kappa}, where $\kappa$ grows to about $7$: the loss is a few bits
only when the eigenphase read-out precision $m$ is comparable to the state size
$n$, and vanishes once $n$ is a few times $m$. Sweeping the ratio at fixed $m=3$
makes the crossover explicit, with $\kappa\approx2.08,\,1.32,\,1.10,\,1.04$ at
$n/m=1,2,3,4$, and the decay of $\log_2(\kappa-1)$ tracks the lemma's
$-n+O(\sqrt{nm})$ envelope through the crossover region, flattening at larger
$n/m$ only against the finite-sample floor. Practically: the feature loses a few
bits of min-entropy at $n=m$ and is within a small fraction of a bit of ideal once
$n\gtrsim3m$, so a deployment that wants the full $m$ bits per instance simply
takes $n$ a few times larger than $m$.
\section{Circuits}
\begin{figure}[ht!]
  \centering

  \begin{subfigure}[t]{0.3\linewidth}
    \centering
    \includegraphics[width=\linewidth]{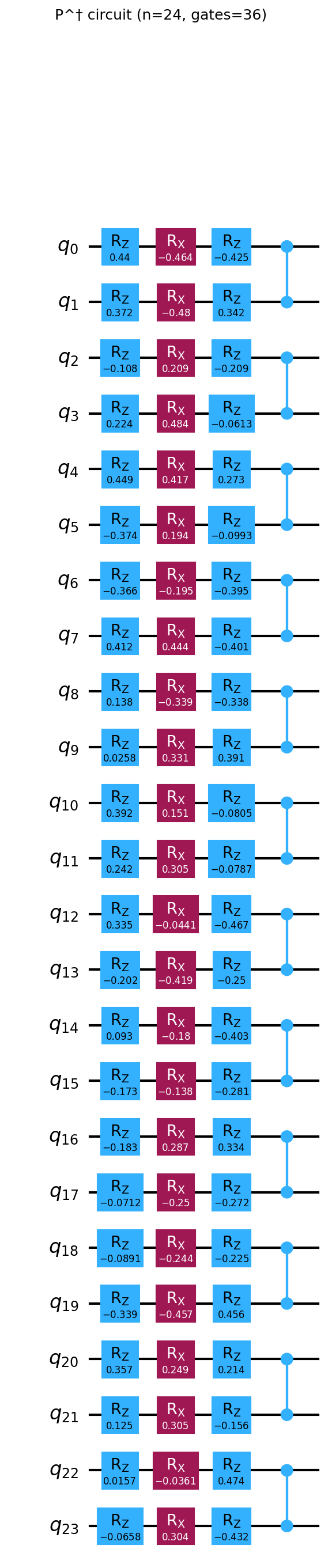}
    \subcaption{$P^{\dagger}$ circuit for a randomly generated $n=24$-qubit state.}
    \label{fig:pdagger-n24}
  \end{subfigure}\hfill
  \begin{subfigure}[t]{0.55\linewidth}
    \centering
    \includegraphics[width=\linewidth]{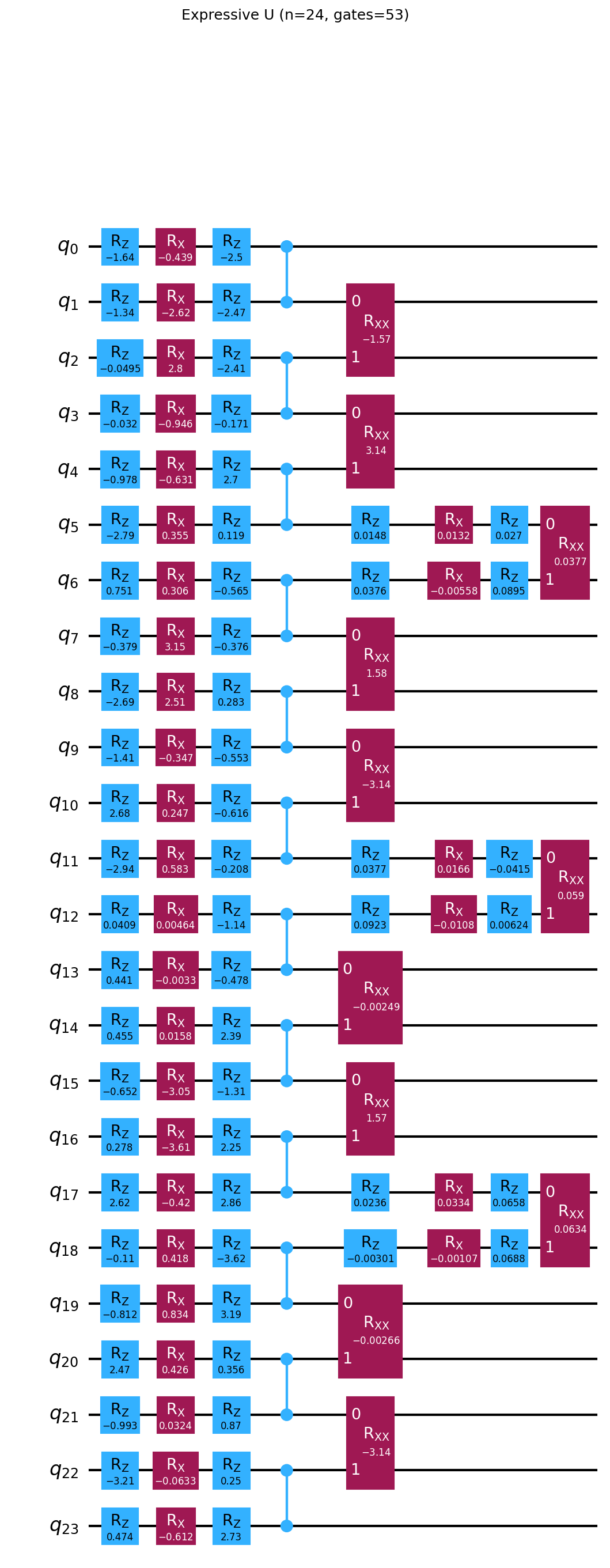}
    \subcaption{Warm-start–generated fully entangling unitary $U$ designed to maintain high overlap with the target state ($n=24$).}
    \label{fig:u-warmstart-n24}
  \end{subfigure}
\caption{}
\end{figure}

\begin{figure}[ht!]
  \centering

  \begin{subfigure}[t]{0.5\linewidth}
    \centering
    \includegraphics[width=\linewidth]{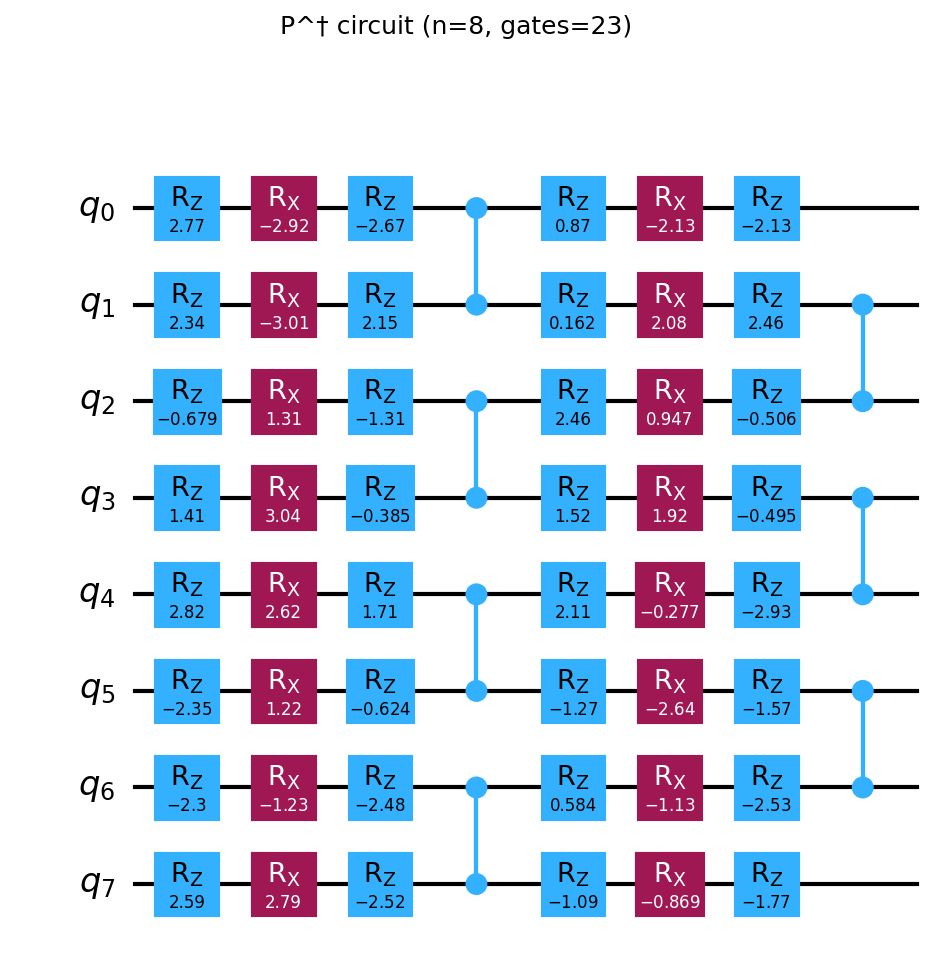}
    \subcaption{$P^{\dagger}$ circuit for a randomly generated $n=8$-qubit state.}
    \label{fig:pdagger-n8}
  \end{subfigure}\hfill
  \begin{subfigure}[t]{0.95\linewidth}
    \centering
    \includegraphics[width=\linewidth]{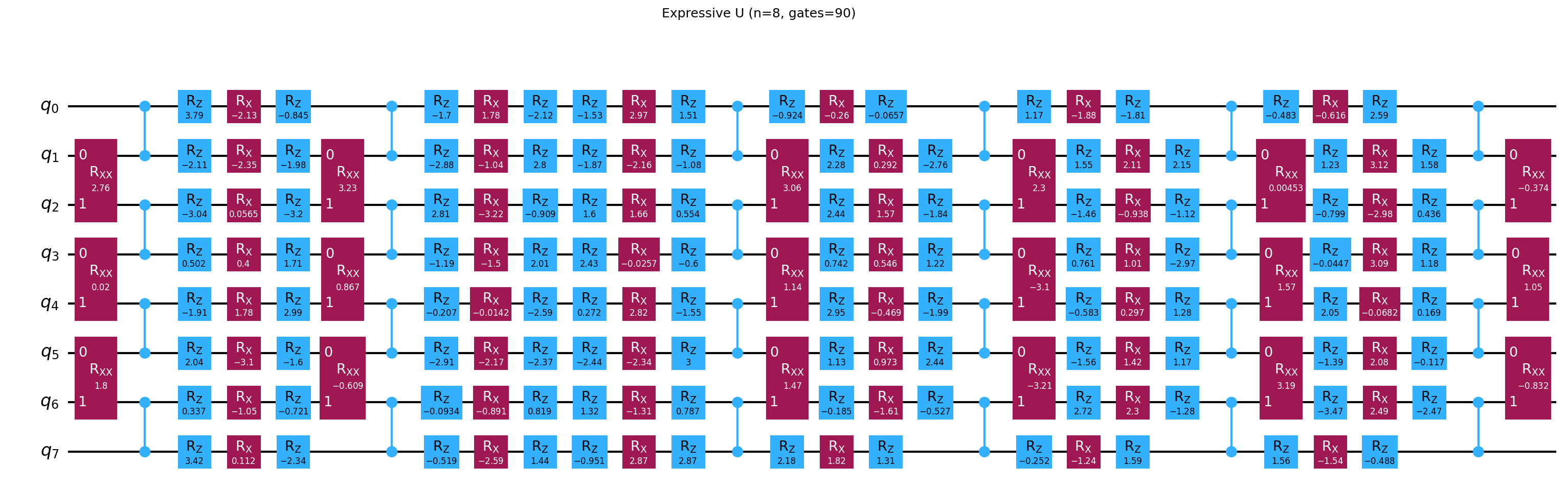}
    \subcaption{Asymmetric-compiler generated fully entangling unitary $U$ designed to maintain high overlap with the target state ($n=8$).}
    \label{fig:u-symmetric-n8}
  \end{subfigure}
\caption{}
\label{fig:n8}
\end{figure}

The circuits shown in this section are illustrative examples of the constructions used in our methods rather than exhaustive templates. Figures~\ref{fig:pdagger-n24} and~\ref{fig:u-warmstart-n24} show, respectively, an example \(P^{\dagger}\) inverse state-preparation circuit for a randomly generated \(n=24\)-qubit state and a warm-start generated fully entangling unitary \(U\) designed to preserve high overlap with the target state. Likewise, Figures~\ref{fig:pdagger-n8} and~\ref{fig:u-symmetric-n8} present the corresponding example circuits for the \(n=8\) case, including the asymmetric-compiler generated fully entangling unitary. These figures are intended to provide representative circuit instances for the relevant constructions discussed in the main text and appendices, and to give a sense of their structure and relative complexity.
\label{circuits}

\label{app:asymmetric}
\section{Prototype performance for small-scale QSA-C} 
\label{subsec:prototype-performance}

To sanity–check feasibility and give a rough sense of concrete costs, we implemented QSA-C in Python on a commodity laptop (single–threaded, no low-level optimisation). The goal here is \emph{not} to compete with conventional KDFs such as HKDF—which are orders of magnitude faster—but to illustrate that the honest evaluation costs of QSA are compatible with low–rate, high–value key-refresh usage.

\section{Complexity-theoretic perspective}
\label{app:complexity-perspective}

The security of QSA is based on a planted state unpredictability assumption: given public descriptions of $\{U_i\}_{i=1}^k$ (and any stated metadata), an adversary should be unable to reproduce the honest eigenphase feature vector $\boldsymbol{\Theta}$, nor output any state that enables doing so with non-negligible advantage over guessing. In this section, we do \emph{not} claim a tight worst-case reduction (e.g.\ from Local Hamiltonian). Instead, we motivate the assumption via an \emph{identifiability} and \emph{information-bottleneck} viewpoint: QSA exposes an extremely lossy, state-dependent functional of a high-dimensional planted state, closely aligned with a ``blind tomography'' inversion problem and with established limits on learning unknown quantum states from restricted information~\cite{aaronson2018shadow,huang2020predicting}.

Subsequently, we argue that the adversary sees a highly compressive, state-dependent map.
Fix a public unitary $U$ with eigenpairs $\{(e^{i\theta_j},\ket{u_j})\}_{j=1}^{2^n}$. For a secret state $\ket{\psi}$, any phase-extraction procedure used by the honest parties (LDQPE or the classical autocorrelation estimator) depends on the induced spectral weights
$
c_j := \langle u_j|\psi\rangle
$
through a small number of low-order moments
\begin{equation}
Z_t(U,\psi) \;=\; \bra{\psi}U^t\ket{\psi}
\;=\; \sum_{j} |c_j|^2 e^{it\theta_j},
\qquad t\in\mathcal{T},
\label{eq:moments}
\end{equation}
followed by a \emph{coarse} decoding map that outputs only $m$ bits (a phase bucket) or a low-precision phase estimate. Thus, even in the idealised noiseless setting, the public transcript (and even a leaked $\boldsymbol{\Theta}$) reveals only $O(km)$ bits about a hidden $2^n$-dimensional object. The mapping $\ket{\psi}\mapsto \boldsymbol{\Theta}$ is therefore generically \emph{many-to-one}, and is far from informationally complete in the sense required for state reconstruction as in standard tomography.

The following ``phase-leakage'' thought experiment formalises the blind-tomography intuition.
To isolate the core difficulty, consider a strong leakage experiment in which Eve is additionally given a \emph{correct} honest feature vector $\boldsymbol{\Theta}$ (or even the underlying dominant phases $\{\theta_i^\star\}$) for each public $U_i$. Eve may then attempt to invert the map by running QPE on $U_i$ to prepare an eigenstate consistent with $\theta_i^\star$, hoping to recover (or correlate with) the hidden planted state(s). Even granting this leakage, inversion remains implausible in the regimes relevant to QSA:

\begin{itemize}
\item \textbf{Near-Haar / mixing regimes (QSA-M/C).}
When each $U_i$ is drawn from an expressive ensemble and is effectively independent of $\ket{\psi}$, the overlap profile $\{|c_j|^2\}$ is delocalised. Conditioning on a particular eigenphase (even the correct dominant one) provides essentially no identifying information about $\ket{\psi}$: for Haar-like eigenbases, the typical squared overlap between a fixed $\ket{\psi}$ and an eigenstate $\ket{u(\theta)}$ is on the order of $2^{-n}$. Consequently, even if Eve can prepare an eigenstate matching $\theta_i^\star$, this eigenstate is, in general, \emph{nearly orthogonal} to $\ket{\psi}$ and behaves like a random direction relative to the hidden basis. In this sense, ``learning $\ket{\psi}$ from leaked phases'' becomes a form of blind inversion that is strictly weaker than having copy access to $\ket{\psi}$.

\item \textbf{Planted compiled regimes (QSA-Q).}
In QSA-Q, each $U_i$ is compiled so that a planted state $\ket{\psi_i}=P_i^\dagger\ket{0^n}$ has a robust dominant-eigenphase signal, enabling reliable low-depth extraction. Even if Eve is given $\theta_i^\star$ and can prepare a matching eigenstate $\ket{u_i^\star}$ of $U_i$, this does not accumulate across $i$ because the planted state changes per instance (via $P_i$ derived from private seed material). Thus, under phase leakage, Eve faces $k$ essentially independent inversion problems rather than a single planted state that can be progressively refined. This is precisely the regime in which ``chaining'' information across public unitaries is designed to fail. 
\end{itemize}

This viewpoint does not change the planted state unpredictability assumption or the key-indistinguishability game used in our security definition. Rather, it explains \emph{why} the assumption is plausible and what kinds of leakage it already contemplates: even unusually strong leakage (e.g.\ revealing the dominant eigenphases themselves) does not obviously furnish an efficient path to reconstruct the hidden planted state(s) or to reproduce $\boldsymbol{\Theta}$ for fresh instances. For a broader worst-case context, note that eigenstate- and witness-search tasks are closely related to QMA-/UniqueQMA-style ground-state search problems~\cite{kempe2004localH,bookatzQMA,aharonov2008unique,anshu2024uniqueqma}. QSA, however, is an average-case planted instance tied to specific ensembles/compilers, and we therefore treat its hardness as an explicit assumption supported by the concrete attack analyses in the main text.

\section{Classically compiled blockwise warm-start compilation of high-overlap public unitaries for QSA-Q}
\label{sec:methods_mps_QSA}

QSA needs, at each unitary, a public circuit $U$ that \emph{looks} generic from its gate list yet is \emph{easy} for honest parties who share a planted state. Let $P$ be any (preferably deep) private unitary, and define the planted state
\[
  \ket{\psi}=P^\dagger\ket{0^n}.
\]

Operationally, we do not require $|\psi\rangle$ to be an exact eigenstate of $U_i$.
Instead, we compile $U_i$ so that a single eigenstate $|\phi_\star\rangle$ of $U_i$
carries almost all of the weight of $|\psi\rangle$ in the eigenbasis, i.e.
\[
  |\psi\rangle = \sum_j \alpha_j |\phi_j\rangle,\qquad
  |\alpha_\star|^2 \ge 1-\delta,\quad
  \sum_{j\neq \star} |\alpha_j|^2 \le \delta,
\]
with a design parameter $\delta \ll 1$ (e.g.\ $\delta\approx 0.05$ in our prototypes).
Ideal phase estimation on $|\psi\rangle$ then returns the ``intended'' eigenphase
$\theta_\star$ with probability at least $1-\delta$ and some other eigenphase with
probability at most $\delta$. 

In the present implementation, we construct $U$ via a \emph{block-wise warm start} followed by a shallow inter-block entangler. For concreteness, our reference example uses $n=12$ qubits, partitioned into two blocks of six qubits each,
\[
\mathcal{H} \cong \mathcal{H}_A \otimes \mathcal{H}_B,
\qquad
\ket{\psi} = \ket{\psi_A}\otimes\ket{\psi_B},
\]
with $\ket{\psi_A}$ and $\ket{\psi_B}$ determined by $P$. On each block, we define a hardware-efficient brickwork ansatz with single-qubit rotations and nearest-neighbour entanglers. Concretely, a depth-1 layer on a block consists of
\begin{itemize}
  \item single-qubit rotations $R_z(\theta^z_{i})R_x(\theta^x_{i})$ on each qubit $i$, and
  \item a pattern of two-qubit entangling gates drawn from
  \(
  \{\,\mathrm{CZ},\, R_{xx}(\theta^{xx}_{(i,j)})\,\}
  \)
  on nearest-neighbour pairs $(i,j)$ within the block.
\end{itemize}
We then parameterise a block unitary $U_A(\vartheta_A)$ on $\mathcal{H}_A$ and $U_B(\vartheta_B)$ on $\mathcal{H}_B$ as the product of one or a few such layers. The first compilation stage independently maximises the overlaps
\[
  f_A(\vartheta_A) = \bigl|\!\langle \psi_A\,|\,U_A(\vartheta_A)\,|\,\psi_A \rangle \bigr|^2,
  \qquad
  f_B(\vartheta_B) = \bigl|\!\langle \psi_B\,|\,U_B(\vartheta_B)\,|\,\psi_B\rangle\bigr|^2,
\]
using a stochastic SPSA loop. Because each block is shallow and low-dimensional, these optimisations converge quickly and can be run in parallel. We stop at the first iterates $(\hat\vartheta_A,\hat\vartheta_B)$ such that $f_A,f_B \ge 1-\delta_{\mathrm{block}}$ for a chosen block-level threshold $\delta_{\mathrm{block}}$.

In the second stage, we freeze $U_A(\hat\vartheta_A)$ and $U_B(\hat\vartheta_B)$ and introduce a small number of expressive inter-block entanglers acting across the cut between the two blocks. The global unitary takes the form
\[
  U(\Theta) \;=\; U_{\mathrm{inter}}(\Theta_{\mathrm{inter}})\,
                 \bigl(U_A(\hat\vartheta_A)\otimes U_B(\hat\vartheta_B)\bigr),
\]
where $U_{\mathrm{inter}}$ is built from one or two layers of gates of the form
\[
  R_z(\phi^z_{c}) R_x(\phi^x_{c}) R_{xx}(\phi^{xx}_{c})
\]
on one or a few chosen cross-block pairs $c$ (e.g.\ between the edge qubits of each block). A short SPSA optimisation over the inter-block parameters $\Theta_{\mathrm{inter}}$, then to promote \emph{eigenstate concentration} aligned with LDQPE, we instead optimize a small set of low-order moments
\[
  Z_t^{(i)} := \langle \psi_i \,|\, U_i^{\,t} \,|\, \psi_i \rangle,
  \qquad t\in\mathcal{T},
\]
for a fixed, small power set $\mathcal{T}$ (e.g.\ $\{1,2,4,8\}$), by minimizing
\[
  \mathcal{L}_{\mathrm{mom}}(U_i;\psi_i)
  =
  \sum_{t\in\mathcal{T}} w_t\bigl(1-|Z_t^{(i)}|^2\bigr),
\]
with weights $w_t>0$ (typically nonincreasing in $t$). Intuitively, enforcing $|Z_t^{(i)}|\approx 1$ for multiple powers suppresses the multi-eigenvector pathology and biases $\ket{\psi_i}$ toward a single dominant eigencomponent, which is precisely the regime where LDQPE returns stable dominant-eigenphase features. Starting from block-wise warm starts means that the optimiser only needs to correct a small misalignment introduced by $U_{\mathrm{inter}}$, rather than discover a high-overlap $n$-qubit unitary from scratch. In practice, we find that the optimiser rapidly reaches values $f(\Theta) \ge 1-\delta$ with $\delta$ in the few-percent range, even with a single inter-block layer.

The resulting public unitary $U$ is thus a shallow, expressive circuit consisting of (i) two locally optimised six-qubit brickwork blocks and (ii) a small number of inter-block entanglers. From the point of view of an adversary who only sees the flattened gate list, the circuit exhibits the gate statistics of a generic hardware-efficient ansatz built from $\{R_z, R_x, R_{xx}, \mathrm{CZ}\}$ and does not reveal the hidden signal structure. Honest parties, who know $P$, can always prepare $\ket{\psi}=P^\dagger\ket{0^n}$ and run the low-depth QPE routine described in the next paragraphs to extract a unitary phase at modest cost.

\begin{figure}[htbp!]
    \centering
    \begin{subfigure}[b]{0.55\textwidth}
        \centering
        \includegraphics[width=\textwidth]{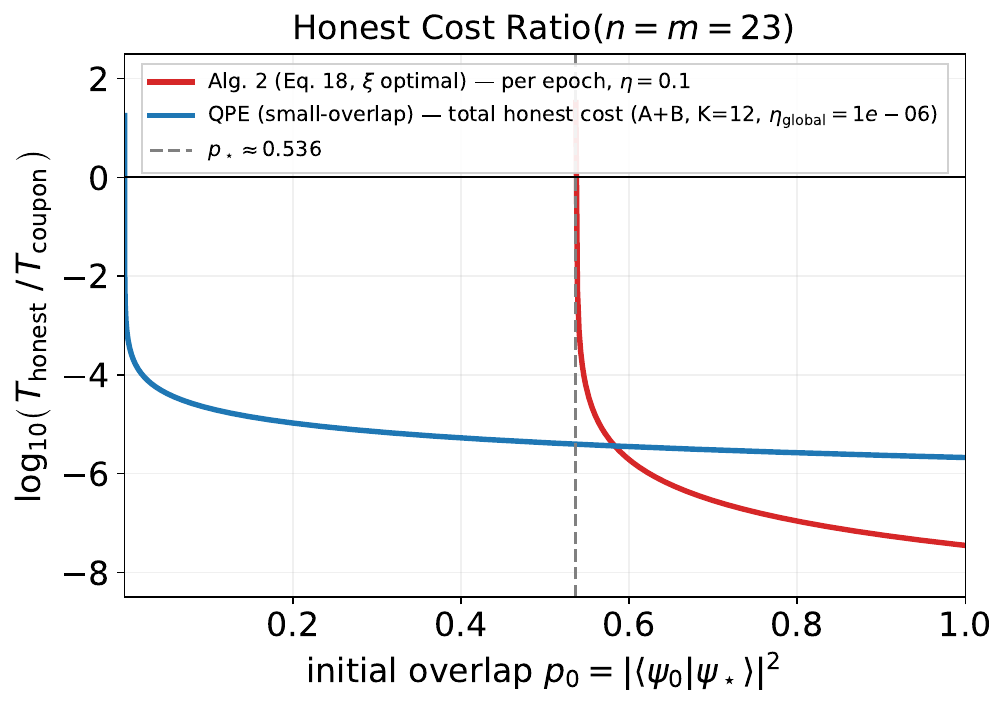}
        \caption{ Ratio of computational cost of honest parties to Eve's attack for $n=m=24$.}
        \label{eveQSA}
    \end{subfigure}
    \caption{}
    \label{fig:cost}
\end{figure}

\emph{Code pointer.} Our reference implementation of this two-block warm-start pipeline is provided in

\texttt{mps\_expressive\_unitary\_optimizer.py}. A typical command-line invocation is
\begin{verbatim}
!python mps_expressive_unitary_optimizer.py --delta 0.05 -n 24 -blocksize 8 -m 8 --layers 1 
--steps 2000 --restarts 200 --seed 12421 --depth_ctrl 1 --plot --plot_pdag --save_pdag Pdag.png
\end{verbatim}

which instantiates three 8-qubit blocks with $R_z$, $R_x$, $R_{xx}$, and CZ gates, maximises the overlap with the separable planted state
$\ket{\psi}=\ket{\psi_1}\ket{\psi_2}\ket{\psi_3}$ on each block using SPSA, and then adds an expressive inter-block entangling layer to maximise the final global overlap. In the present work, we use this compiled $U$ as the public unitary for each QSA unitary; more general partitions into multiple blocks are straightforward.

In the circuit-based, classically evaluated variant QSA-C, we do not rely on physical measurements, but on full state-vector simulation of a low-depth phase-estimation routine. Given the public circuit $U_i$ and the planted state $|\psi\rangle = P^\dagger |0^n\rangle$, the honest algorithm simulates the QPE-style circuit on $U_i$ and computes the Born probabilities over all phase-register bitstrings. QSA-C then \emph{deterministically} outputs the phase bitstring (and hence eigenphase) with maximum probability (with a fixed tie-breaking rule if needed). Thus, even if $|\psi\rangle$ has support on several eigenstates of $U_i$, as long as one eigenphase has strictly larger weight than all others (for example $|\alpha_\star|^2 > 1-\delta_{max}$, and in practice we target $|\alpha_\star|^2 \gtrsim 0.9$ via the compilation pipeline), the extracted eigenphase is a well-defined deterministic function of $(U_i,P)$. In particular, QSA-C does not require a reconciliation layer for key agreement: both honest parties who know $P$ and simulate the same circuit obtain exactly the same eigenphase vector, despite the underlying quantum picture involving a superposition over multiple eigenstates.

We implement the per-unitary protocol using Algorithm~2 of Ni-Li-Ying (low-depth Quantum Phase Estimation, QPE)~\cite{ni2023lowdepthqpe}. Algorithm~2 estimates a dominant eigenphase from power moments
\[
Z_{2^j} \;=\; \bra{\psi}U^{2^j}\ket{\psi}, \qquad j=0,\dots,J,
\]
with $J=\lceil\log_2(\xi/2^{-m})\rceil$ set by a target precision parameter $\xi$ and the desired number of bits $m$. The algorithm then reconstructs the phase by most-significant-bit to least-significant-bit unwrapping of the complex-valued sequence $\{Z_{2^j}\}$.

In a \emph{simulated} evaluation path, we estimate $Z_{2^j}$ by repeatedly applying $U^{(\mathrm{pub})}_e$ to the planted state $\ket{\psi}$ in a state-vector simulator (or, for larger $n$, a tensor-network / MPS simulator) and computing the inner product
\[
Z_{2^j} \approx \langle \psi | U^{2^j}|\psi \rangle
\]
directly from the simulated state. The total cost scales with the chosen circuit depth and the simulator backend; for the shallow, block-wise compiled unitaries used here and moderate $n$, this is dominated by $O(J)$ applications of $U^{(\mathrm{pub})}_e$.

In a \emph{hardware} evaluation path, we estimate each $Z_{2^j}$ via Hadamard tests, i.e., using a single control qubit, a controlled-$U^{2^j}$ on $\ket{\psi}$, and measurements of the control in suitable bases to extract the real and imaginary parts of $\bra{\psi}U^{2^j}\ket{\psi}$. We follow the sampling prescriptions of~\cite{ni2023lowdepthqpe} to choose the number of shots per moment so that the overall phase reconstruction error is below $2^{-m}$ with high confidence. By design, $|\!\langle\psi|U|\psi\rangle\!|^2 \ge 1-\delta$, so $\ket{\psi}$ has large overlap with one eigenvector of $U$; the moment sequence $\{Z_{2^j}\}$ is therefore dominated by a single eigenphase, and the unwrapping procedure recovers $m$ phase bits reliably in both simulated and hardware modes. For this low-depth QPE algorithm, $\delta \le 2\sqrt{3}-3$.

From a hardware perspective, the main challenge is implementing controlled powers of an expressive unitary at low depth. A na\"ive decomposition of controlled-$U^{2^j}$ into native one- and two-qubit gates multiplies the depth roughly by $2^j$ and can amplify coherent errors. In our setting, the QSA parameters are chosen so that (i) the base unitary $U^{(\mathrm{pub})}_e$ is shallow by construction, and (ii) only a small number of powers $2^j$ are required to reach the target precision. Controlled powers can then be realised either as repeated applications of a shallow controlled-$U$ block, or via iterative / semi-classical QPE variants that recycle a single control qubit and avoid large controlled powers altogether. For the $n$ and $m$ regimes we target, the resulting gate counts and depths are compatible with near-term devices, while still being too demanding to support brute-force spectral attacks across many independent seeds. Shown in Fig. \ref{fig:cost} by the red line is the ratio of the low-depth QPE cost of honest parties to the cost of Eve's full-spectrum QPE attack (Appendix Attack A.2). We note that for the previous example invocation, $\delta=0.05$, which means the initial overlap is $p_0=0.95$, meaning this cost ratio is approximately $10^{-8}$. 

Per unitary, the recovered $m$ bits are concatenated across $k=\lceil L/m\rceil$ unitaries to reach total length $\ge L$ and then hashed to a session key using a classical Key Derivation Function (KDF), e.g.\ an HMAC-based KDF (HKDF) with extract-expand. The only public artifact is the flattened circuit for $U^{(\mathrm{pub})}_e$; honest evaluation can proceed either by classical simulation of QPE on $\ket{\psi}$, which is also shown (blue line) in Fig. \ref{fig:cost}, or by executing this low-depth QPE routine on quantum hardware.

\end{appendix}

%\end{appendix}

\section{Teleported QSA variant}
\label{app:teleported-qsa}
\label{subsec:teleported-qsa}
\begin{figure}[ht!]
\centering
\resizebox{0.75\textwidth}{!}{%
\begin{quantikz}[row sep=0.32cm, column sep=0.1cm]
\lstick{$\ket{0}$}
  & \gate[wires=2]{P^\dagger}
  & \qw{}
  & \ctrl{2}
  & \gate{H}
  & \meter{} \vcw{4}
  & \setwiretype{n}
  & 
  & 
  & 
  & 
  &
  &
  &
\\
\lstick{$\ket{0}$}
  & \qw{}
  & \linethrough
  & \linethrough
  & \linethrough
  & \qw{}
  & \ctrl{2}
  & \gate{H}
  & \meter{} \vcw{4}
  & \setwiretype{n}
&
&
&
&
\\
\lstick{$\ket{0}_{A_1}$}
  & \gate{H}
& \ctrl{2}
  & \targ{}
  & \meter{} \vcw{2}
  & \setwiretype{n}
  & 
  & 
  & 
  & 
&
&
&
&
\\
\lstick{$\ket{0}_{A_2}$}
  & \gate{H}
  & \linethrough
  & \ctrl{2}
  & \qw
   & \qw
  & \targ{}
  & \meter{} \vcw{2}
  & \setwiretype{n}
  & 
 &
 &
 &
 &
\\
\lstick{$\ket{0}_{B_1}$}
  & \qw
  & \targ{}
   & \qw
  & \gate{X^{b_1}}
  & \gate{Z^{a_1}}
  & \qw
  & \qw
  & \qw
  & \qw
  & \gate[wires=2]{U^{2^j}}
  &\setwiretype{n}
  &
  &
\\
\lstick{$\ket{0}_{B_2}$}
  & \qw
  & \qw
  & \targ{}
   & \qw
  & \qw
  & \qw
  & \gate{X^{b_2}}
  & \gate{Z^{a_2}}
  & \qw
  & \qw
  & \setwiretype{n}
  &
  &
\\
\lstick{$\ket{0}_{\mathrm{anc}}$}
  & \qw
  & \qw
  & \qw
  & \qw
   & \qw
  & \qw
  & \qw
  & \qw
  & \gate{H}
  & \ctrl{-2}
  & \gate{I/S^\dagger}
  & \gate{H}
  & \meter{}
\end{quantikz}%
}
\caption{\textbf{Teleported QSA for the \(n=2\) case.}
Alice prepares the planted state \(\ket{\psi}=P^\dagger\ket{00}\) and teleports its two qubits to Bob using two shared Bell pairs, with Alice holding \(A_1,A_2\) and Bob holding \(B_1,B_2\). For each qubit \(i\), Alice performs a Bell-basis measurement by applying a CNOT from the data qubit to \(A_i\), then a Hadamard on the data qubit, followed by computational-basis measurements. The two classical bits \((a_i,b_i)\) are sent to Bob, who applies the Pauli correction \(X^{b_i}Z^{a_i}\) to \(B_i\). After both corrections, Bob holds a copy of \(\ket{\psi}\). The public challenge \(U\) is supplied classically, and Bob evaluates the LDQPE moment \(Z_j=\langle\psi|U^{2^j}|\psi\rangle\) using a Hadamard test, with the ancilla controlling the two-qubit operation \(U^{2^j}\). The ancilla measurement estimates \(\Re Z_j\), or \(\Im Z_j\) when the optional phase gate \(S^\dagger\) is inserted before the final Hadamard. The same protocol extends directly to arbitrary \(n\): for an \(n\)-qubit planted state \(\ket{\psi}=P^\dagger\ket{0}^{\otimes n}\), Alice teleports each qubit using one shared Bell pair and a local Bell-basis measurement, and Bob applies the corresponding single-qubit Pauli corrections \(X^{b_i}Z^{a_i}\) to recover the full \(n\)-qubit state, including all internal entanglement. Each teleported \(n\)-qubit shot consumes \(n\) Bell pairs. Estimating both real and imaginary parts for all \(m\) moments with \(N_s\) shots requires \(2N_snm\) Bell pairs per challenge instance, or \(2N_snmk\) Bell pairs across \(k\) independent challenge instances; for the symmetric compiler, the Methods fast path replaces the Hadamard test with a single \(V^\dagger\)-and-measure per instance, lowering this to \(\approx nk\) (the teleported variant is state mode, \(\delta=0\)).}
\label{fig:teleported-qsa-n2}
\end{figure}
A natural distributed variant of QSA arises when Alice and Bob already share entanglement in the form of Bell pairs. Suppose Alice holds the planted state
$\ket{\psi}=P^\dagger\ket{0^n}$
while Bob is the party that evaluates the public unitary challenge. Instead of requiring Bob to prepare \(\ket{\psi}\) locally, Alice can teleport the state to Bob using \(n\) shared Bell pairs. For each qubit of \(\ket{\psi}\), Alice performs a Bell-basis measurement by applying a CNOT from the data qubit to her half of the Bell pair, followed by a Hadamard on the data qubit and computational-basis measurements of both qubits. If the resulting classical bits are \(a_i,b_i\in\{0,1\}\), then Bob applies the Pauli correction \(X^{b_i}Z^{a_i}\) to his half of the corresponding Bell pair. After all \(n\) qubits are teleported and corrected, Bob holds \(\ket{\psi}\) and can evaluate the QSA challenge exactly as in the local version.

For the LDQPE-style evaluation, Bob applies a Hadamard test for
$
Z_j=\langle \psi|U^{2^j}|\psi\rangle,$
where the public challenge \(U\) is sent classically and the required controlled power \(U^{2^j}\) is implemented on Bob's reconstructed state. Estimating both \(\Re Z_j\) and \(\Im Z_j\) for \(j=0,\dots,m-1\) with \(N_s\) shots per setting requires, in the straightforward teleported implementation, a Bell-pair budget of
$
N_{\mathrm{Bell}} = 2N_snm
$
per challenge instance. Across \(k\) independent challenge instances, this becomes
$N_{\mathrm{Bell}}^{\mathrm{tot}} = 2N_snmk.$
The prefactor \(2\) accounts for separate real- and imaginary-part estimation in the Hadamard test.

For the symmetric compiler, Bob may instead run the honest fast path of Methods: he applies \(V^\dagger\) to the teleported copy and measures in the computational basis to recover the label \(b\), with no Hadamard test and no power moments. The teleported variant is a state-mode deployment by construction: Bob holds a delivered physical register rather than a seed, so the planted state satisfies \(\ket{\psi}=V\ket{b}\) with \(\delta=0\), and \(V^\dagger\ket{\psi}=\ket{b}\) exactly, so a single teleported copy recovers \(b\) in one measurement. Removing the factor of \(2\) (no separate real/imaginary estimation) and the factor of \(m\) (no per-moment circuits) then reduces the budget to \(N_{\mathrm{Bell}}^{\mathrm{fast}}=n\) Bell pairs per instance, and across \(k\) instances
\[
N_{\mathrm{Bell}}^{\mathrm{fast,tot}} = n\,k,
\]
up to a small constant overhead if a few extra shots are taken to suppress hardware readout error. For the representative choice \(k=36\), \(n=m=8\), this lowers the LDQPE figure of \(\simeq 500{,}000\) Bell pairs to \(\approx nk\sim 300\). We stress that this saving still requires Bob to hold and measure a genuine quantum register. Sending the phase, or an \(m\)-bit \(R_z\) description of it, as classical data would remove the teleportation cost entirely, but also the quantum evaluation, so acceptance would no longer evidence a QPU and the protocol would reduce to a classical challenge-response, forfeiting the evaluability and the state-mode no-copy security that are the entire point of delivering the register (Table~\ref{tab:compare}).

\end{document}